\documentclass[aps,prx,twocolumn,superscriptaddress,longbibliography]{revtex4-2}

\usepackage{graphicx}
\usepackage{dcolumn}
\usepackage{bm}
\usepackage{siunitx}
\usepackage{xcolor}
\usepackage{mathtools,amssymb}
\usepackage{dsfont}
\usepackage[capitalize]{cleveref}
\usepackage{upgreek}
\usepackage{comment}
\usepackage{physics}
\usepackage[normalem]{ulem}
\usepackage[utf8]{inputenc}

\newcommand{\added}[1]{\textcolor{black}{#1}}

\begin{document}

\preprint{APS/123-QED}

\title{Laser-induced spectral diffusion and excited-state mixing of silicon T centres}%

\author{Camille Bowness}
\affiliation{Simon Fraser University, Department of Physics, Burnaby, British Columbia, Canada}%
\affiliation{Photonic Inc., Coquitlam, British Columbia, Canada}%
\altaffiliation{These authors contributed equally.}
\author{Simon A. Meynell}
\affiliation{Simon Fraser University, Department of Physics, Burnaby, British Columbia, Canada}%
\affiliation{Photonic Inc., Coquitlam, British Columbia, Canada}%
\altaffiliation{These authors contributed equally.}
\author{Michael Dobinson}
\affiliation{Simon Fraser University, Department of Physics, Burnaby, British Columbia, Canada}%
\affiliation{Photonic Inc., Coquitlam, British Columbia, Canada}%
\author{Chloe Clear}
\affiliation{Photonic Inc., Coquitlam, British Columbia, Canada}%
\author{Kais Jooya}
\affiliation{Simon Fraser University, Department of Physics, Burnaby, British Columbia, Canada}%
\affiliation{Photonic Inc., Coquitlam, British Columbia, Canada}%
\author{Nicholas Brunelle}
\affiliation{Simon Fraser University, Department of Physics, Burnaby, British Columbia, Canada}%
\affiliation{Photonic Inc., Coquitlam, British Columbia, Canada}%
\author{Mehdi Keshavarz}
\affiliation{Simon Fraser University, Department of Physics, Burnaby, British Columbia, Canada}%
\affiliation{Photonic Inc., Coquitlam, British Columbia, Canada}%
\author{Katarina Boos}
\affiliation{Walter Schottky Institut, Department of Electrical and Computer Engineering and MCQST, Technische Universität München, 85748 Garching, Germany}
\author{Melanie Gascoine}
\affiliation{Simon Fraser University, Department of Physics, Burnaby, British Columbia, Canada}%
\affiliation{Photonic Inc., Coquitlam, British Columbia, Canada}%
\author{Shahrzad Taherizadegan}
\affiliation{University of Calgary, Department of Physics and Astronomy, Calgary, Alberta, Canada}
\affiliation{Institute for Quantum Science and Technology, University of Calgary, Calgary, Alberta, Canada}%
\author{Christoph Simon}
\affiliation{University of Calgary, Department of Physics and Astronomy, Calgary, Alberta, Canada}
\affiliation{Institute for Quantum Science and Technology, University of Calgary, Calgary, Alberta, Canada}%
\author{Mike L. W. Thewalt}
\affiliation{Simon Fraser University, Department of Physics, Burnaby, British Columbia, Canada}%
\author{Stephanie Simmons}
\affiliation{Simon Fraser University, Department of Physics, Burnaby, British Columbia, Canada}%
\affiliation{Photonic Inc., Coquitlam, British Columbia, Canada}%
\author{Daniel B. Higginbottom}
\affiliation{Simon Fraser University, Department of Physics, Burnaby, British Columbia, Canada}%
\affiliation{Photonic Inc., Coquitlam, British Columbia, Canada}%
\altaffiliation{Corresponding author: daniel\_higginbottom@sfu.ca}

\date{\today}

\begin{abstract}

To find practical application as photon sources for entangled optical resource states or as spin-photon interfaces in entangled networks, semiconductor emitters must produce indistinguishable photons with high efficiency and spectral stability. 
Nanophotonic cavity integration increases efficiency and bandwidth, but it also introduces environmental charge instability and spectral diffusion. 
Among various candidates, silicon colour centres have emerged as compelling platforms for integrated-emitter quantum technologies. 
Here we investigate the dynamics of spectral wandering in nanophotonics-coupled, individual silicon T centres using  spectral correlation measurements. 
We observe that spectral fluctuations are driven predominantly by the near-infrared excitation laser, consistent with a power-dependent Ornstein-Uhlenbeck process, and show that the spectrum is stable for up to $1.5$~ms in the dark.
We demonstrate a $35\times$ narrowing of the emitter linewidth to $110$~MHz using a resonance-check scheme and  discuss the advantage for pairwise entanglement rates and optical resource state generators. 
Finally, we report laser-induced spin-mixing in the excited state and discuss potential mechanisms common to both phenomena. 
These effects must be considered in calibrating T centre devices for high-performance entanglement generation. 
\end{abstract}

\maketitle

Semiconductor point defects are a versatile platform for quantum technologies \cite{Wolfowicz2021}; they can function as single-photon sources for quantum communication and as resource state generators for optical quantum computers \cite{Michaels2021}. 
If the emitter couples single photons to a local spin qubit, then it may also function as a spin-photon interface (SPI) with applications including quantum repeater networks \cite{Stas2022robust,Knaut2024EntanglementNetwork} and hybrid spin-photon quantum computing \cite{Simmons_2024_ScalableFaultTolerant,Wei2024UniversalDistributed,Gliniasty2024SpinOptical}. 
Semiconductor emitters may be integrated directly into nanophotonic optical resonators to enhance their emission rate and coherence through the Purcell effect \cite{Gritsch2023Purcell,Merkel2020CoherentPurcell}. 
In turn, these optical resonators can be connected by photonic integrated circuits to prepare large entangled states on-chip \cite{Siew2021ReviewSilicon}.  
However, integration also exposes the emitter to material interfaces and crystal damage that produce charge noise and cause the emission frequency to fluctuate in time \cite{Wolters_2013_UltrafastSDDiamondNV, Heiler_2024_SiCSDthicknessdependence}. 
Such spectral diffusion (SD, also called spectral wandering) is a critical challenge for practical quantum technologies.

Recently, silicon colour centres have emerged as an intriguing category of emitters harnessing the advanced material and nanofabrication capabilities of silicon \cite{Higginbottom_2022_OpticalSpinsInSilicon,Johnston_2024_CavityCoupledTcentre,Lefaucher2024BrightSingle,Redjem2024AllSilicon,Saggio2024CavityEnhanced,komza_2025_multiplexedcolorcenterssilicon}. 
The silicon T centre, for example, is the focus of ongoing commercial efforts to build a spin-photon quantum computer \cite{Simmons_2024_ScalableFaultTolerant,Afzal2024}. 
Silicon combines global microelectronics and nanophotonics foundry capabilities with extremely high chemical and isotopic purities \cite{Saeedi2013}. 
Owing to the mass homogeneity, strain-free environment, and high-quality lattice of their ``semiconductor vacuum” host, the optical transitions of silicon colour centres in bulk $^{28}$Si crystals can have remarkably low inhomogeneous broadening and spectral diffusion.
Popular silicon colour centres including C and T have demonstrated inhomogeneous linewidths below $60$~MHz and the G centre can be nearly lifetime-limited \cite{Chartrand_2018_CGW, Bergeron:2020_PRX}. 
However, spectral diffusion has proven to be significant in the first generation of silicon colour centre devices \cite{Higginbottom_2022_OpticalSpinsInSilicon,DeAbreu_2023_WaveguideIntegratedCenters, Afzal2024,komza_2025_multiplexedcolorcenterssilicon} and increasing the spectral stability closer to their performance in bulk is a critical challenge. 

Interference-based entanglement protocols require spectrally matched, stable emitters \cite{Barrett2005,Cabrillo1999}.
Resonant excitation using a shared, narrow-linewidth laser acts as a frequency filter, selecting only emitters that are spectrally aligned.
However, spectral diffusion reduces the probability of such alignment by a factor of $\approx \eta_\mathrm{SD}^2$, where $\eta_\mathrm{SD} = \Gamma_\mathrm{hom}/\Gamma_\mathrm{SD}$ is the ratio of the SD and homogeneous linewidths.
For example, SD decreases the entanglement rate between remote T centre devices by three orders of magnitude in Ref.~\cite{Afzal2024}. 
Techniques for reducing or mitigating spectral diffusion have been widely explored for other semiconductor emitters. 
Resonance-check (RC) techniques can confirm both the charge state of diamond colour centres (e.g. NV, SiV and SnV) and mutual resonance between remote emitters \cite{Brevoord_2024_heraldinitializationDiamondSnV,Hermans2023EntanglingRemote}. 
In wide-bandgap semiconductors, spectral diffusion is relatively slow ($\sim$s) \cite{Schmidgall_2018_frequencyControlNVSD,Acosta_2012_DynamicStabilizationNVcenterOpticalResonances}, and hundreds of entanglement attempts may follow a single RC \cite{Pompili2021a}. 
In contrast, resonantly driven silicon colour centres exhibit emission correlations typical of short SD timescales of $5$--$10$~$\upmu$s \cite{Higginbottom_2022_OpticalSpinsInSilicon,Afzal2024}.

\begin{figure}[ht]
  \centering
  \includegraphics[width = 3.4in]{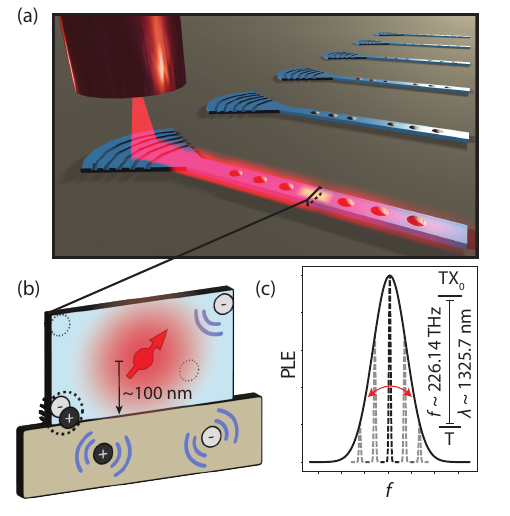}
  \caption{Spectral diffusion in integrated colour centre devices. 
  (a) A 1D silicon photonic crystal cavity is addressed by a resonant laser through an above-chip optical fibre and a grating coupler. 
  (b) Cross-sectional view showing an integrated T centre at the device centre. 
  The laser resonates in the single-mode cavity (red). 
  Charge traps at surfaces, interfaces, impurities, and vacancies within the silicon device layer and oxide insulator generate electric fields that perturb the colour centre emission frequency. 
  \added{We expect the closest interface to a T centre in a cavity to be less than \SI{100}{\nano\meter} away.}
  (c) An illustration of spectral diffusion. 
  The T centre emission frequency fluctuates as the charge environment is reconfigured. }
  \label{fig:intro}
\end{figure}

In this paper, we investigate spectral diffusion processes in single, cavity-coupled silicon T centres, shown schematically in \cref{fig:intro}. 
We show through spectrally-resolved two-photon correlations in the device luminescence that spectral diffusion of the TX$_0$ transition is driven predominantly by the interaction of the NIR laser and the material. 
We show that resonance checks can dramatically narrow the effective linewidth of T centres in devices, and that they remain effective for up to \SI{1.5}{\milli\second} of wait-time in the dark. 
We explore the conditions under which they can achieve a practical speedup for entangled state generation. 
Finally, we investigate the effect of NIR laser on the spin of the optically-excited bound-exciton state TX$_0$. 
Together, these results illustrate the mechanism of laser-induced spectral diffusion and demonstrate a promising approach to mitigate spectral diffusion in T centre devices. 
These results have immediate relevance for near-term T centre quantum processors and network demonstrations. 

\section{\label{sec:spectraldiffusion} Spectral diffusion of device-integrated T centres}

\begin{figure}[t!]
  \centering
  \includegraphics[width = 3.4in]{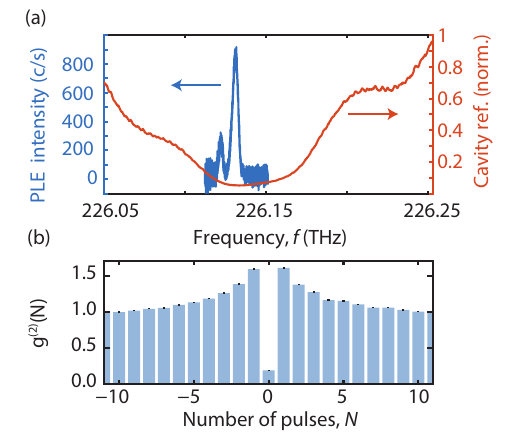}
  \caption{A representative integrated T centre device (T Centre II). 
  (a) The photoluminescence excitation (PLE) spectrum at zero magnetic field (blue), and the cavity spectrum measured in reflection (orange). 
  (b) Correlation measurements of the centre luminescence under pulsed resonant excitation exhibit a high degree of antibunching with $g^{(2)}(0) = \SI{0.186\pm0.004}{} \ll1$, confirming the emission is dominated by a single centre. 
  $g^{(2)}(N)$ shows a positive correlation at low pulse separation number $N$, \added{consistent with }spectral diffusion \cite{Sallen_2010_subnanosecondSDquantumdots}.}
  \label{fig:spectra}
\end{figure}

\begin{figure*}[t!]
  \centering
  \includegraphics[width=6.8in]{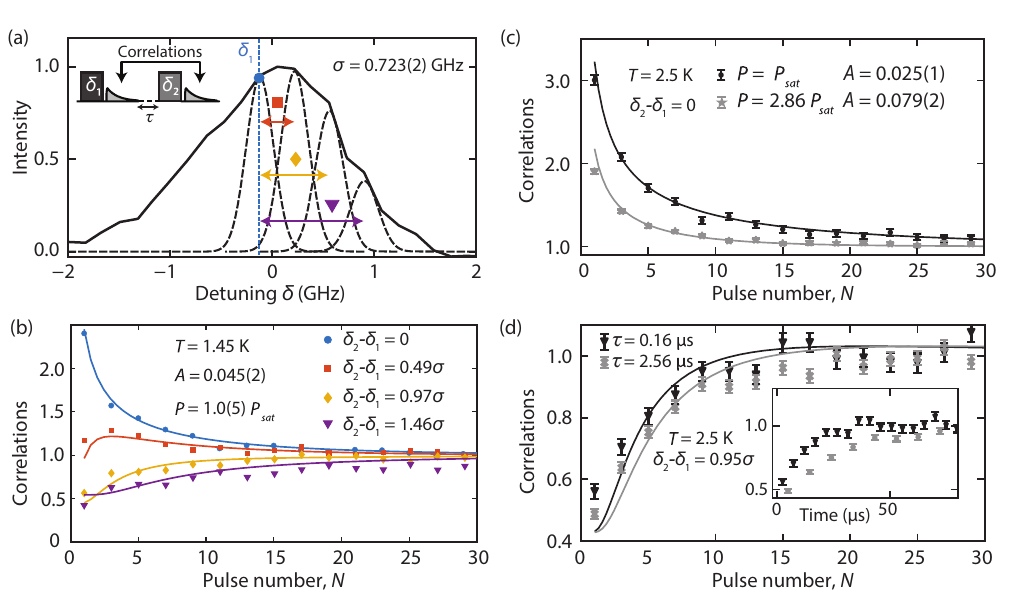}
  \caption{Two-colour correlation measurement.
  (a) Laser detunings $\delta_{1,2}$ from the centre of the inhomogeneous line (solid) and modelled homogeneous lines (dashed). 
  The vertical blue line indicates $\delta_1$.
  (Inset) Pulse sequence showing consecutive excitation pulses at detunings $\delta_{1,2}$. 
  Correlations are calculated between photons emitted in detection windows indicated by the grey luminescence transients. 
  (b) Correlations between frequencies $\delta_{1,2}$ as a function of the number of intermediate pulses. 
  Solid lines display a joint single-parameter ($A$) fit to an O-U model of spectral diffusion.
  (c) Power dependence of the two colour correlation for two different powers. 
  The correlation amplitude reduces with higher power.
  (d) Dark-time dependence of the two-colour correlation. 
  The curves show closer agreement when plotted as a function of pulse number compared to when they are plotted as a function of time (inset).
}
  \label{fig:spectraldiffusion}
\end{figure*}

We characterize the spectral diffusion dynamics of integrated, single T centres with and without near-infrared (NIR) excitation. 
The T centre is a silicon colour centre combining telecommunications-band optical emission, a ground-state electron spin qubit, and an intrinsic register of 1--3 nuclear spin qubits depending on isotopic composition \cite{Bergeron:2020_PRX}. 
Since it was first proposed for quantum technologies in 2020, it has been integrated with silicon nanophotonics \cite{Higginbottom_2022_OpticalSpinsInSilicon,Lee2023HighEfficiency}, Purcell-enhanced in nanocavities \cite{Islam2023cavityenhanced,Johnston_2024_CavityCoupledTcentre,Afzal2024,komza_2025_multiplexedcolorcenterssilicon,Dobinson_2025_electricallytriggeredspinphotondevicessilicon}, and embedded in p-i-n diode devices for electrical control \cite{Dobinson_2025_electricallytriggeredspinphotondevicessilicon,Day_2025_probingnegativedifferentialresistance}.
Moreover, these devices exhibit strong qubit performance: one- and two-qubit gates have been realized using the electron and hydrogen nuclear spins; nuclear coherence times exceeding 200~ms have been demonstrated; and indistinguishable emission has been observed between devices in separate cryostats \cite{Afzal2024}. 
In high purity, isotopically-enriched bulk $^{28}$Si samples the T centre's inhomogeneous linewidth (including static spatial inhomogeneity and SD) is as low as $33$~MHz \cite{Bergeron:2020_PRX}, just 15 times the lifetime-limit of Purcell-enhanced T centres in devices \cite{Afzal2024}. 
However, the typical SD linewidth of integrated T centres observed to date is much larger, $0.4$--$4$~GHz \cite{Higginbottom_2022_OpticalSpinsInSilicon,MacQuarrie_2021_GeneratingT,Johnston_2024_CavityCoupledTcentre,komza_2025_multiplexedcolorcenterssilicon}. 

The centres in our study are integrated in 1D silicon photonic crystal cavities on a silicon-on-insulator chip and resonantly excited to the bound-exciton ground state TX$_0$. 
We report results from two devices, T Centre I and II, with moderate cavity quality factors $Q_\mathrm{I}=$880(40) and $Q_\mathrm{II}=$1,743(2) such that the cavities can be excited on- or off-resonance from the T centres.
The devices are operated at either\SI{0.27}{\kelvin}-- \SI{0.4}{\kelvin} or \SI{1.45}{\kelvin}--\SI{2.5}{\kelvin} in one of two cryostats, and addressed by cryogenic optical fibre arrays (see Methods for more details). 
\Cref{fig:spectra} shows the photoluminescence excitation (PLE) spectrum and $g^{(2)}$ of T Centre II.
Details of the devices reported in this study are provided in the Supplementary Material (SM)~\cite{Supplement}.

\subsection{\label{subsec:twocolourcorrelation}Two-colour correlation}

We measure spectral dynamics using frequency-resolved photon correlation measurements. 
For this experiment, we choose a device (T Centre I) with a PLE full width at half maximum linewidth $\Gamma = \SI{1.694(5)}{\giga\hertz}$ (standard deviation $\sigma = \SI{0.723(2)}{\giga\hertz}$). 
Spectral hole-burning measurements of the homogeneous linewidth, $\Gamma_\text{hom}$ (see SM \cite{Supplement}), confirm a temporal inhomogeneous broadening factor of $1/\eta_\mathrm{SD} = \Gamma/\Gamma_\text{hom} = \SI{100\pm15}{}$.
With the sample at \SI{1.45}{\kelvin}, we detune two sub-MHz linewidth excitation lasers from the PLE peak by frequencies $\delta_1$ and $\delta_2$. 
We measure correlations between photons emitted in detection bins corresponding to each of the two excitation frequencies, and separated by inter-pulse delay $\tau = \SI{160}{\nano\second}$, using the pulse sequence in \cref{fig:spectraldiffusion}(a).
Each pulse is $\SI{900}{\nano\second}$ in duration and followed by a \SI{1375}{\nano\second} collection window.
We fix $\delta_1$ at the centre frequency and vary $\delta_2$ across four values shown in \cref{fig:spectraldiffusion}(a). 
When the detuning difference $\delta_2 - \delta_1 = 0$ (\emph{i.e.}, single laser PLE), detections in consecutive windows are positively correlated, as observed in the $g^{(2)}$ measurement. 
The emitter has a spectral diffusion time larger than the pulse sequence length, and if it is resonant for the first pulse it is more likely to be resonant for the second. 
On the other hand, if the two lasers are separated by $0.97 \,\sigma$, emission into consecutive time bins is anticorrelated, since a finite time is required for the emitter to wander from frequency $\delta_1$ to $\delta_2$. 
We model the spectral diffusion as an Ornstein-Uhlenbeck (O-U) process active during the excitation pulse (solid lines in \cref{fig:spectraldiffusion}(b)--(d)), parametrized solely by the SD rate, $\alpha$, and the ratio of background counts to total counts, $\beta$ \cite{Delteil_2024_photonstatisticsdiffusiveemittersOUMarkov}.
We use a closed form expression for the correlations and connect it to a discrete charge microscopic model (details in SM~\cite{Supplement}).
We obtain $\beta$ by comparing the uncorrelated PLE count rate as a function of frequency to our background rate. 
Finally, we extract the SD rate $\alpha \added{= \SI{0.050(2)}{\per\micro\second}}$ by jointly fitting the correlation data taken at \SI{1.45}{\kelvin} across multiple detuning values, $\delta_2 - \delta_1 = {0, 0.49\sigma, 0.97\sigma, 1.46\sigma}$ and extract a per-pulse rate of $A = \int_\mathrm{pulse}\alpha\, \dd t= 0.045(2)$. \added{In the short time limit, $\alpha$ can be understood as an effective diffusion constant \cite{Supplement}}.
We repeat the measurements at \SI{2.5}{\kelvin} and find a similarly good match to the model (see SM \cite{Supplement}). 

To test the SD mechanism, we next investigate how the correlation measurements depend on laser power (\cref{fig:spectraldiffusion}(c)). 
Here, we keep $\delta_2 - \delta_1 = 0$ and change between two powers, $P_\text{sat}$ and $2.86P_\text{sat}$ where $P_\text{sat}$ is the calibrated saturation power. 
We observe that the correlation amplitude drops with increasing power, consistent with an increase in SD rate.
Hole-burning measurements (see SM \cite{Supplement}) verify that the homogeneous linewidth is effectively constant over this power range. 
The O-U model (solid lines) agrees well with the ratio of fitted charge reconfiguration rates, $A(P_\text{sat})/A(2.86P_\text{sat}) = \SI{0.025(1)}{}/\SI{0.079(2)}{}=0.32(2)$, matching the ratio of powers ($0.35$), which suggests that the fraction of charges reconfigured per pulse scales with the incident intensity. 
We further observe that adding a near-resonant laser has the same effect as increasing the resonant power (see SM \cite{Supplement}), indicating that the effect is consistent with a broadband interaction with the host material. 

Finally, we increase the delay $\tau$ between the two-colour pulses from $\tau = \SI{160}{\nano\second}$ to $\SI{2560}{\nano\second}$. 
The resulting correlation curves, shown in \cref{fig:spectraldiffusion}(d), differ when plotted as a function of absolute time but show close agreement when the data are plotted as a function of pulse number, indicating that spectral wandering is negligible during the dark inter-pulse time. 
Fitting each dataset with a fixed value of $\beta$, we extract $A(\tau = \SI{160}{\nano\second}) = \SI{0.024(2)}{}/\text{pulse}$ and $A(\tau = \SI{2560}{\nano\second}) = \SI{0.019(2)}{}/\text{pulse}$.
That $A$ scales with pulse number and power, rather than time, indicates that the diffusion is driven by the energy delivered by each excitation pulse.
This is consistent with the findings of van de Stolpe \emph{et al.}, which identified laser-induced spectral diffusion in emitters in silicon carbide \cite{vandeStolpe_2025_checkprobeSiC}, and analogous behaviour observed in SnV centres \cite{Gorlitz_2022_SnVSDunderlaserillumination}.
The slow spectral diffusion of the T centre in the absence of optical excitation suggests a strategy for multi-emitter alignment: after a successful resonance check, waiting in the dark may preserve the emitter's spectral position long enough to synchronize multiple T centres.

\subsection{\label{subsec:resonancecheck} Resonance-check spectroscopy}
\begin{figure}[b!]
  \centering
  \includegraphics[width=3.4in]{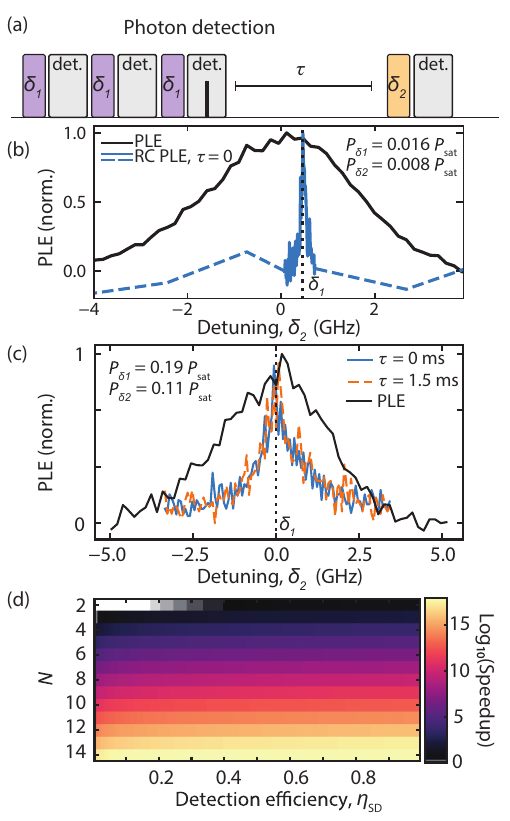}
  \caption{Resonance check (RC) PLE scheme.
  (a) The experimental pulse sequence. 
  We excite at check frequency $\delta_1$ until a photon is detected, which triggers a probe pulse at frequency $\delta_2$ after delay $\tau$.
  (b) \added{Background subtracted} RC PLE \added{(solid blue: high resolution, dashed blue: low resolution)} compared to regular PLE \added{(black)} showing the linewidth narrowing to $\Gamma = \SI{110\pm10}{\mega\hertz}$.
  (c) RC PLE with dark wait time, $\tau$.
  We measure broadening in the absence of laser excitation as a function of wait time and observe no significant broadening out to $\SI{1.5}{\milli\second}$.
  (d) Simulated $N$-qubit entangled state preparation using parallel RC-PLE, with parameters corresponding to T centre II.
  The speedup, defined as $\tau^*/\tau_{\textrm{RC}}$, where $\tau_{\textrm{RC}}$ is the time required to obtain $N$ resonant photons from $N$ independent T centres prepared with RC PLE and $\tau^*$ is the time it takes to obtain $N$ resonant photons from $N$ independent T centres prepared conventionally. 
  White indicates the region where no speedup is expected.
  }
  \label{fig:figure3_checkprobespectroscopy}
\end{figure}
We investigate a RC scheme \cite{Brevoord_2024_heraldinitializationDiamondSnV,vandeStolpe_2025_checkprobeSiC} as an initialization step for preceding resource state generation or bipartite entanglement attempts. 
We choose a second device on the same chip (T centre II, see SM \cite{Supplement}) at temperature $T=$ \SI{0.4}{\kelvin} and apply a series of pulses detuned by $\delta_1$ to prepare the T centre at a target optical frequency.
Following a photon detection that heralds resonance with the `check' laser, we proceed with the target application.
In this case, we perform PLE measurements conditioned on the resonance check by sending a low-power `probe' pulse at frequency $\delta_2$ after a delay $\tau$, as illustrated in \cref{fig:figure3_checkprobespectroscopy}(a). 
If SD between the check and probe pulses is negligible, the RC-conditioned PLE linewidth measured by the probe pulse should be narrower than the raw PLE spectrum. 
Indeed, at the lowest-power setting, check (probe) power  $P_{\delta_1} = 0.016P_\textrm{sat}$ ($P_{\delta_2} = 0.008P_\textrm{sat}$), we observe a linewidth of \SI{110(10)}{\mega\hertz}, $35\times$ narrower than the raw PLE linewidth  as shown in \cref{fig:figure3_checkprobespectroscopy}(b). 
Following the resonance check, the photon-detection probability increases by $6\times$ (see SM \cite{Supplement}). 
\added{It is possible that with improvements in detection efficiency, an improvement in resonance check linewidths may be realized by increasing the number of prerequisite successful check pulses~\cite{vandeStolpe_2025_checkprobeSiC}.}

To test for spectral diffusion in the absence of excitation, we vary the dark interval $\tau$ and monitor the RC linewidth at low temperatures, $T = \SI{0.27}{\kelvin}$. 
To improve count rates, we increase the powers for the check and probe lasers to $P_{\delta_1} = 0.19 P_\textrm{sat}$ and $P_{\delta_2} = 0.11P_\textrm{sat}$.
At these powers, the $\tau=0$ RC-conditioned linewidth is \SI{1.2(1)}{\giga\hertz} and shows no significant broadening out to $\tau\SI{ = 1.5}{\milli\second}$ (\Cref{fig:figure3_checkprobespectroscopy}(c)). 
However we observe a small increase in linewidth at $\tau \gtrsim\SI{2.5}{\milli\second}$ (see the SM \cite{Supplement}).
Independent T centre devices can therefore be prepared by resonance checks, and remain stable for milliseconds---sufficient time to prepare additional devices.

RC schemes offer considerable advantages in scaling SPIs to generate large optical resource states \cite{Azuma_2015_AllPhotonicQuantumRepeaters,Hilaire_2021_ErrorCorrectingEntanglement,Zhang_2022_AllPhotonicRepeater} by preparing $N$ qubits with identical optical frequencies. 
Interference-based remote entanglement schemes such as the Barrett-Kok protocol \cite{Barrett2005}, demonstrated with T centre devices in Ref.~\cite{Afzal2024}, require $N=2$ spectrally indistinguishable qubits, but optical resource state generator schemes can require many simultaneously resonant emitters. 
In the example scheme proposed by Wein \emph{et al.}, $12$ SPIs are first entangled by pairwise operations, before all $N=12$ are required to emit simultaneously at the target frequency \cite{Wein_2024_minimizingresourceoverheadfusionbased}. 

We derive the expected number of attempts $T$ required to prepare $N$ qubits by RC
\begin{equation}
    \mathds{E}[T] = \sum_{j = 1}^N (-1)^{j + 1} {N\choose j} \frac{1}{1 - (1 - \eta_{\mathrm{SD}}\eta_{\mathrm{det}})^j} \,,
\end{equation}
where $\eta_{\mathrm{det}}$ is the detection efficiency.
If the effective linewidth ratio under the RC scheme is given by $\eta^{\prime}_\mathrm{SD}$, the resulting speedup in preparing $N$ qubits is

\begin{equation}
    \mathrm{Speedup} = \frac{\left(\eta^{\prime}_\mathrm{SD} / \eta_\mathrm{SD}\right)^N}{1 + \mathds{E}[T]} \,,
\end{equation}
which is shown as a function of $N$ and $\eta_\mathrm{det}$ in \cref{fig:figure3_checkprobespectroscopy}(d). 
Even for small $N$, the improvement can be many orders of magnitude. Moreover, the RC scheme also offers benefits in protocols that involve long spin control sequences or sequential photon generation, where each step depends on the success of prior emission events \cite{Borregaard_2020_OneWayQuantumRepeater}. 

\section{\label{sec:lasermixing}Laser-induced excited state spin mixing}
\begin{figure}[b t h]
  \centering
  \includegraphics[width=3.4in]{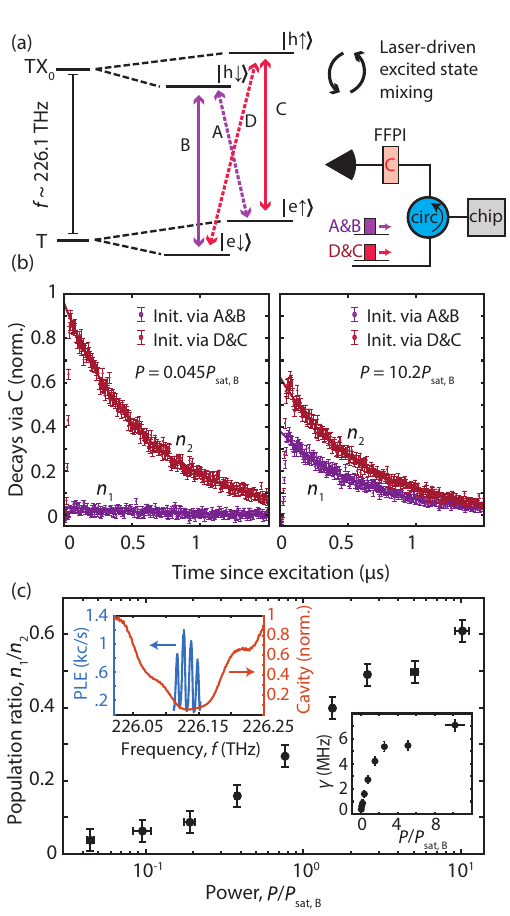}
  \caption{TX$_0$ spin populations as mixed by laser power.
  (a) The level diagram of T centre II under an applied field and experiment schematic.
  (b) Time-resolved decay measurements showing the populations converging at higher excitation powers, indicating increased spin mixing.
  Powers are quoted in the at-field saturation power, $P_\text{sat,~B}$.
  (c) Population ratio $n_1 / n_2$ as a function of excitation power. The insets show the at-field PLE spectrum for this T centre (top) and the extracted mixing rate $\gamma$ versus power (bottom).
  }
  \label{fig:figure4_lasermixing}
\end{figure}

We observe a second laser-driven effect causing TX$_0$ spin mixing, which disrupts the cyclicity of otherwise spin-conserving transitions and reduces the fidelity of repeated spin-photon entanglement cycles.
We apply a magnetic field $B = \SI{755}{\milli\tesla}$ along $[110]$ to T centre II with a superconducting Helmholtz coil pair to lift the spin degeneracy, as shown in \Cref{fig:figure4_lasermixing}(a), and operate at $T = \SI{0.4}{\kelvin}$ to reduce post-pulse thermal effects.
We prepare either of the two TX$_0$ states $\ket{\mathrm{h}\uparrow}$ (red) and $\ket{\mathrm{h}\downarrow}$ (purple) by spin-selective excitation and align a fibre Fabry–Pérot interferometer to filter spin-selective luminescence from the C transition, which allows us to measure the population in $\ket{h \uparrow}$.
\Cref{fig:figure4_lasermixing}(b) shows the $\ket{\mathrm{h} \uparrow}$ population as a function of time for both spin preparation schemes, where $n_1$ and $n_2$ are the population following preparation of $\ket{h \downarrow}$ and $\ket{h \uparrow}$ respectively. 
The left panel shows relative populations for low-power preparation ($P = 0.045P_{\textrm{sat,B}}$), while the right shows high-power preparation ($P = 10.2P_{\textrm{sat,B}}$), where $P_\text{sat,B}$ is the at-field saturation power.
Single exponential fits match the decay with no measurable indication of post-pulse thermalization.
\added{
The two closest transitions, B and C, have a peak-to-peak separation of \SI{11.05\pm0.03}{\giga\hertz} compared to the Gaussian inhomogeneous linewidth of \SI{4.1\pm0.01}{\giga\hertz} at $P= 11.2 P_\mathrm{sat,B}$.
We rule out confounding effects of off-resonant driving by measuring PLE spectra at powers higher than 10.2 $P_{\mathrm{sat,B}}$ (see the SM~\cite{Supplement} for details).}

The ratio $n_1/n_2$ corresponds to the degree of spin mixing (given equal mixing rates as shown in SM~\cite{Supplement}) and increases from 0 (no mixing) to $\approx0.6$ with increasing excitation power as shown in \cref{fig:figure4_lasermixing}(c).
Assuming perfect spin-state initialization and uniform excitation during the pulse, we define an effective mixing rate
\begin{equation}
    \gamma = \frac{1}{2T}\ln(\frac{n_2+n_1}{n_2-n_1}) \,,
\end{equation}
plotted in the inset of \cref{fig:figure4_lasermixing}(c).
Finally, we find that excited-state spin mixing is not a resonant effect. 
When the experiment is repeated at low resonant powers with additional high-power off-resonant light, we observe similar levels of mixing (see SM~\cite{Supplement}).
Electric field coupling to hole g-factors has been observed in quantum dots~\cite{Pingenot_2011_EfieldCouplingHolegFactor,Kuhlmann_2013_ChargeNoiseSpinNoise,Studenikin_2019_TunablegfactorHole}.
We speculate that analogous mechanisms may be present in the bound T centre exciton, where local charges that reconfigure the spectral diffusion environment may also couple to the hole g-factor, inducing a finite spin flip rate.
This mixing effect reduces the optical cyclicity, $C$, the average number of optical cycles before the spin state flips of T centre devices below the dipole branching ratio limit~\cite{Clear_2024_OpticalTransitionParameters}.
\added{This means the spin flip error channel probability per emitter per $\pi$-pulse is $\frac{1}{C}$ with} consequences for quantum non-demolition readout by resonant excitation and for remote entanglement schemes requiring cyclic transitions~\cite{Barrett2005}.

\section{Discussion}
\label{sec:discussion}

Our results establish the significance of laser-induced effects for silicon colour centre emitters and SPIs. 
We find that the spectral diffusion of T centres in nanophotonic devices is primarily induced by the NIR excitation laser and, using a resonance-check scheme, we achieve a $35\times$ reduction in effective linewidth to \SI{110(10)}{\mega\hertz}. 
We expect that the RC linewidth can be further reduced towards the homogeneous limit ($32$~MHz, see SM \added{Table I for information on the homogeneous linewidth} \cite{Supplement}) by optimizing the check/probe pulses. We find that the RC linewidth remains stable for up to $\sim1.5$ ms, long enough for practical advantage. 
Along with concurrent work in Ref.~\cite{Zhang_2025_SD}, this is the first demonstration of a functional RC scheme for silicon colour centres to our knowledge. 

Effective RCs enhance the feasibility of interference-based silicon colour centre quantum technologies, increasing emission efficiencies for resource state generation and other protocols requiring spectrally matched emitters or photons. 
Under a wide range of conditions, RCs can improve the performance of even pairwise ($N=2$) entanglement schemes \cite{taherizadegan_2025_gates}. 
We anticipate that the scheme introduced here can be applied to improve the entanglement rate of prototype T centre repeaters, network hubs, and spin-photon processors \cite{Simmons_2024_ScalableFaultTolerant,Afzal2024}.

Finally, we observe laser-induced spin mixing in the optically-excited bound-exciton state TX$_0$.
We suggest that these two laser-induced effects have the same root cause. 
A broad, low intensity band, likely a superposition of donor-acceptor-pair bands, is common to radiation-damaged silicon photoluminescence and underlies the sharp emitter lines of T and other silicon colour centres. 
This band may be due to donor- and acceptor-like damage centres. 
We observe that the NIR excitation laser, detuned from the T centre or any other sharp lines, drives luminescence from this broad band. 
This indicates that the laser is weakly photoneutralizing these ionized defects so they can recombine and luminesce. 
Such laser-driven charge fluctuations would cause T centre spectral wandering, and could also couple to the TX$_0$ hole state to induce spin-state mixing.

Photo-induced spectral diffusion has been extensively studied in quantum dots~\cite{Robinson_2000_laser-inducedSD,Holmes_2015_SDemissionLWGaN,Gao_2017_NanosecondGaNQD-Photo-inducedSD}.
These studies identify surface traps and charge traps in surrounding oxide layers as major contributors to spectral diffusion \cite{Ha_2015_SizeDependenceLineBroadening,Manna_2020_SurfacePassivationQD} and the same mechanisms may be present at the interfaces and oxide layers of our devices.
Efforts to passivate quantum dot surfaces have led to significant improvements in spectral stability \cite{Manna_2020_SurfacePassivationQD}, and the same engineering approaches may, in the longer term, reduce the silicon colour centre spectral diffusion closer to their performance in bulk, and perhaps make resonance checks redundant.
\added{Whether surface charges or volume charges dominate the spectral diffusion may be answered as smaller mode volume cavities are explored: as mode volumes become smaller, the power density at surfaces may also decrease.}
\added{The} results \added{described in this manuscript} are critical for understanding and mitigating laser-driven processes that limit the fidelity and success rate of multi-qubit entanglement sequences with silicon T centres and should be relevant to a broad class of emerging devices, as well as electronically similar silicon colour centres \cite{Xiong2024Discovery}.

\begin{acknowledgments}
This work was supported by the Natural Sciences and Engineering Research Council of Canada (NSERC) through Discovery Grants held by D.B.H, S.S., M.L.W.T. and C.S. and the ARAQNE Quantum Alliance Consortium, the
New Frontiers in Research Fund (NFRF), the Canada Research Chairs program (CRC), the Canada Foundation for Innovation (CFI), the B.C. Knowledge Development Fund (BCKDF), and the Canadian Institute for Advanced Research (CIFAR) Quantum Information Science program. 
S.A.M. acknowledges support from NSERC Postdoctoral Fellowship 571590123.
S.T and C.S. acknowledge support by the National Research Council through its High-Throughput Secure Networks (HTSN) Challenge Program.

The chip was provided by our industry partner Photonic Inc. We are grateful to Amin Khorshidahmad, Francis Afzal, Iain MacGilp, and Mohsen Akhlaghi and the entire Integrated Photonics team for their support of this project. We would also like to thank Evan MacQuarrie and Ata Ulhaq for their thorough review of the manuscript and helpful perspective throughout the project.

During the preparation of this manuscript, we became aware of related work by Zhang \emph{et al.}~\cite{Zhang_2025_SD}. 

\end{acknowledgments}

\appendix

\section{{\label{sec:methods}}Methods}

\subsection{{\label{sec:methods-sample}}Sample}

T centres are generated in a float-zone p-type SOI chip using a multi-stage implant and anneal process following MacQuarrie \emph{et al.} \cite{MacQuarrie_2021_GeneratingT}. 
Photonic devices are patterned by e-beam lithography and etched into the chip. 
We use quadratically tapered L0 nanobeam cavities (\cite{Loncar_2011_quadraticL0nanobeam}) with the cavity-axis along $[110]$ to enhance our T centre emission rate through the Purcell effect. 
We find quality factors of $Q= 600$--2,000. 
Single-mode waveguides couple the nanobeam cavities to on-chip diffraction grating couplers with \SI{6}{\tera\hertz} bandwidths centred between \SI{223.7}{\tera\hertz}--\SI{228.9}{\tera\hertz} that couple light into and out of the chip via optical fibres.

\subsection{\label{sec:methods-experimentalsetup-two-colourcorrelation}Apparatus}

For the two-colour spectroscopy, two fibre-coupled, tuneable, continuous-wave lasers (Toptica CTL) are pulsed independently and combined. 
Each laser is frequency-locked to a Bristol wavemeter. 
Pulsing is achieved using a booster optical amplifier (BOA) and an electro-optical modulator (EOM). 
We adjust the power of each laser using a combination of manual variable optical attenuators (VOA) and electrically variable optical attenuators (EVOA). 
The two lasers are combined using a fibre-coupled 50/50 splitter, then a circulator connects them to polarization maintaining fibres which feed into a custom cryogenic fibre array to deliver light to the sample held at temperatures between $1.45$~K--$2.5$~K in an ICE Oxford He flow cryostat.
Reflected laser light and T centre emission are collected through the same fibre path and routed to superconducting nanowire single photon detectors (SNSPDs) housed in a separate cryostat (IDQ 281).
The excitation pulse length ($t_{\text{exc}}$) for the corelation measurements taken at \SI{1.45}{\kelvin} was \SI{800}{\nano\second}. At \SI{2.5}{\kelvin}, $t_{\text{exc}} = \SI{800}{\nano\second}$.

The configuration for the resonance-check measurements is the same as for the two-colour correlation measurements with two additional components: 
An AOM is placed in the excitation path between the EOM and the circulator to improve the pulse extinction ratio, and an AOM is inserted in the collection path between the circulator and the SNSPDs to prevent detector saturation during the excitation pulse.
In this experiment, the samples are housed in a Bluefors LD400 dilution refrigerator, with the devices thermally anchored to a mixing plate held at \SI{0.27}{\kelvin}.
For the RC measurements $t_{\text{exc}} = \SI{100}{\nano\second}$.

The configuration for the $g^{(2)}$ measurement is the same as the RC measurements except an additional 50/50 splitter is placed in the collection path after the AOM. 
Each output of the 50/50 splitter is sent to separate SNSPD channels in the same cryostat.
For T centre I, II, and III, we used $t_{\text{exc}} = 900, 100$ and $\SI{800}{\nano\second}$ respectively.

The configuration for the laser-induced hole-spin mixing is the same as the resonance-check measurements without the AOM in the excitation path and operating at \SI{0.4}{\kelvin}.
A 50/50 splitter is placed in the collection path after the collection AOM. One output of the 50/50 passes through a fibre-coupled, temperature-tuneable Fabry-Perot interferometer with a bandwidth of $0.985$~GHZ and free spectral range of $102.4$~GHz (LUNA Innovations).
Two optical isolators are placed before and after the interferometer to suppress reflections.
The signal at the output of the isolator is then sent to one SNSPD. 
The other output of the 50/50 is connected directly to a second SNSPD in the same cryostat.
For the laser-induced hole spin mixing, $t_{\text{exc}} = \SI{100}{\nano\second}$.
 
\providecommand{\noopsort}[1]{}\providecommand{\singleletter}[1]{#1}%

\clearpage
\onecolumngrid  

\begin{center}
\large{\textbf{Supplemental Material for: ``Laser-induced spectral diffusion and excited-state mixing of silicon T centres''}}%
\end{center}

\setcounter{equation}{0}
\setcounter{figure}{0}
\renewcommand{\thefigure}{S\arabic{figure}}
\renewcommand{\theequation}{S\arabic{equation}}
\renewcommand{\thesection}{S\Roman{section}}
\newcommand{\simon}[1]{\textcolor{blue}{SAM: #1}}
\newcommand{\camille}[1]{\textcolor{magenta}{CAM: #1}}



\section{Summary of measured devices}

\Cref{table:tcentres} summarizes the key experimental parameters and figures of merit for the two T centre devices (T Centre I and II) presented in the manuscript and a third device (T Centre III) included here for additional verification.
T centre I was measured in an ICE Oxford He-flow cryostat at $1.5$ and $2.5$~K, while T centres II and III were measured in a Bluefors LD400 dilution refrigerator at $\SI{0.27}{\kelvin}$ and $\SI{0.4}{\kelvin}$.

\begin{table}[ht]
\begin{center}
    \begin{tabular}{|| l | l | l | l || }
    \hline
    & T centre I & T centre II & T centre III\\
    \hline
    Zero-field saturation power ($P_\text{sat}$) &  \SI{0.08\pm0.01}{\micro\watt} ($t_{\text{exc}} = \SI{800}{\nano\second}$)& \SI{1.3\pm0.1}{\micro\watt} ($t_{\text{exc}} = \SI{100}{\nano\second}$) & - \\
    At-field saturation power ($P_\text{sat,B}$) &  -& \SI{2.5\pm0.2}{\micro\watt} ($t_{\text{exc}} = \SI{100}{\nano\second}$) & \SI{1.3\pm0.1}{\micro\watt} ($t_{\text{exc}} = \SI{100}{\nano\second}$) \\
    Resonance frequency ($P_\text{sat}$) & \SI{226.128042\pm0.000002}{\tera\hertz} & \SI{226.13061\pm0.000001}{\tera\hertz} & \SI{226.13582\pm0.00002}{\tera\hertz}\\
    Inhomo. linewidth ($P_\text{sat}$) & \SI{1.694\pm0.005}{\giga\hertz} & \SI{3.8\pm0.1}{\giga\hertz}& \SI{3.9\pm0.1}{\giga\hertz}\\
    Holeburning linewidth ($P_\text{sat}$)  & \SI{328\pm5}{\mega\hertz}& \SI{880\pm50}{\mega\hertz}& -\\
    Holeburning linewidth (low power)  & \SI{33\pm5}{\mega\hertz} & \SI{64\pm3}{\mega\hertz} & - \\
    $g^{(2)}$(0) & \SI{0.26\pm0.02}{} & \SI{0.186\pm0.004}{}& \SI{0.372\pm0.009}{}\\
    Cavity centre frequency & \SI{226.136\pm0.004}{THz} & \SI{226.13512\pm0.00006}{\tera\hertz} &\SI{226.124\pm0.001}{\tera\hertz}\\
    Cavity Q factor & \SI{880\pm40}{} & \SI{1743\pm2}{} & \SI{630\pm90}{}\\
    Purcell enhanced lifetime & $\SI{691\pm1}{\nano\second}$ & $\SI{602\pm9}{\nano\second}$ & $\SI{530\pm10}{\nano\second}$\\
    Operating temperature & 1.45--2.5 K & 0.27 -- 0.4 K & 0.27 K\\
    Pressure environment & \SI{20}{\milli\bar} (Helium vapor) & \SI{10e-6}{\milli\bar}& \SI{10e-6}{\milli\bar} \\
    Applied $B$-field & 0 & 0 or $\SI{0.755}{\tesla}$ & \SI{0.755}{\tesla} \\
    Hole g-factor, $g_h$ & - & \SI{0.96\pm0.05}{} & \SI{1.26\pm0.05}{}\\
    \hline
    \end{tabular}
\end{center}
\caption{Measured optical and cavity parameters for the three T centres investigated in this work.
The inhomogeneous saturation power, $P_{\text{sat}}$ is obtained from PLE spectra at zero-field. 
The at-field saturation power, $P_{\text{sat,B}}$ is obtained from PLE under equal excitation of both ground states (\emph{e.g.} PLE while pumping the A and B transitions).
The power is calibrated with a measurement before the circulator that connects to the fibre array going to the device.
For T centre I we calibrate the power with an excitation pulse length of  $t_\text{exc} = \SI{800}{\nano\second}$ and for T centres II and III we measure saturation with $t_\text{exc} = \SI{100}{\nano\second}$ which accounts for the difference in saturation power between the centres.
We estimate $50\%$ grating coupler efficiency, leading to the same reduction in final power at the T centre.
The resonance frequency and the inhomogeneous and saturation holeburning linewidths were taken at the saturation power $P_\text{sat}$.
The lowest measured saturation holeburning linewidths were taken at $P =$ \SI{0.001\pm0.002}{} $P_\text{sat}$ and $P =\SI{0.007\pm0.002}{}P_\text{sat}$ for T centres I and II respectively.
For the spectral diffusion correlation measurements made on T centre I, we measure the holeburning linewidth at both $P_\text{sat}$ and $2.86P_\text{sat}$  $\Gamma_{P_\text{sat}} =  \SI{328\pm5}{\mega\hertz}$ and $\Gamma_{2.86 P_\text{sat}} = \SI{341\pm5}{\mega\hertz}$.
Homogeneous linewidths at low powers can be estimated as half the holeburning linewidth \cite{Deabreu2023waveguide}.
The zero-delay $g^{(2)}$ values indicate that these centres have emission primarily due to single centres and have been normalized to the long-time uncorrelated $g^{(2)}(\tau)$ rate.
We performed $g^{(2)}(\tau)$ measurements using a Hanbury–Brown–Twiss setup.
For T centre I, $g^{(2)}(\tau)$ measurements were taken at $T = \SI{2.5}{\kelvin}$.
For T centre II, we included an additional AOM in the excitation path, and performed measurements at $T = \SI{0.27}{\kelvin}$.
Measurements for T centre III were performed at $T = \SI{0.4}{\kelvin}$.
The cavity centre frequency and cavity $Q$ factor were obtained by measuring reflection spectra.
The applied field for the RC-PLE and the two-colour correlation measurements was $\SI{0}{\tesla}$.
For the laser-induced excited-state mixing in TX$_0$ measurements, the applied field was $\SI{0.755}{\tesla}$.}
\label{table:tcentres}
\end{table}

\added{\section{Bounding off-resonant crosstalk}}

\begin{figure}[h]
  \centering
  \includegraphics{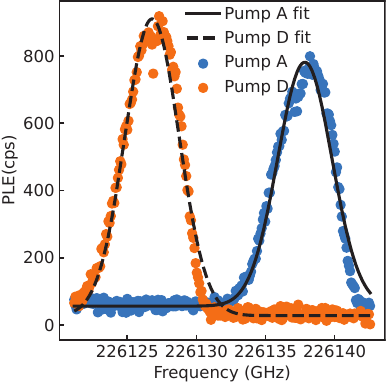}
  \caption{\added{PLE measured at 0.755T, pumping on either the A (blue, solid black) or D (orange, dotted black) transition to relieve hyperpolarization.}}
  \label{fig:highpple}
\end{figure}

\added{In principle, the excited state mixing measurements reported in this paper could be confounded by off-resonant excitation crosstalk between adjacent optical transitions. 
We explicitly rule out contributions from this confounding effect in this section.
The two closest transition frequencies being driven in this measurement, B and C, have a peak-to-peak separation of \SI{11.05\pm0.03}{GHz}. 
The inhomogeneous line at $P=$ \SI{11.2}{} $P_\mathrm{sat,B}$ is Gaussian with FWHM linewidth \SI{4.1\pm0.1}{\giga\hertz} and we bound the Rabi frequency at this power to below \SI{440}{\mega\hertz}.
We measure high power PLE at field shown in Fig.~{\ref{fig:highpple}} where we pump on either the A or D transition to relax hyperpolarization and sweep over both B and C transitions. 
We use our fitted Gaussians to extract relative excitation probabilities and find the relative probability of exciting B when driving C is \SI{1e-7}, negligible compared to the degree of mixing we observe.}

\section{Off-resonant effects}

In this section we explore the effects of off-resonant laser excitation in both the spectral diffusion and the TX$_0$ excited-state spin mixing phenomena.

\begin{figure}[t]
  \centering
  \includegraphics[width=6.8in]{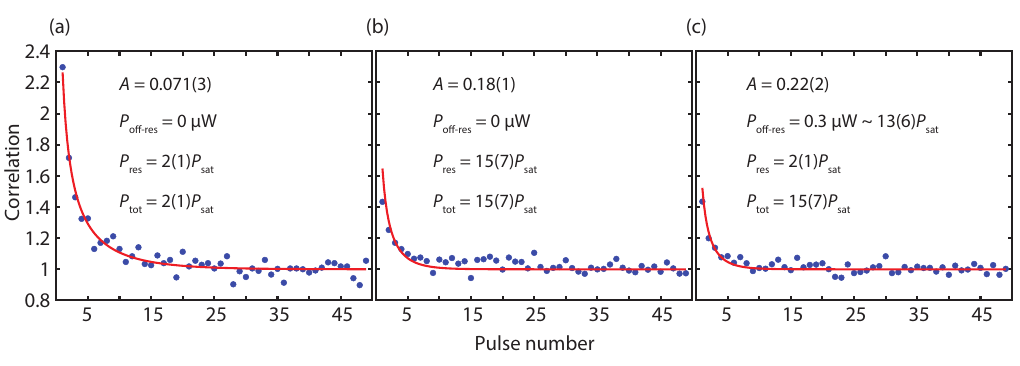}
  \caption{
  The effect of off-resonant light on the spectral diffusion of a T centre.
  Markers are measured $g^{(2)}$ correlation (normalized to the long-time value) and the solid lines are one-parameter fits with the per-pulse SD rate, $A$ as the free parameter.
  The value of $\beta$ can be determined by the $g^{(2)}(0)$ value (not plotted).
  (a) A $g^{(2)}(\tau)$ measurement of T centre I with a resonant power of $P_\text{res} \approx2P_\text{sat}$.
  (b) Same measurement with higher resonant excitation power of $P_\text{res} \approx15P_\text{sat}$ showing a faster diffusion rate.
  (c) A measurement with the same resonant power as (a), but with an additional off-resonant laser $P_\text{off-res}\approx 13P_\text{sat}$, bringing the total power to $P_\text{tot} \approx15P_\text{sat}$. 
  The extracted spectral diffusion rate $A$ in (c) closely matches that in (b), supporting the interpretation that laser-induced spectral diffusion is driven by the total excitation energy rather than by purely resonant processes.
  }
  \label{fig:OffresSDMixing}
\end{figure}

\subsection{Off-resonant excitation and spectral diffusion}

We examine the role of off-resonant excitation by introducing a second laser at $f = \SI{226.140354}{\tera\hertz}$, which pulses simultaneously with the resonant excitation laser at $f = \SI{226.127902}{\tera\hertz}$ during a ${g^{(2)}}(\tau)$ correlation measurement. 
A $g^{(2)}$ measurement is conceptually equivalent to the two-colour correlation measurements discussed in the main text when $\delta_2 - \delta_1 = 0$. 
There are two minor differences between $g^{(2)}$ and the two-colour correlation. 
First, $g^{(2)}$ measures correlations between every pulse, whereas the two-colour measurement only correlates alternating pulses with different detunings. 
Second, the two-colour correlation uses separate detectors for each spectral bin. This only affects the $\tau = 0$ bin, where photon antibunching suppresses coincidences. 
At nonzero delays, the effect is a constant scaling of the count rate, which we normalize out using the long-time data.

\Cref{fig:OffresSDMixing} presents a $g^{(2)}$ measurement on T centre I for three different power settings taken at a temperature of $T = \SI{1.45}{\kelvin}$. 
We vary the resonant laser power between $P_\text{res} = \SI{2\pm1}{}P_\text{sat}$ and $P_\text{res} = \SI{15\pm1}{}P_\text{sat}$. 
To isolate the effect of off-resonant light, we perform a third measurement with $P_\text{res} = \SI{2\pm1}{}P_\text{sat}$ and an additional off-resonant laser at $P_{\text{off-res}} = \SI{13\pm6}{}P_{\text{sat}}$, yielding a total power equivalent to the higher-resonant-power case. 
We find that the mixing rate $A$ under the off-resonant condition closely resembles that of the high-power resonant case, rather than the low-power case with equivalent resonant excitation. 
This observation supports the hypothesis that laser-induced SD arises from broadband excitation of a surrounding charge environment, rather than a purely resonant interaction with the emitter.

\subsection{Off-resonant laser-induced excited state mixing}

\begin{figure}[t]
  \centering
  \includegraphics[width=3.4in]{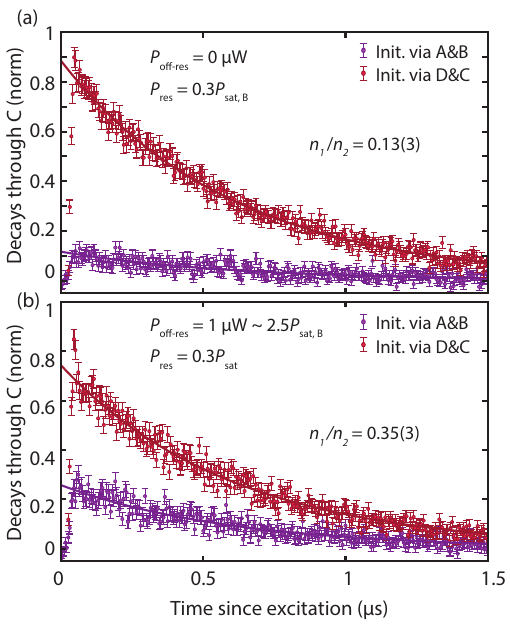}
  \caption{
  The effect of off-resonant laser on the laser-induced excited state mixing.
 (a) Spin mixing measured at low resonant excitation power, $P = 0.3P_{\text{sat, B}}$, without off-resonant illumination. The extracted mixing fraction is $n_1/n_2 = \SI{0.13\pm0.03}{}$.
 (b) Spin mixing under the same resonant conditions, but with the addition of an off-resonant laser at $P_{\text{off-res}} = \SI{1}{\micro\watt}$, corresponding approximately to $2.5P_{\text{sat, B}}$. 
 The mixing fraction increases to $n_1/n_2 = \SI{0.35\pm0.03}{}$.
 We express the off-resonant power in units of $P_{\text{sat, B}}$ to facilitate direct comparison with the power levels used for resonant excitation in the main text, even though saturation is not strictly defined for off-resonant excitation.
 The observed increase in spin mixing with off-resonant power suggests that the mechanism driving excited-state spin transitions cannot be attributed solely to resonant optical excitation.
 }
  \label{fig:OffresMixing}
\end{figure}

We probe whether the laser-induced excited state mixing results from resonant excitation or whether off-resonant excitation has a similar effect.
We compare the populations measured after resonant excitation with the addition of a high-power, red-detuned laser ($f = \SI{226.104878}{\tera\hertz}$) which pulses simultaneously with our resonant initialization lasers. 
The frequency is chosen to be within the cavity but far from any of the T centre optical transitions. On its own, this laser negligibly excites the emitter. 
By comparing the degree of spin mixing with and without simultaneous off-resonant light, we show that the mechanism driving the laser-induced spin mixing is not purely a resonant mechanism, but rather non-resonant light can also mix during the excitation pulse.

\Cref{fig:filterOnBTcentreIII}~(b) summarizes our findings, showing the ratio of populations $n_1/n_2$ (as in the main text) for our two different excitation conditions. 
The calculated population ratio for the off-resonant experiment ($n_1/n_2 = \SI{0.35\pm0.03}{})$ is higher than the ratio without off-resonant light ($n_1/n_2 = \SI{0.13\pm0.03}{}$) but lower than the average of measurements at equivalent power for T centre II ($n_1/n_2 = \SI{0.46\pm0.04}{}$). 
This discrepancy may be due to differences in the coupling to the cavity at different positions (leading to a slightly different effective power) or may indicate that the mixing rate displays some frequency dependence. 
In either case, these results rule out the hypothesis that the laser-induced mixing is the result of purely resonant effects.

\section{Evidence for TX$_0$ spin state independence of laser mixing and tests on T Centre III}

\begin{figure}[t]
  \centering
  \includegraphics[width=3.4in]{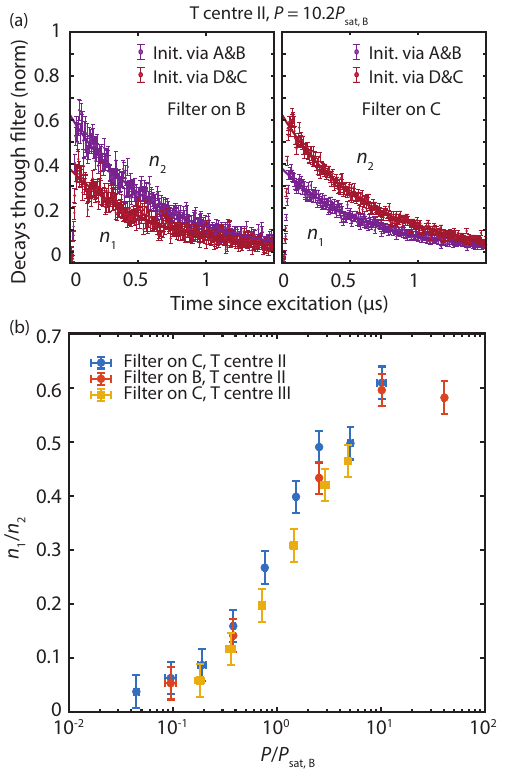}
  \caption{
 (a) Comparison of laser-induced mixing observed when filtering emission through transitions B (left) and C (right) for T centre II. 
 While populations associated with each initialization scheme swap when the filter is changed, the ratio of populations between prepared and unprepared states remains consistent, confirming symmetry in spin-state mixing rates.
 (b) Comparison of the population ratio $n_1/n_2$ for T centre II using filters on B and C, alongside measurements for another device (T centre III). 
 All three data sets agree, indicating consistent mixing dynamics across multiple devices.
 }
  \label{fig:filterOnBTcentreIII}
\end{figure}

In this section, we demonstrate that the extracted post-pulse populations—and thus the degree of spin mixing—are independent of the initially prepared state. 
Verifying that $n_1/n_2$ is independent of the detection pathway helps establish that spin mixing is a fundamental effect of the excitation process, rather than a measurement artifact. 
It also supports that the ratio of populations under different initialization conditions serves as a reliable proxy for the relative populations of the two spin states under identical excitation. 
Specifically, we prepare the system in either $\ket{h\uparrow}$ or $\ket{h\downarrow}$ and measure post-pulse populations using filters aligned to both the C and B transitions, thereby probing both hole spin states.

\Cref{fig:filterOnBTcentreIII}(a) shows the results for T centre II under all four combinations of initialization (either $\ket{h\uparrow}$ or $\ket{h\downarrow}$) and detection (via the C or B transition), at $P = 10.2P_{\text{sat, B}}$.
The initial rise in signal is attributed to the collection path’s acousto-optic modulator turning on. 
Here, $n_1$ ($n_2$) denotes the population in the detected state when the opposite (same) hole state is prepared.
As expected, switching the detection filter to the B transition reverses the relative signal strengths for the two preparation states.
Consequently, the ratio $n_1/n_2$ remains consistent regardless of whether we filter on the B or C transition.
\Cref{fig:filterOnBTcentreIII}~(b) plots $n_1/n_2$ versus excitation power for both filter configurations in T centre II, as well as for a separate device (T centre III) using the C transition.
The agreement between the C and B data for T centre II confirms symmetric mixing between the hole states, validating $n_1/n_2$ as a reliable measure of mixing strength.
The similar mixing behaviour observed in T centre III suggests that laser-induced spin mixing is not device-specific and may be a general characteristic of T centres under similar conditions.

\section{Photon-rate improvement using RC-PLE}

\begin{figure}[t]
  \centering
  \includegraphics[width=3.4in]{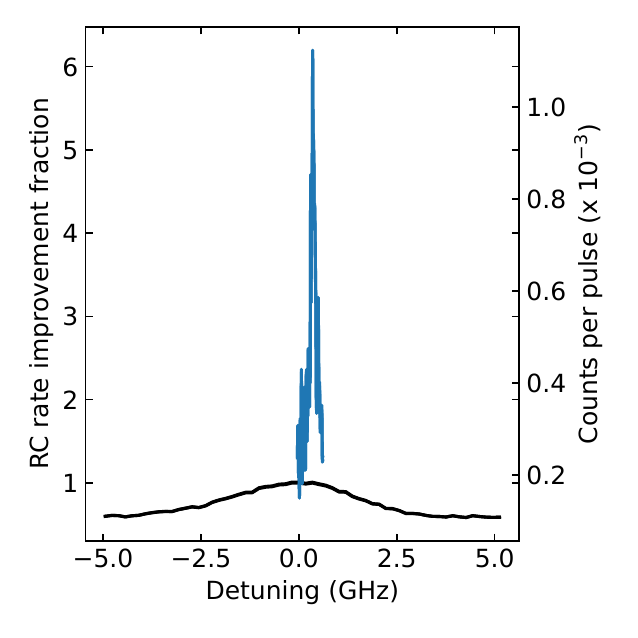}
  \caption{RC-PLE measurement at the lowest power setting ($P_{\text{probe}} = 0.008 P_{\text{sat}}$, $P_{\text{pump}} = 0.016 P_{\text{sat}}$).
  The count rate per pulse, normalized to that of non-heralded PLE, demonstrates a sixfold improvement.
  Unnormalized counts per pulse are shown for reference.}
  \label{fig:rc-countrate-improvement}
\end{figure}

We measure the improvement in count rate per pulse in RC PLE compared to the non-heralded PLE count rate. 
Figure~\ref{fig:rc-countrate-improvement} shows the $6\times$ count rate improvement at the lowest-measured power ($P_{\text{probe}} = 0.008 P_{\text{sat}}$  , $P_{\text{pump}} = 0.016P_{\text{sat}}$, $P_{\text{PLE}} = P_{\text{probe}} = 0.008 P_{\text{sat}}$).

\section{Long dark time resonance check PLE}

\begin{figure}[t]
  \centering
  \includegraphics[width=3.4in]{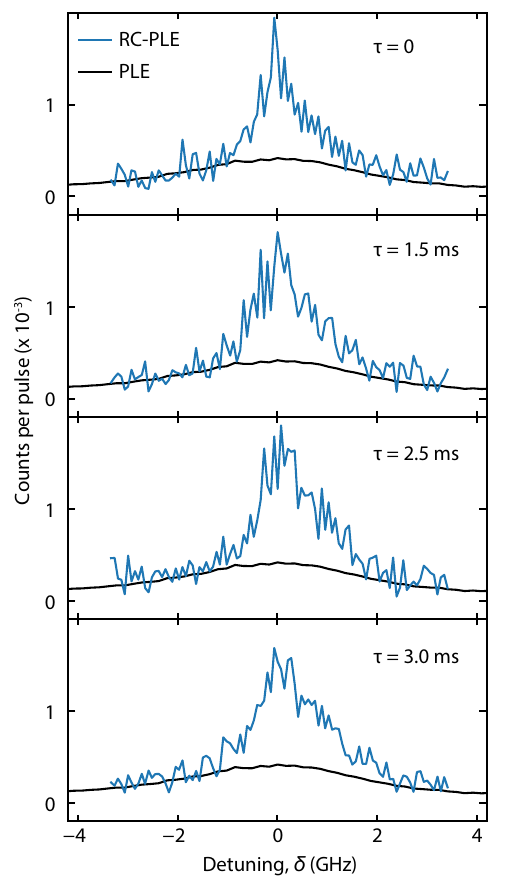}
  \caption{
  RC-PLE linewidth broadening for T centre II as a function of dark time, $\tau$. 
  After detecting a ‘check pulse’, we introduce a variable delay before applying the ‘probe pulse’.
  For $\tau = \SI{1.5}{\milli\second}$, no statistically significant broadening is observed.
  However, for longer delays ($\tau \approx 2.5–3.0$~ms), we observe spectral broadening—on the order of $200–500$ MHz—particularly on the high-frequency side of the spectrum.
  The RC-PLE data are normalized to the number of counts per probe pulse, representing the conditional probability of detection given a preceding check pulse. 
  For the regular PLE, we plot the unconditional detection probability per shot. 
  The ratio between these two traces indicates the enhancement achieved through the resonance-check scheme.
  }
  \label{fig:darktimedependence}
\end{figure}

\begin{figure}[t]
  \centering
  \includegraphics[width=3.4in]{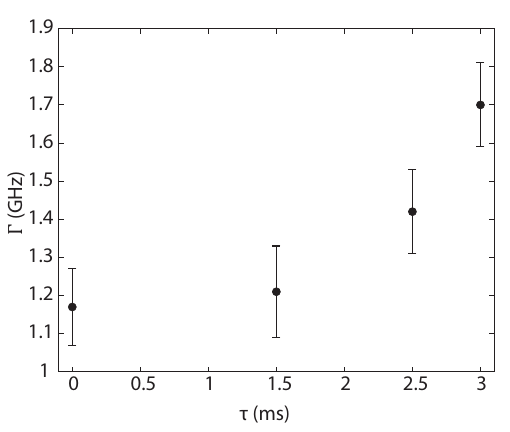}
  \caption{
  Linewidths are extracted by fitting Lorentzians to the RC-PLE spectra shown in \cref{fig:darktimedependence}.
  The resulting linewidths, $\Gamma$, are plotted here as a function of the dark interval $\tau$ between the check and probe pulses.
  No statistically significant broadening is observed for $\tau = \SI{1.5}{\milli\second}$; however, for longer delays ($\tau \gtrsim \SI{2.5}{\milli\second}$), a gradual increase in linewidth is apparent.
  }
  \label{fig:gammavstau}
\end{figure}

Here, we explore the linewidths of resonance-check photoluminescence excitation (RC-PLE) in the limit of long dark time, $\tau$. 
We vary the dark time $\tau = {0, 1.5, 2.5, 3.0},\mathrm{ms}$ for T centre II to investigate how spectral broadening evolves over extended intervals. 
We observe negligible broadening up to $\tau = \SI{1.5}{\milli\second}$, but find measurable linewidth increases at longer delays. 
\Cref{fig:darktimedependence} shows the resonance-check PLE spectra as a function of detuning for each value of $\tau$.
Counts are normalized to the number of probe pulses, corresponding to the probability of detecting a photon per probe pulse. 
For comparison, we also plot the unconditioned PLE spectrum, normalized in the same way.
The ratio between the two spectra indicates the enhancement factor achieved through resonance-check preparation.

We fit Lorentzian functions to the resonance-check PLE spectra to determine the corresponding linewidths.
No statistically significant broadening is observed for $\tau \leq \SI{1.5}{\milli\second}$; however, at longer dark times, a measurable increase in linewidth appears.
This broadening is most pronounced on the high-frequency side of the spectra, suggesting asymmetry in the diffusion process.
\Cref{fig:gammavstau} shows the extracted linewidth $\Gamma$ as a function of the dark interval $\tau$.
The observed dependence of $\Gamma(\tau)$ deviates from the Ornstein–Uhlenbeck prediction: the second-derivative is positive, in contrast to the negative second-derivative expected for O-U dynamics.
This suggests that a distinct physical mechanism may be responsible for the long-time spectral broadening.
\added{Given the nature of slow spectral diffusion in other semi-conductor host materials like diamond, it is possible that with further materials development, the already long timescale of this slow spectral diffusion could be extended \cite{Taminiau_2019_NVmatdevSDimprove}.}

\section{Temperature dependence of spectral diffusion}

\begin{figure}[t]
  \centering
  \includegraphics[width=3.4in]{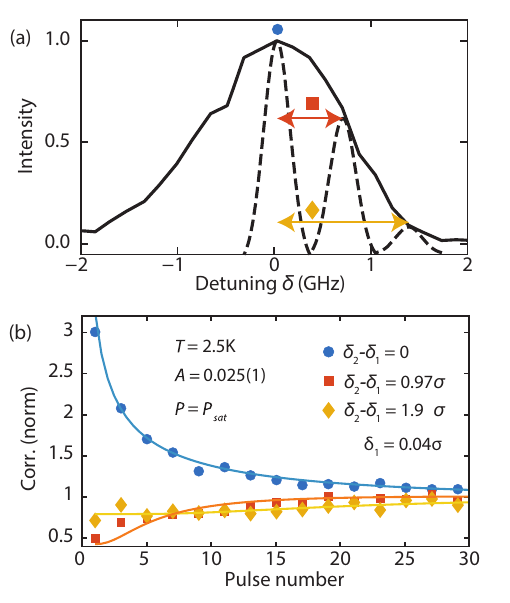}
  \caption{
  (a) Laser detunings from the centre of the inhomogeneous line (solid) and modelled homogeneous lines at $P_{sat}$ (dashed)
  (b) Two-colour correlation data for $T = \SI{2.5}{\kelvin}$ normalized to the long-time value.
  Solid lines are fits to the O-U model described in the main text where $A$ is the sole fitting parameter.
  The fitted $\alpha$ is larger than that which was found at $T = \SI{2.5}{\kelvin}$.
  This difference may result from a systematic change in delivered excitation power, introduced during fibre realignment required for temperature switching.
  }
  \label{fig:1p5Kdata}
\end{figure}

We increase the temperature of our device to $T = \SI{2.5}{\kelvin}$ and perform the same measurements at detuning values $\delta_2 -\delta_1 = \{0,0.97,1.9\}\sigma$.
We jointly fit these correlation measurements using the same method as the main text.
\Cref{fig:1p5Kdata} shows the two-colour correlation data at $T = \SI{2.5}{\kelvin}$ with an excitation power of $P = P_\text{sat}$
We find a value of $A = \SI{0.025\pm0.001}{}$ that is somewhat lower than the value found at $T = \SI{1.45}{\kelvin}$.
We note that $A$ cannot be meaningfully compared between these experiments due to reconfiguring the apparatus which may produce systematic variation in delivered power.


\section{Ornstein-Uhlenbeck Model}

In this section, we derive the Ornstein–Uhlenbeck model fitted to our spectral diffusion measurements from a microscopic discrete-charge model~\cite{Uhlenbeck_1930_TheoryOfBrownianMotion}.
For related discussions, see Refs.~\cite{Delteil_2024_photonstatisticsdiffusiveemittersOUMarkov,Gardiner_2004_HandbookOfStochasticMethods,vanKampen_2007_StochasticProcesses}.
We model our local environment as an ensemble of $N$ charges, each equally coupled to the T centre, and each taking time-dependent values of $\pm1$ such that the $i^{\text{th}}$ charge will be $c_i(t) \in \{-1,+1\}$ at time $t$.
If each charge contributes a small perturbation to the transition frequency, then the centre frequency (up to an overall scaling factor) will be given by

\begin{equation}
    \delta = \sum_{i = 1}^Nc_i/N \,.
\end{equation}

At each time step, we scramble $X$ charges—meaning each selected charge is randomly assigned a value of $\pm1$.
We introduce the variable $\alpha^* = \frac{X}{N}$ which is the fraction of charges in the ensemble that are scrambled each time step.
We consider the limit of large $N$ but finite $\alpha^*$.
The change in $\delta$ for each time step will be,

\begin{equation}
    \mathds{E}[\Delta \delta | \delta(t)] = -\alpha^{*} \delta(t) \,.
\end{equation}

As the overall sum of charges moves away from zero, the odds of flipping a charge that brings the distribution back to zero becomes higher.
The variance per time step will be proportional to the number of charges scrambled, i.e., $X$.
Therefore, the problem is mapped to a random walk with a restoring force,

\begin{equation}
    \Delta \delta(t) = -\alpha^* \delta(t) + \sqrt{X}\xi_t \,,
\end{equation}

where $\xi_t$ is a Gaussian random variable with mean of $0$ and standard deviation $1$.
We would like to find the correlations of the frequency as a function of time.
We note the conceptual similarities to the problem of magnetization fluctuations in a paramagnet as a function of time.

Next we move to the continuous limit of this problem using

\begin{equation}
    \frac{d\delta}{dt} = -\alpha \delta(t) + \sqrt{2D}~\eta(t) \,, 
\end{equation}

where $\eta(t) = \frac{dW}{dt}$ is a white noise term (derivative of the Wiener process, $W(t)$), $\alpha$ is the spectral diffusion rate and $D$ is a diffusion constant. 

The conditional probability that the centre frequency takes the value $\delta(t) = \delta_2$ at time $t$, given $\delta(0) = \delta_1$, is given by

\begin{equation}
    P(\delta_2,~t~|~\delta_1, ~0) = \frac{1}{\sqrt{2\pi~(1-e^{-2\alpha t})}}\exp\left[ 
    -\frac{(\delta_2 - \delta_1 e^{-\alpha t})^2}
         {2\left(1 - e^{-2\alpha t}\right)} 
 \right] \,,
 \label{eqn:condprobOU}
\end{equation}

\begin{figure}[t]
  \centering
  \includegraphics[width=3.4in]{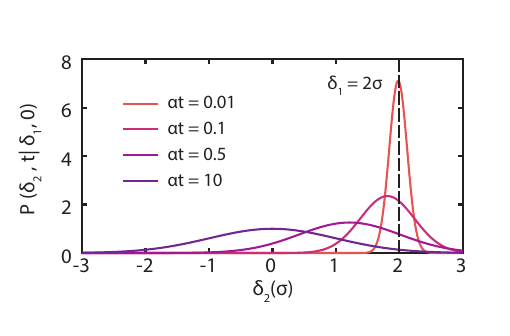}
  \caption{
 Plot of \cref{eqn:condprobOU} showing the conditional probability amplitude distribution for an initial detuning $\delta_1 = 2\sigma$.
 The probability amplitudes have been normalized to the maximum of the long-time inhomogeneous distribution.
 At early times ($\alpha t \ll1$) the distribution remains localized near the initial frequency $\delta_1$.
 As time evolves the distribution broadens toward the steady-state inhomogeneous distribution with mean $\delta_2 = 0$.
  }
  \label{fig:OUconfitionalProb}
\end{figure}

where we have made the rescaling $D = \alpha$ for convenience (equivalent to rescaling $\delta(t)$ in units of the inhomogeneous or long-time standard deviation).
\added{In the limit where $\alpha t \ll 1$ and $\delta_1 \ll \delta_2$,  Eq.~\ref{eqn:condprobOU} simplifies to}

\added{
\begin{equation}
    P(\delta_2,~t~|~\delta_1, ~0) = 
    \frac{1}{\sqrt{2\pi 2\alpha t}}
    \exp\left[ -\frac{(\delta_2)^2}{2\alpha t} 
 \right] \,,
 \label{eqn:OUshorttime}
\end{equation}
}

\added{and we can treat $\alpha$ as an effective diffusion constant.}

\Cref{fig:OUconfitionalProb} displays $P(\delta_2, t \mid \delta_1, 0)$ as a function of time for $\delta_1 = 2\sigma$.
Over time, the narrow peak initially centred near $2\sigma$ broadens, shifts and converges to the inhomogeneous distribution.

To model the correlation data, we include a finite dark count term, $\beta$, and normalize the correlation such that $C \to 1$ at long times yielding

\begin{equation}
    C = \beta + \sqrt{2\pi}~e^{ \delta_2^2/2}[1-\beta]P(\delta_2, t \mid \delta_1, 0) \,.
\end{equation}

We define a per-pulse mixing rate by integrating over the duration of the excitation pulse: 
\begin{equation}
    A = \int_\mathrm{pulse}\alpha\,\dd t = \alpha t_\text{pulse}\,,
\label{eq:A}
\end{equation}
 where $t_\text{pulse}$ is the excitation pulse duration. 
For low rates of diffusion, we can interpret $A$ as the fraction of the local charge environment that is reconfigured during each pulse. 

We determine $\beta$ from three sources of background contributions to the correlation signal.
These include: $P(\delta_2|\text{noise})$ (true second-bin signal with a first-bin non-signal [noise] count), $P(\text{noise}|\delta_1)$, and $P(\text{noise}|\text{noise})$.
We measure the background subtracted uncorrelated count rates corresponding to the different spectral positions.
Let $R_1$ and $R_2$ denote the T centre count rates at $f = \delta_1$ and $f = \delta_2$, respectively, and let $R_N$ denote the noise count rate.
The resulting expression for $\beta$ is:

\begin{equation}
    \beta = \frac{R_N(R_N + R_1 + R_2)}{R_N(R_N + R_1 + R_2) + R_1R_2} \,,
\end{equation}

which we compute from our measured PLE spectrum.

\added{We can use the extracted A parameter (Eq.~\ref{eq:A}) to estimate the spectral diffusion during an optical $\pi$-pulse.}
\added{For T centre I at \SI{1.5}{\kelvin} and $P=P_\mathrm{sat}$, we extract $\mathrm{A} = 0.045$. We estimate the length of a $\pi$-pulse  from the holeburning linewidth ($\sigma_{\mathrm{hole}}=$\SI{328(5)}{\mega\hertz}) on T centre I at $P=P_\mathrm{sat}$, shown in Table~\ref{table:tcentres}.}
\added{Assuming this hole linewidth is dominated by power broadening, we determine $T_{\pi}=$\SI{3}{\nano\second} and a lower-bound per-$\pi$-pulse SD of 29 MHz.
Similarly, assuming the hole linewidth is entirely SD, we determine an upper bound per-$\pi$-pulse SD of 93 MHz.}

\section{Resonance check for multi-emitter entanglement}

In this section, we analyze the statistics of a resonance check in the context of multi-qubit preparation.
We consider $N$ qubits, each with independent excitation and photon detection probabilities. The probability of detecting a photon on any given pulse is

\begin{equation}
    \eta_{\mathrm{SD}}\eta_{\mathrm{det}} = \frac{1}{M},
\end{equation}

where $\eta_\mathrm{SD}$ is the ratio of the homogeneous to inhomogeneous linewidths, and $\eta_{\mathrm{det}}$ is the detection efficiency.
For simplicity, we assume coherent excitation by an optical $\pi$-pulse with unit excitation efficiency, $\eta_{\mathrm{exc}} = 1$.
For non-ideal excitation efficiency, $\eta_{\mathrm{exc}}$ enters the analysis in the same way as $\eta_{\mathrm{det}}$, effectively rescaling the detection probability.

We analyze the use of charge resonance checks across all qubits in parallel, continuing until each qubit is spectrally prepared. In this setting, the total preparation time is limited by the slowest qubit in the ensemble.
We consider regimes in which a charge resonance check will result in overall sequence speedup.

Let $X_i$ denote the number of attempts required to observe a photon detection in the check-probe sequence for the $i^\text{th}$ qubit.
Then, $X_i$ is a geometric random variable with mean $\mathds{E}[X_i] = M$, a probability density function (PDF) of
\begin{equation}
    \mathrm{Pr}(X_i = k) = (1 - \frac{1}{M})^{k-1}\frac{1}{M} \,,
\end{equation}
and a cumulative density function (CDF) of
\begin{equation}
    \mathrm{Pr}(X_i \leq k) = 1 - (1- \frac{1}{M})^{k} \,.
\end{equation}

For $N$ qubits, the total number of attempts required to spectrally prepare all qubits is given by

\begin{equation}
    T = \mathrm{max}\{X_1,X_2,...\} \,.
\end{equation}

We now compute the expected value of $T$ using the tail-sum formula:

\begin{equation}
    \mathds{E}[T] = \sum_{k = 1}^\infty P(T \geq k) = \sum_{k = 1}^\infty \left[1 - \mathrm{Pr}(T < k )\right] = \sum_{k = 1}^\infty \left[1 - \prod_{i = 1}^N P(X_i \leq k - 1)\right] \,.
\end{equation}

The expression in the product is the CDF for our geometric random variable and hence,

\begin{equation}
    \mathds{E}[T] = \sum_{k = 1}^\infty  \left[1 - \{1 - (1 - \frac{1}{M})^{k-1}\}^N \right] \,,
\end{equation}

which has a closed form representation

\begin{equation}
    \mathds{E}[T] = \sum_{j = 1}^N (-1)^{j + 1} {N\choose j} \frac{1}{1 - (1 - 1/M)^j} \,.
\end{equation}

We next compare two approaches for spectrally preparing $N$ qubits: parallel preparation with resonance checks versus repeated unconditioned attempts.
Let $\eta_{\mathrm{SD}}^*$ denote the ratio of the homogeneous to check-probe linewidth, and let $\tau$ be the duration of a single excitation-check attempt.
In the first scheme, all $N$ qubits are excited simultaneously, and their photon outputs are independently checked for spectral alignment.
The average amount of time required to successfully generate these photons will be
\begin{equation}
    \mathrm{Time~(independent~check)} = \tau\left(\frac{1}{\eta_{\mathrm{SD}}\eta_{\mathrm{Det}}}\right)^N \,.
\end{equation}

For the resonance-checked approach, the average preparation time is

\begin{equation}
    \mathrm{Time~(resonance~check)} = \tau~(1 + \mathds{E}[T])\left(\frac{1}{\eta_{\mathrm{SD}}^*\eta_{\mathrm{Det}}}\right)^N.
\end{equation}

The resulting speedup is

\begin{equation}
    \mathrm{Speedup} = \left(\frac{\eta_{\mathrm{SD}}^*}{\eta_\mathrm{SD}}\right)^N /(1 + \mathds{E}[T]).
\end{equation}

Note that $\mathds{E}[T]$ scales as $\log N$, whereas the numerator grows exponentially with $N$. 
Thus, the expected speedup always exceeds 1 for sufficiently large $N$.

\providecommand{\noopsort}[1]{}\providecommand{\singleletter}[1]{#1}%


\begin{thebibliography}{62}%
\makeatletter
\providecommand \@ifxundefined [1]{%
 \@ifx{#1\undefined}
}%
\providecommand \@ifnum [1]{%
 \ifnum #1\expandafter \@firstoftwo
 \else \expandafter \@secondoftwo
 \fi
}%
\providecommand \@ifx [1]{%
 \ifx #1\expandafter \@firstoftwo
 \else \expandafter \@secondoftwo
 \fi
}%
\providecommand \natexlab [1]{#1}%
\providecommand \enquote  [1]{``#1''}%
\providecommand \bibnamefont  [1]{#1}%
\providecommand \bibfnamefont [1]{#1}%
\providecommand \citenamefont [1]{#1}%
\providecommand \href@noop [0]{\@secondoftwo}%
\providecommand \href [0]{\begingroup \@sanitize@url \@href}%
\providecommand \@href[1]{\@@startlink{#1}\@@href}%
\providecommand \@@href[1]{\endgroup#1\@@endlink}%
\providecommand \@sanitize@url [0]{\catcode `\\12\catcode `\$12\catcode `\&12\catcode `\#12\catcode `\^12\catcode `\_12\catcode `\%12\relax}%
\providecommand \@@startlink[1]{}%
\providecommand \@@endlink[0]{}%
\providecommand \url  [0]{\begingroup\@sanitize@url \@url }%
\providecommand \@url [1]{\endgroup\@href {#1}{\urlprefix }}%
\providecommand \urlprefix  [0]{URL }%
\providecommand \Eprint [0]{\href }%
\providecommand \doibase [0]{https://doi.org/}%
\providecommand \selectlanguage [0]{\@gobble}%
\providecommand \bibinfo  [0]{\@secondoftwo}%
\providecommand \bibfield  [0]{\@secondoftwo}%
\providecommand \translation [1]{[#1]}%
\providecommand \BibitemOpen [0]{}%
\providecommand \bibitemStop [0]{}%
\providecommand \bibitemNoStop [0]{.\EOS\space}%
\providecommand \EOS [0]{\spacefactor3000\relax}%
\providecommand \BibitemShut  [1]{\csname bibitem#1\endcsname}%
\let\auto@bib@innerbib\@empty
\bibitem [{\citenamefont {Wolfowicz}\ \emph {et~al.}(2021)\citenamefont {Wolfowicz}, \citenamefont {Heremans}, \citenamefont {Anderson}, \citenamefont {Kanai}, \citenamefont {Seo}, \citenamefont {Gali}, \citenamefont {Galli},\ and\ \citenamefont {Awschalom}}]{Wolfowicz2021}%
  \BibitemOpen
  \bibfield  {author} {\bibinfo {author} {\bibfnamefont {G.}~\bibnamefont {Wolfowicz}}, \bibinfo {author} {\bibfnamefont {F.~J.}\ \bibnamefont {Heremans}}, \bibinfo {author} {\bibfnamefont {C.~P.}\ \bibnamefont {Anderson}}, \bibinfo {author} {\bibfnamefont {S.}~\bibnamefont {Kanai}}, \bibinfo {author} {\bibfnamefont {H.}~\bibnamefont {Seo}}, \bibinfo {author} {\bibfnamefont {A.}~\bibnamefont {Gali}}, \bibinfo {author} {\bibfnamefont {G.}~\bibnamefont {Galli}},\ and\ \bibinfo {author} {\bibfnamefont {D.~D.}\ \bibnamefont {Awschalom}},\ }\bibfield  {title} {\bibinfo {title} {{Quantum guidelines for solid-state spin defects}},\ }\href {https://doi.org/10.1038/s41578-021-00306-y} {\bibfield  {journal} {\bibinfo  {journal} {Nature Reviews Materials}\ }\textbf {\bibinfo {volume} {6}},\ \bibinfo {pages} {906} (\bibinfo {year} {2021})}\BibitemShut {NoStop}%
\bibitem [{\citenamefont {Michaels}\ \emph {et~al.}(2021)\citenamefont {Michaels}, \citenamefont {Martinez}, \citenamefont {Debroux}, \citenamefont {Parker}, \citenamefont {Stramma}, \citenamefont {Huber}, \citenamefont {Purser}, \citenamefont {Atature},\ and\ \citenamefont {Gangloff}}]{Michaels2021}%
  \BibitemOpen
  \bibfield  {author} {\bibinfo {author} {\bibfnamefont {C.~P.}\ \bibnamefont {Michaels}}, \bibinfo {author} {\bibfnamefont {J.~A.}\ \bibnamefont {Martinez}}, \bibinfo {author} {\bibfnamefont {R.}~\bibnamefont {Debroux}}, \bibinfo {author} {\bibfnamefont {R.~A.}\ \bibnamefont {Parker}}, \bibinfo {author} {\bibfnamefont {A.~M.}\ \bibnamefont {Stramma}}, \bibinfo {author} {\bibfnamefont {L.~I.}\ \bibnamefont {Huber}}, \bibinfo {author} {\bibfnamefont {C.~M.}\ \bibnamefont {Purser}}, \bibinfo {author} {\bibfnamefont {M.}~\bibnamefont {Atature}},\ and\ \bibinfo {author} {\bibfnamefont {D.~A.}\ \bibnamefont {Gangloff}},\ }\bibfield  {title} {\bibinfo {title} {{Multidimensional cluster states using a single spin-photon interface coupled strongly to an intrinsic nuclear register}},\ }\href {https://doi.org/10.22331/q-2021-10-19-565} {\bibfield  {journal} {\bibinfo  {journal} {Quantum}\ }\textbf {\bibinfo {volume} {5}},\ \bibinfo {pages} {565} (\bibinfo {year} {2021})}\BibitemShut {NoStop}%
\bibitem [{\citenamefont {Stas}\ \emph {et~al.}(2022)\citenamefont {Stas}, \citenamefont {Huan}, \citenamefont {Machielse}, \citenamefont {Knall}, \citenamefont {Suleymanzade}, \citenamefont {Pingault}, \citenamefont {Sutula}, \citenamefont {Ding}, \citenamefont {Knaut}, \citenamefont {Assumpcao}, \citenamefont {Wei}, \citenamefont {Bhaskar}, \citenamefont {Riedinger}, \citenamefont {Sukachev}, \citenamefont {Park}, \citenamefont {Lon{\v{c}}ar}, \citenamefont {Levonian},\ and\ \citenamefont {Lukin}}]{Stas2022robust}%
  \BibitemOpen
  \bibfield  {author} {\bibinfo {author} {\bibfnamefont {P.~J.}\ \bibnamefont {Stas}}, \bibinfo {author} {\bibfnamefont {Y.~Q.}\ \bibnamefont {Huan}}, \bibinfo {author} {\bibfnamefont {B.}~\bibnamefont {Machielse}}, \bibinfo {author} {\bibfnamefont {E.~N.}\ \bibnamefont {Knall}}, \bibinfo {author} {\bibfnamefont {A.}~\bibnamefont {Suleymanzade}}, \bibinfo {author} {\bibfnamefont {B.}~\bibnamefont {Pingault}}, \bibinfo {author} {\bibfnamefont {M.}~\bibnamefont {Sutula}}, \bibinfo {author} {\bibfnamefont {S.~W.}\ \bibnamefont {Ding}}, \bibinfo {author} {\bibfnamefont {C.~M.}\ \bibnamefont {Knaut}}, \bibinfo {author} {\bibfnamefont {D.~R.}\ \bibnamefont {Assumpcao}}, \bibinfo {author} {\bibfnamefont {Y.~C.}\ \bibnamefont {Wei}}, \bibinfo {author} {\bibfnamefont {M.~K.}\ \bibnamefont {Bhaskar}}, \bibinfo {author} {\bibfnamefont {R.}~\bibnamefont {Riedinger}}, \bibinfo {author} {\bibfnamefont {D.~D.}\ \bibnamefont {Sukachev}}, \bibinfo {author} {\bibfnamefont {H.}~\bibnamefont {Park}}, \bibinfo {author}
  {\bibfnamefont {M.}~\bibnamefont {Lon{\v{c}}ar}}, \bibinfo {author} {\bibfnamefont {D.~S.}\ \bibnamefont {Levonian}},\ and\ \bibinfo {author} {\bibfnamefont {M.~D.}\ \bibnamefont {Lukin}},\ }\bibfield  {title} {\bibinfo {title} {{Robust multi-qubit quantum network node with integrated error detection}},\ }\href {https://doi.org/10.1126/SCIENCE.ADD9771/SUPPL{\_}FILE/SCIENCE.ADD9771{\_}SM.PDF} {\bibfield  {journal} {\bibinfo  {journal} {Science}\ }\textbf {\bibinfo {volume} {378}},\ \bibinfo {pages} {557} (\bibinfo {year} {2022})}\BibitemShut {NoStop}%
\bibitem [{\citenamefont {Knaut}\ \emph {et~al.}(2024)\citenamefont {Knaut}, \citenamefont {Suleymanzade}, \citenamefont {Wei}, \citenamefont {Assumpcao}, \citenamefont {Stas}, \citenamefont {Huan}, \citenamefont {Machielse}, \citenamefont {Knall}, \citenamefont {Sutula}, \citenamefont {Baranes}, \citenamefont {Sinclair}, \citenamefont {De-Eknamkul}, \citenamefont {Levonian}, \citenamefont {Bhaskar}, \citenamefont {Park}, \citenamefont {Lon{\v{c}}ar},\ and\ \citenamefont {Lukin}}]{Knaut2024EntanglementNetwork}%
  \BibitemOpen
  \bibfield  {author} {\bibinfo {author} {\bibfnamefont {C.~M.}\ \bibnamefont {Knaut}}, \bibinfo {author} {\bibfnamefont {A.}~\bibnamefont {Suleymanzade}}, \bibinfo {author} {\bibfnamefont {Y.-C.}\ \bibnamefont {Wei}}, \bibinfo {author} {\bibfnamefont {D.~R.}\ \bibnamefont {Assumpcao}}, \bibinfo {author} {\bibfnamefont {P.-J.}\ \bibnamefont {Stas}}, \bibinfo {author} {\bibfnamefont {Y.~Q.}\ \bibnamefont {Huan}}, \bibinfo {author} {\bibfnamefont {B.}~\bibnamefont {Machielse}}, \bibinfo {author} {\bibfnamefont {E.~N.}\ \bibnamefont {Knall}}, \bibinfo {author} {\bibfnamefont {M.}~\bibnamefont {Sutula}}, \bibinfo {author} {\bibfnamefont {G.}~\bibnamefont {Baranes}}, \bibinfo {author} {\bibfnamefont {N.}~\bibnamefont {Sinclair}}, \bibinfo {author} {\bibfnamefont {C.}~\bibnamefont {De-Eknamkul}}, \bibinfo {author} {\bibfnamefont {D.~S.}\ \bibnamefont {Levonian}}, \bibinfo {author} {\bibfnamefont {M.~K.}\ \bibnamefont {Bhaskar}}, \bibinfo {author} {\bibfnamefont {H.}~\bibnamefont {Park}}, \bibinfo {author}
  {\bibfnamefont {M.}~\bibnamefont {Lon{\v{c}}ar}},\ and\ \bibinfo {author} {\bibfnamefont {M.~D.}\ \bibnamefont {Lukin}},\ }\bibfield  {title} {\bibinfo {title} {{Entanglement of nanophotonic quantum memory nodes in a telecom network}},\ }\href {https://doi.org/10.1038/s41586-024-07252-z} {\bibfield  {journal} {\bibinfo  {journal} {Nature}\ }\textbf {\bibinfo {volume} {629}},\ \bibinfo {pages} {573} (\bibinfo {year} {2024})}\BibitemShut {NoStop}%
\bibitem [{\citenamefont {Simmons}(2024)}]{Simmons_2024_ScalableFaultTolerant}%
  \BibitemOpen
  \bibfield  {author} {\bibinfo {author} {\bibfnamefont {S.}~\bibnamefont {Simmons}},\ }\bibfield  {title} {\bibinfo {title} {Scalable fault-tolerant quantum technologies with silicon color centers},\ }\href {https://doi.org/10.1103/PRXQuantum.5.010102} {\bibfield  {journal} {\bibinfo  {journal} {PRX Quantum}\ }\textbf {\bibinfo {volume} {5}},\ \bibinfo {pages} {010102} (\bibinfo {year} {2024})}\BibitemShut {NoStop}%
\bibitem [{\citenamefont {Wei}\ \emph {et~al.}(2024)\citenamefont {Wei}, \citenamefont {Stas}, \citenamefont {Suleymanzade}, \citenamefont {Baranes}, \citenamefont {Machado}, \citenamefont {Huan}, \citenamefont {Knaut}, \citenamefont {Ding}, \citenamefont {Merz}, \citenamefont {Knall}, \citenamefont {Yazlar}, \citenamefont {Sirotin}, \citenamefont {Wang}, \citenamefont {Machielse}, \citenamefont {Yelin}, \citenamefont {Borregaard}, \citenamefont {Park}, \citenamefont {Loncar},\ and\ \citenamefont {Lukin}}]{Wei2024UniversalDistributed}%
  \BibitemOpen
  \bibfield  {author} {\bibinfo {author} {\bibfnamefont {Y.-C.}\ \bibnamefont {Wei}}, \bibinfo {author} {\bibfnamefont {P.-J.}\ \bibnamefont {Stas}}, \bibinfo {author} {\bibfnamefont {A.}~\bibnamefont {Suleymanzade}}, \bibinfo {author} {\bibfnamefont {G.}~\bibnamefont {Baranes}}, \bibinfo {author} {\bibfnamefont {F.}~\bibnamefont {Machado}}, \bibinfo {author} {\bibfnamefont {Y.~Q.}\ \bibnamefont {Huan}}, \bibinfo {author} {\bibfnamefont {C.~M.}\ \bibnamefont {Knaut}}, \bibinfo {author} {\bibfnamefont {W.~S.}\ \bibnamefont {Ding}}, \bibinfo {author} {\bibfnamefont {M.}~\bibnamefont {Merz}}, \bibinfo {author} {\bibfnamefont {E.~N.}\ \bibnamefont {Knall}}, \bibinfo {author} {\bibfnamefont {U.}~\bibnamefont {Yazlar}}, \bibinfo {author} {\bibfnamefont {M.}~\bibnamefont {Sirotin}}, \bibinfo {author} {\bibfnamefont {I.~W.}\ \bibnamefont {Wang}}, \bibinfo {author} {\bibfnamefont {B.}~\bibnamefont {Machielse}}, \bibinfo {author} {\bibfnamefont {S.~F.}\ \bibnamefont {Yelin}}, \bibinfo {author} {\bibfnamefont
  {J.}~\bibnamefont {Borregaard}}, \bibinfo {author} {\bibfnamefont {H.}~\bibnamefont {Park}}, \bibinfo {author} {\bibfnamefont {M.}~\bibnamefont {Loncar}},\ and\ \bibinfo {author} {\bibfnamefont {M.~D.}\ \bibnamefont {Lukin}},\ }\bibfield  {title} {\bibinfo {title} {{Universal distributed blind quantum computing with solid-state qubits}},\ }\href {https://arxiv.org/abs/2412.03020v2} {\  (\bibinfo {year} {2024})}\BibitemShut {NoStop}%
\bibitem [{\citenamefont {de~Gliniasty}\ \emph {et~al.}(2024)\citenamefont {de~Gliniasty}, \citenamefont {Hilaire}, \citenamefont {Emeriau}, \citenamefont {Wein}, \citenamefont {Salavrakos},\ and\ \citenamefont {Mansfield}}]{Gliniasty2024SpinOptical}%
  \BibitemOpen
  \bibfield  {author} {\bibinfo {author} {\bibfnamefont {G.}~\bibnamefont {de~Gliniasty}}, \bibinfo {author} {\bibfnamefont {P.}~\bibnamefont {Hilaire}}, \bibinfo {author} {\bibfnamefont {P.~E.}\ \bibnamefont {Emeriau}}, \bibinfo {author} {\bibfnamefont {S.~C.}\ \bibnamefont {Wein}}, \bibinfo {author} {\bibfnamefont {A.}~\bibnamefont {Salavrakos}},\ and\ \bibinfo {author} {\bibfnamefont {S.}~\bibnamefont {Mansfield}},\ }\bibfield  {title} {\bibinfo {title} {{A Spin-Optical Quantum Computing Architecture}},\ }\href {https://doi.org/10.22331/q-2024-07-24-1423} {\bibfield  {journal} {\bibinfo  {journal} {Quantum}\ }\textbf {\bibinfo {volume} {8}},\ \bibinfo {pages} {1423} (\bibinfo {year} {2024})}\BibitemShut {NoStop}%
\bibitem [{\citenamefont {Gritsch}\ \emph {et~al.}(2023)\citenamefont {Gritsch}, \citenamefont {Ulanowski},\ and\ \citenamefont {Reiserer}}]{Gritsch2023Purcell}%
  \BibitemOpen
  \bibfield  {author} {\bibinfo {author} {\bibfnamefont {A.}~\bibnamefont {Gritsch}}, \bibinfo {author} {\bibfnamefont {A.}~\bibnamefont {Ulanowski}},\ and\ \bibinfo {author} {\bibfnamefont {A.}~\bibnamefont {Reiserer}},\ }\bibfield  {title} {\bibinfo {title} {{Purcell enhancement of single-photon emitters in silicon}},\ }\href {https://doi.org/10.1364/optica.486167} {\bibfield  {journal} {\bibinfo  {journal} {Optica}\ }\textbf {\bibinfo {volume} {10}},\ \bibinfo {pages} {783} (\bibinfo {year} {2023})}\BibitemShut {NoStop}%
\bibitem [{\citenamefont {Merkel}\ \emph {et~al.}(2020)\citenamefont {Merkel}, \citenamefont {Ulanowski},\ and\ \citenamefont {Reiserer}}]{Merkel2020CoherentPurcell}%
  \BibitemOpen
  \bibfield  {author} {\bibinfo {author} {\bibfnamefont {B.}~\bibnamefont {Merkel}}, \bibinfo {author} {\bibfnamefont {A.}~\bibnamefont {Ulanowski}},\ and\ \bibinfo {author} {\bibfnamefont {A.}~\bibnamefont {Reiserer}},\ }\bibfield  {title} {\bibinfo {title} {{Coherent and Purcell-Enhanced Emission from Erbium Dopants in a Cryogenic High- Q Resonator}},\ }\href {https://doi.org/10.1103/PHYSREVX.10.041025/COHERENT{\_}EMISSION{\_}OF{\_}ERBIUM{\_}DOPANTS{\_}IN{\_}A{\_}HIGH{\_}Q{\_}RESONATOR{\_}FINAL{\_}SUPPLEMENT{\_}.PDF} {\bibfield  {journal} {\bibinfo  {journal} {Physical Review X}\ }\textbf {\bibinfo {volume} {10}},\ \bibinfo {pages} {041025} (\bibinfo {year} {2020})}\BibitemShut {NoStop}%
\bibitem [{\citenamefont {Li}\ \emph {et~al.}(2021)\citenamefont {Li}, \citenamefont {Guo}, \citenamefont {Xie}, \citenamefont {Luo}, \citenamefont {Siew}, \citenamefont {Luo}, \citenamefont {Dong}, \citenamefont {Zheng}, \citenamefont {Zhang}, \citenamefont {Gao}, \citenamefont {Lo}, \citenamefont {Song},\ and\ \citenamefont {Li}}]{Siew2021ReviewSilicon}%
  \BibitemOpen
  \bibfield  {author} {\bibinfo {author} {\bibfnamefont {C.}~\bibnamefont {Li}}, \bibinfo {author} {\bibfnamefont {P.}~\bibnamefont {Guo}}, \bibinfo {author} {\bibfnamefont {S.~W.}\ \bibnamefont {Xie}}, \bibinfo {author} {\bibfnamefont {L.~W.}\ \bibnamefont {Luo}}, \bibinfo {author} {\bibfnamefont {S.~Y.}\ \bibnamefont {Siew}}, \bibinfo {author} {\bibfnamefont {X.}~\bibnamefont {Luo}}, \bibinfo {author} {\bibfnamefont {B.}~\bibnamefont {Dong}}, \bibinfo {author} {\bibfnamefont {H.~Y.}\ \bibnamefont {Zheng}}, \bibinfo {author} {\bibfnamefont {W.}~\bibnamefont {Zhang}}, \bibinfo {author} {\bibfnamefont {F.}~\bibnamefont {Gao}}, \bibinfo {author} {\bibfnamefont {G.-Q.}\ \bibnamefont {Lo}}, \bibinfo {author} {\bibfnamefont {A.}~\bibnamefont {Song}},\ and\ \bibinfo {author} {\bibfnamefont {B.}~\bibnamefont {Li}},\ }\bibfield  {title} {\bibinfo {title} {{Review of Silicon Photonics Technology and Platform Development}},\ }\href {https://doi.org/10.1109/JLT.2021.3066203} {\bibfield  {journal} {\bibinfo  {journal}
  {Journal of Lightwave Technology, Vol. 39, Issue 13, pp. 4374-4389}\ }\textbf {\bibinfo {volume} {39}},\ \bibinfo {pages} {4374} (\bibinfo {year} {2021})}\BibitemShut {NoStop}%
\bibitem [{\citenamefont {Wolters}\ \emph {et~al.}(2013)\citenamefont {Wolters}, \citenamefont {Sadzak}, \citenamefont {Schell}, \citenamefont {Schr\"oder},\ and\ \citenamefont {Benson}}]{Wolters_2013_UltrafastSDDiamondNV}%
  \BibitemOpen
  \bibfield  {author} {\bibinfo {author} {\bibfnamefont {J.}~\bibnamefont {Wolters}}, \bibinfo {author} {\bibfnamefont {N.}~\bibnamefont {Sadzak}}, \bibinfo {author} {\bibfnamefont {A.~W.}\ \bibnamefont {Schell}}, \bibinfo {author} {\bibfnamefont {T.}~\bibnamefont {Schr\"oder}},\ and\ \bibinfo {author} {\bibfnamefont {O.}~\bibnamefont {Benson}},\ }\bibfield  {title} {\bibinfo {title} {Measurement of the ultrafast spectral diffusion of the optical transition of nitrogen vacancy centers in nano-size diamond using correlation interferometry},\ }\href {https://doi.org/10.1103/PhysRevLett.110.027401} {\bibfield  {journal} {\bibinfo  {journal} {Phys. Rev. Lett.}\ }\textbf {\bibinfo {volume} {110}},\ \bibinfo {pages} {027401} (\bibinfo {year} {2013})}\BibitemShut {NoStop}%
\bibitem [{\citenamefont {Heiler}\ \emph {et~al.}(2024)\citenamefont {Heiler}, \citenamefont {K{\"o}rber}, \citenamefont {Hesselmeier}, \citenamefont {Kuna}, \citenamefont {St{\"o}hr}, \citenamefont {Fuchs}, \citenamefont {Ghezellou}, \citenamefont {Ul-Hassan}, \citenamefont {Knolle}, \citenamefont {Becher}, \citenamefont {Kaiser},\ and\ \citenamefont {Wrachtrup}}]{Heiler_2024_SiCSDthicknessdependence}%
  \BibitemOpen
  \bibfield  {author} {\bibinfo {author} {\bibfnamefont {J.}~\bibnamefont {Heiler}}, \bibinfo {author} {\bibfnamefont {J.}~\bibnamefont {K{\"o}rber}}, \bibinfo {author} {\bibfnamefont {E.}~\bibnamefont {Hesselmeier}}, \bibinfo {author} {\bibfnamefont {P.}~\bibnamefont {Kuna}}, \bibinfo {author} {\bibfnamefont {R.}~\bibnamefont {St{\"o}hr}}, \bibinfo {author} {\bibfnamefont {P.}~\bibnamefont {Fuchs}}, \bibinfo {author} {\bibfnamefont {M.}~\bibnamefont {Ghezellou}}, \bibinfo {author} {\bibfnamefont {J.}~\bibnamefont {Ul-Hassan}}, \bibinfo {author} {\bibfnamefont {W.}~\bibnamefont {Knolle}}, \bibinfo {author} {\bibfnamefont {C.}~\bibnamefont {Becher}}, \bibinfo {author} {\bibfnamefont {F.}~\bibnamefont {Kaiser}},\ and\ \bibinfo {author} {\bibfnamefont {J.}~\bibnamefont {Wrachtrup}},\ }\bibfield  {title} {\bibinfo {title} {Spectral stability of v2 centres in sub-micron 4h-sic membranes},\ }\href {https://doi.org/10.1038/s41535-024-00644-4} {\bibfield  {journal} {\bibinfo  {journal} {npj Quantum Materials}\
  }\textbf {\bibinfo {volume} {9}},\ \bibinfo {pages} {34} (\bibinfo {year} {2024})}\BibitemShut {NoStop}%
\bibitem [{\citenamefont {Higginbottom}\ \emph {et~al.}(2022)\citenamefont {Higginbottom}, \citenamefont {Kurkjian}, \citenamefont {Chartrand}, \citenamefont {Kazemi}, \citenamefont {Brunelle}, \citenamefont {MacQuarrie}, \citenamefont {Klein}, \citenamefont {Lee-Hone}, \citenamefont {Stacho}, \citenamefont {Ruether}, \citenamefont {Bowness}, \citenamefont {Bergeron}, \citenamefont {DeAbreu}, \citenamefont {Harrigan}, \citenamefont {Kanaganayagam}, \citenamefont {Marsden}, \citenamefont {Richards}, \citenamefont {Stott}, \citenamefont {Roorda}, \citenamefont {Morse}, \citenamefont {Thewalt},\ and\ \citenamefont {Simmons}}]{Higginbottom_2022_OpticalSpinsInSilicon}%
  \BibitemOpen
  \bibfield  {author} {\bibinfo {author} {\bibfnamefont {D.~B.}\ \bibnamefont {Higginbottom}}, \bibinfo {author} {\bibfnamefont {A.~T.~K.}\ \bibnamefont {Kurkjian}}, \bibinfo {author} {\bibfnamefont {C.}~\bibnamefont {Chartrand}}, \bibinfo {author} {\bibfnamefont {M.}~\bibnamefont {Kazemi}}, \bibinfo {author} {\bibfnamefont {N.~A.}\ \bibnamefont {Brunelle}}, \bibinfo {author} {\bibfnamefont {E.~R.}\ \bibnamefont {MacQuarrie}}, \bibinfo {author} {\bibfnamefont {J.~R.}\ \bibnamefont {Klein}}, \bibinfo {author} {\bibfnamefont {N.~R.}\ \bibnamefont {Lee-Hone}}, \bibinfo {author} {\bibfnamefont {J.}~\bibnamefont {Stacho}}, \bibinfo {author} {\bibfnamefont {M.}~\bibnamefont {Ruether}}, \bibinfo {author} {\bibfnamefont {C.}~\bibnamefont {Bowness}}, \bibinfo {author} {\bibfnamefont {L.}~\bibnamefont {Bergeron}}, \bibinfo {author} {\bibfnamefont {A.}~\bibnamefont {DeAbreu}}, \bibinfo {author} {\bibfnamefont {S.~R.}\ \bibnamefont {Harrigan}}, \bibinfo {author} {\bibfnamefont {J.}~\bibnamefont {Kanaganayagam}}, \bibinfo
  {author} {\bibfnamefont {D.~W.}\ \bibnamefont {Marsden}}, \bibinfo {author} {\bibfnamefont {T.~S.}\ \bibnamefont {Richards}}, \bibinfo {author} {\bibfnamefont {L.~A.}\ \bibnamefont {Stott}}, \bibinfo {author} {\bibfnamefont {S.}~\bibnamefont {Roorda}}, \bibinfo {author} {\bibfnamefont {K.~J.}\ \bibnamefont {Morse}}, \bibinfo {author} {\bibfnamefont {M.~L.~W.}\ \bibnamefont {Thewalt}},\ and\ \bibinfo {author} {\bibfnamefont {S.}~\bibnamefont {Simmons}},\ }\bibfield  {title} {\bibinfo {title} {Optical observation of single spins in silicon},\ }\href {https://doi.org/10.1038/s41586-022-04821-y} {\bibfield  {journal} {\bibinfo  {journal} {Nature}\ }\textbf {\bibinfo {volume} {607}},\ \bibinfo {pages} {266} (\bibinfo {year} {2022})}\BibitemShut {NoStop}%
\bibitem [{\citenamefont {Johnston}\ \emph {et~al.}(2024)\citenamefont {Johnston}, \citenamefont {Felix-Rendon}, \citenamefont {Wong},\ and\ \citenamefont {Chen}}]{Johnston_2024_CavityCoupledTcentre}%
  \BibitemOpen
  \bibfield  {author} {\bibinfo {author} {\bibfnamefont {A.}~\bibnamefont {Johnston}}, \bibinfo {author} {\bibfnamefont {U.}~\bibnamefont {Felix-Rendon}}, \bibinfo {author} {\bibfnamefont {Y.-E.}\ \bibnamefont {Wong}},\ and\ \bibinfo {author} {\bibfnamefont {S.}~\bibnamefont {Chen}},\ }\bibfield  {title} {\bibinfo {title} {Cavity-coupled telecom atomic source in silicon},\ }\href {https://doi.org/10.1038/s41467-024-46643-8} {\bibfield  {journal} {\bibinfo  {journal} {Nature Communications}\ }\textbf {\bibinfo {volume} {15}},\ \bibinfo {pages} {2350} (\bibinfo {year} {2024})}\BibitemShut {NoStop}%
\bibitem [{\citenamefont {Lefaucher}\ \emph {et~al.}(2025)\citenamefont {Lefaucher}, \citenamefont {Baron}, \citenamefont {Jager}, \citenamefont {Calvo}, \citenamefont {Els{\"{a}}sser}, \citenamefont {Coppola}, \citenamefont {Mazen}, \citenamefont {Kerdil{\`{e}}s}, \citenamefont {Cache}, \citenamefont {Dr{\'{e}}au}, \citenamefont {G{\'{e}}rard},\ and\ \citenamefont {Grenoble}}]{Lefaucher2024BrightSingle}%
  \BibitemOpen
  \bibfield  {author} {\bibinfo {author} {\bibfnamefont {B.}~\bibnamefont {Lefaucher}}, \bibinfo {author} {\bibfnamefont {Y.}~\bibnamefont {Baron}}, \bibinfo {author} {\bibfnamefont {J.-B.}\ \bibnamefont {Jager}}, \bibinfo {author} {\bibfnamefont {V.}~\bibnamefont {Calvo}}, \bibinfo {author} {\bibfnamefont {C.}~\bibnamefont {Els{\"{a}}sser}}, \bibinfo {author} {\bibfnamefont {G.}~\bibnamefont {Coppola}}, \bibinfo {author} {\bibfnamefont {F.}~\bibnamefont {Mazen}}, \bibinfo {author} {\bibfnamefont {S.}~\bibnamefont {Kerdil{\`{e}}s}}, \bibinfo {author} {\bibfnamefont {F.}~\bibnamefont {Cache}}, \bibinfo {author} {\bibfnamefont {A.}~\bibnamefont {Dr{\'{e}}au}}, \bibinfo {author} {\bibfnamefont {J.-M.}\ \bibnamefont {G{\'{e}}rard}},\ and\ \bibinfo {author} {\bibfnamefont {â.}~\bibnamefont {Grenoble}},\ }\bibfield  {title} {\bibinfo {title} {{Bright single-photon source in a silicon chip by nanoscale positioning of a color center in a microcavity}},\ }\href {http://arxiv.org/abs/2501.12744} {\  (\bibinfo {year}
  {2025})}\BibitemShut {NoStop}%
\bibitem [{\citenamefont {Redjem}\ \emph {et~al.}(2023)\citenamefont {Redjem}, \citenamefont {Zhiyenbayev}, \citenamefont {Qarony}, \citenamefont {Ivanov}, \citenamefont {Papapanos}, \citenamefont {Liu}, \citenamefont {Jhuria}, \citenamefont {Al~Balushi}, \citenamefont {Dhuey}, \citenamefont {Schwartzberg}, \citenamefont {Tan}, \citenamefont {Schenkel},\ and\ \citenamefont {Kant{\'{e}}}}]{Redjem2024AllSilicon}%
  \BibitemOpen
  \bibfield  {author} {\bibinfo {author} {\bibfnamefont {W.}~\bibnamefont {Redjem}}, \bibinfo {author} {\bibfnamefont {Y.}~\bibnamefont {Zhiyenbayev}}, \bibinfo {author} {\bibfnamefont {W.}~\bibnamefont {Qarony}}, \bibinfo {author} {\bibfnamefont {V.}~\bibnamefont {Ivanov}}, \bibinfo {author} {\bibfnamefont {C.}~\bibnamefont {Papapanos}}, \bibinfo {author} {\bibfnamefont {W.}~\bibnamefont {Liu}}, \bibinfo {author} {\bibfnamefont {K.}~\bibnamefont {Jhuria}}, \bibinfo {author} {\bibfnamefont {Z.~Y.}\ \bibnamefont {Al~Balushi}}, \bibinfo {author} {\bibfnamefont {S.}~\bibnamefont {Dhuey}}, \bibinfo {author} {\bibfnamefont {A.}~\bibnamefont {Schwartzberg}}, \bibinfo {author} {\bibfnamefont {L.~Z.}\ \bibnamefont {Tan}}, \bibinfo {author} {\bibfnamefont {T.}~\bibnamefont {Schenkel}},\ and\ \bibinfo {author} {\bibfnamefont {B.}~\bibnamefont {Kant{\'{e}}}},\ }\bibfield  {title} {\bibinfo {title} {{All-silicon quantum light source by embedding an atomic emissive center in a nanophotonic cavity}},\ }\href
  {https://doi.org/10.1038/s41467-023-38559-6} {\bibfield  {journal} {\bibinfo  {journal} {Nature Communications 2023 14:1}\ }\textbf {\bibinfo {volume} {14}},\ \bibinfo {pages} {1} (\bibinfo {year} {2023})}\BibitemShut {NoStop}%
\bibitem [{\citenamefont {Saggio}\ \emph {et~al.}(2024)\citenamefont {Saggio}, \citenamefont {Errando-Herranz}, \citenamefont {Gyger}, \citenamefont {Panuski}, \citenamefont {Prabhu}, \citenamefont {De~Santis}, \citenamefont {Christen}, \citenamefont {Ornelas-Huerta}, \citenamefont {Raniwala}, \citenamefont {Gerlach}, \citenamefont {Colangelo},\ and\ \citenamefont {Englund}}]{Saggio2024CavityEnhanced}%
  \BibitemOpen
  \bibfield  {author} {\bibinfo {author} {\bibfnamefont {V.}~\bibnamefont {Saggio}}, \bibinfo {author} {\bibfnamefont {C.}~\bibnamefont {Errando-Herranz}}, \bibinfo {author} {\bibfnamefont {S.}~\bibnamefont {Gyger}}, \bibinfo {author} {\bibfnamefont {C.}~\bibnamefont {Panuski}}, \bibinfo {author} {\bibfnamefont {M.}~\bibnamefont {Prabhu}}, \bibinfo {author} {\bibfnamefont {L.}~\bibnamefont {De~Santis}}, \bibinfo {author} {\bibfnamefont {I.}~\bibnamefont {Christen}}, \bibinfo {author} {\bibfnamefont {D.}~\bibnamefont {Ornelas-Huerta}}, \bibinfo {author} {\bibfnamefont {H.}~\bibnamefont {Raniwala}}, \bibinfo {author} {\bibfnamefont {C.}~\bibnamefont {Gerlach}}, \bibinfo {author} {\bibfnamefont {M.}~\bibnamefont {Colangelo}},\ and\ \bibinfo {author} {\bibfnamefont {D.}~\bibnamefont {Englund}},\ }\bibfield  {title} {\bibinfo {title} {{Cavity-enhanced single artificial atoms in silicon}},\ }\href {https://doi.org/10.1038/s41467-024-49302-0} {\bibfield  {journal} {\bibinfo  {journal} {Nature Communications 2024
  15:1}\ }\textbf {\bibinfo {volume} {15}},\ \bibinfo {pages} {1} (\bibinfo {year} {2024})}\BibitemShut {NoStop}%
\bibitem [{\citenamefont {Komza}\ \emph {et~al.}(2025)\citenamefont {Komza}, \citenamefont {Zhang}, \citenamefont {Song}, \citenamefont {Tang}, \citenamefont {Wei},\ and\ \citenamefont {Sipahigil}}]{komza_2025_multiplexedcolorcenterssilicon}%
  \BibitemOpen
  \bibfield  {author} {\bibinfo {author} {\bibfnamefont {L.}~\bibnamefont {Komza}}, \bibinfo {author} {\bibfnamefont {X.}~\bibnamefont {Zhang}}, \bibinfo {author} {\bibfnamefont {H.}~\bibnamefont {Song}}, \bibinfo {author} {\bibfnamefont {Y.-L.}\ \bibnamefont {Tang}}, \bibinfo {author} {\bibfnamefont {X.}~\bibnamefont {Wei}},\ and\ \bibinfo {author} {\bibfnamefont {A.}~\bibnamefont {Sipahigil}},\ }\href {https://arxiv.org/abs/2501.17339} {\bibinfo {title} {Multiplexed color centers in a silicon photonic cavity array}} (\bibinfo {year} {2025}),\ \Eprint {https://arxiv.org/abs/2501.17339} {arXiv:2501.17339} \BibitemShut {NoStop}%
\bibitem [{\citenamefont {Afzal}\ \emph {et~al.}(2024)\citenamefont {Afzal}, \citenamefont {Akhlaghi}, \citenamefont {Beale}, \citenamefont {Bedroya}, \citenamefont {Bell}, \citenamefont {Bergeron}, \citenamefont {Bonsma-Fisher}, \citenamefont {Bychkova}, \citenamefont {Chaisson}, \citenamefont {Chartrand}, \citenamefont {Clear}, \citenamefont {Darcie}, \citenamefont {DeAbreu}, \citenamefont {DeLisle}, \citenamefont {Duncan}, \citenamefont {Smith}, \citenamefont {Dunn}, \citenamefont {Ebrahimi}, \citenamefont {Evetts}, \citenamefont {Pinheiro}, \citenamefont {Fuentes}, \citenamefont {Georgiou}, \citenamefont {Guha}, \citenamefont {Haenel}, \citenamefont {Higginbottom}, \citenamefont {Jackson}, \citenamefont {Jahed}, \citenamefont {Khorshidahmad}, \citenamefont {Shandilya}, \citenamefont {Kurkjian}, \citenamefont {Lauk}, \citenamefont {Lee-Hone}, \citenamefont {Lin}, \citenamefont {Litynskyy}, \citenamefont {Lock}, \citenamefont {Ma}, \citenamefont {MacGilp}, \citenamefont {MacQuarrie}, \citenamefont {Mar},
  \citenamefont {Khah}, \citenamefont {Matiash}, \citenamefont {Meyer-Scott}, \citenamefont {Michaels}, \citenamefont {Motira}, \citenamefont {Noori}, \citenamefont {Ospadov}, \citenamefont {Patel}, \citenamefont {Patscheider}, \citenamefont {Paulson}, \citenamefont {Petruk}, \citenamefont {Ravindranath}, \citenamefont {Reznychenko}, \citenamefont {Ruether}, \citenamefont {Ruscica}, \citenamefont {Saxena}, \citenamefont {Schaller}, \citenamefont {Seidlitz}, \citenamefont {Senger}, \citenamefont {Lee}, \citenamefont {Sevoyan}, \citenamefont {Simmons}, \citenamefont {Soykal}, \citenamefont {Stott}, \citenamefont {Tran}, \citenamefont {Tserkis}, \citenamefont {Ulhaq}, \citenamefont {Vine}, \citenamefont {Weeks}, \citenamefont {Wolfowicz},\ and\ \citenamefont {Yoneda}}]{Afzal2024}%
  \BibitemOpen
  \bibfield  {author} {\bibinfo {author} {\bibfnamefont {F.}~\bibnamefont {Afzal}}, \bibinfo {author} {\bibfnamefont {M.}~\bibnamefont {Akhlaghi}}, \bibinfo {author} {\bibfnamefont {S.~J.}\ \bibnamefont {Beale}}, \bibinfo {author} {\bibfnamefont {O.}~\bibnamefont {Bedroya}}, \bibinfo {author} {\bibfnamefont {K.}~\bibnamefont {Bell}}, \bibinfo {author} {\bibfnamefont {L.}~\bibnamefont {Bergeron}}, \bibinfo {author} {\bibfnamefont {K.}~\bibnamefont {Bonsma-Fisher}}, \bibinfo {author} {\bibfnamefont {P.}~\bibnamefont {Bychkova}}, \bibinfo {author} {\bibfnamefont {Z.~M.~E.}\ \bibnamefont {Chaisson}}, \bibinfo {author} {\bibfnamefont {C.}~\bibnamefont {Chartrand}}, \bibinfo {author} {\bibfnamefont {C.}~\bibnamefont {Clear}}, \bibinfo {author} {\bibfnamefont {A.}~\bibnamefont {Darcie}}, \bibinfo {author} {\bibfnamefont {A.}~\bibnamefont {DeAbreu}}, \bibinfo {author} {\bibfnamefont {C.}~\bibnamefont {DeLisle}}, \bibinfo {author} {\bibfnamefont {L.~A.}\ \bibnamefont {Duncan}}, \bibinfo {author} {\bibfnamefont
  {C.~D.}\ \bibnamefont {Smith}}, \bibinfo {author} {\bibfnamefont {J.}~\bibnamefont {Dunn}}, \bibinfo {author} {\bibfnamefont {A.}~\bibnamefont {Ebrahimi}}, \bibinfo {author} {\bibfnamefont {N.}~\bibnamefont {Evetts}}, \bibinfo {author} {\bibfnamefont {D.~F.}\ \bibnamefont {Pinheiro}}, \bibinfo {author} {\bibfnamefont {P.}~\bibnamefont {Fuentes}}, \bibinfo {author} {\bibfnamefont {T.}~\bibnamefont {Georgiou}}, \bibinfo {author} {\bibfnamefont {B.}~\bibnamefont {Guha}}, \bibinfo {author} {\bibfnamefont {R.}~\bibnamefont {Haenel}}, \bibinfo {author} {\bibfnamefont {D.}~\bibnamefont {Higginbottom}}, \bibinfo {author} {\bibfnamefont {D.~M.}\ \bibnamefont {Jackson}}, \bibinfo {author} {\bibfnamefont {N.}~\bibnamefont {Jahed}}, \bibinfo {author} {\bibfnamefont {A.}~\bibnamefont {Khorshidahmad}}, \bibinfo {author} {\bibfnamefont {P.~K.}\ \bibnamefont {Shandilya}}, \bibinfo {author} {\bibfnamefont {A.~T.~K.}\ \bibnamefont {Kurkjian}}, \bibinfo {author} {\bibfnamefont {N.}~\bibnamefont {Lauk}}, \bibinfo {author}
  {\bibfnamefont {N.~R.}\ \bibnamefont {Lee-Hone}}, \bibinfo {author} {\bibfnamefont {E.}~\bibnamefont {Lin}}, \bibinfo {author} {\bibfnamefont {R.}~\bibnamefont {Litynskyy}}, \bibinfo {author} {\bibfnamefont {D.}~\bibnamefont {Lock}}, \bibinfo {author} {\bibfnamefont {L.}~\bibnamefont {Ma}}, \bibinfo {author} {\bibfnamefont {I.}~\bibnamefont {MacGilp}}, \bibinfo {author} {\bibfnamefont {E.~R.}\ \bibnamefont {MacQuarrie}}, \bibinfo {author} {\bibfnamefont {A.}~\bibnamefont {Mar}}, \bibinfo {author} {\bibfnamefont {A.~M.}\ \bibnamefont {Khah}}, \bibinfo {author} {\bibfnamefont {A.}~\bibnamefont {Matiash}}, \bibinfo {author} {\bibfnamefont {E.}~\bibnamefont {Meyer-Scott}}, \bibinfo {author} {\bibfnamefont {C.~P.}\ \bibnamefont {Michaels}}, \bibinfo {author} {\bibfnamefont {J.}~\bibnamefont {Motira}}, \bibinfo {author} {\bibfnamefont {N.~K.}\ \bibnamefont {Noori}}, \bibinfo {author} {\bibfnamefont {E.}~\bibnamefont {Ospadov}}, \bibinfo {author} {\bibfnamefont {E.}~\bibnamefont {Patel}}, \bibinfo {author}
  {\bibfnamefont {A.}~\bibnamefont {Patscheider}}, \bibinfo {author} {\bibfnamefont {D.}~\bibnamefont {Paulson}}, \bibinfo {author} {\bibfnamefont {A.}~\bibnamefont {Petruk}}, \bibinfo {author} {\bibfnamefont {A.~L.}\ \bibnamefont {Ravindranath}}, \bibinfo {author} {\bibfnamefont {B.}~\bibnamefont {Reznychenko}}, \bibinfo {author} {\bibfnamefont {M.}~\bibnamefont {Ruether}}, \bibinfo {author} {\bibfnamefont {J.}~\bibnamefont {Ruscica}}, \bibinfo {author} {\bibfnamefont {K.}~\bibnamefont {Saxena}}, \bibinfo {author} {\bibfnamefont {Z.}~\bibnamefont {Schaller}}, \bibinfo {author} {\bibfnamefont {A.}~\bibnamefont {Seidlitz}}, \bibinfo {author} {\bibfnamefont {J.}~\bibnamefont {Senger}}, \bibinfo {author} {\bibfnamefont {Y.~S.}\ \bibnamefont {Lee}}, \bibinfo {author} {\bibfnamefont {O.}~\bibnamefont {Sevoyan}}, \bibinfo {author} {\bibfnamefont {S.}~\bibnamefont {Simmons}}, \bibinfo {author} {\bibfnamefont {O.}~\bibnamefont {Soykal}}, \bibinfo {author} {\bibfnamefont {L.}~\bibnamefont {Stott}}, \bibinfo {author}
  {\bibfnamefont {Q.}~\bibnamefont {Tran}}, \bibinfo {author} {\bibfnamefont {S.}~\bibnamefont {Tserkis}}, \bibinfo {author} {\bibfnamefont {A.}~\bibnamefont {Ulhaq}}, \bibinfo {author} {\bibfnamefont {W.}~\bibnamefont {Vine}}, \bibinfo {author} {\bibfnamefont {R.}~\bibnamefont {Weeks}}, \bibinfo {author} {\bibfnamefont {G.}~\bibnamefont {Wolfowicz}},\ and\ \bibinfo {author} {\bibfnamefont {I.}~\bibnamefont {Yoneda}},\ }\bibfield  {title} {\bibinfo {title} {{Distributed Quantum Computing in Silicon}},\ }\href {http://arxiv.org/abs/2406.01704} {\bibfield  {journal} {\bibinfo  {journal} {arXiv 2406.01704}\ } (\bibinfo {year} {2024})}\BibitemShut {NoStop}%
\bibitem [{\citenamefont {Saeedi}\ \emph {et~al.}(2013)\citenamefont {Saeedi}, \citenamefont {Simmons}, \citenamefont {Salvail}, \citenamefont {Dluhy}, \citenamefont {Riemann}, \citenamefont {Abromosimov}, \citenamefont {Becker}, \citenamefont {Pohl}, \citenamefont {Morton},\ and\ \citenamefont {THewalt}}]{Saeedi2013}%
  \BibitemOpen
  \bibfield  {author} {\bibinfo {author} {\bibfnamefont {K.}~\bibnamefont {Saeedi}}, \bibinfo {author} {\bibfnamefont {S.}~\bibnamefont {Simmons}}, \bibinfo {author} {\bibfnamefont {J.~Z.}\ \bibnamefont {Salvail}}, \bibinfo {author} {\bibfnamefont {P.}~\bibnamefont {Dluhy}}, \bibinfo {author} {\bibfnamefont {H.}~\bibnamefont {Riemann}}, \bibinfo {author} {\bibfnamefont {N.~V.}\ \bibnamefont {Abromosimov}}, \bibinfo {author} {\bibfnamefont {P.}~\bibnamefont {Becker}}, \bibinfo {author} {\bibfnamefont {H.-J.}\ \bibnamefont {Pohl}}, \bibinfo {author} {\bibfnamefont {J.~J.~L.}\ \bibnamefont {Morton}},\ and\ \bibinfo {author} {\bibfnamefont {M.~L.~W.}\ \bibnamefont {THewalt}},\ }\bibfield  {title} {\bibinfo {title} {{Room-temperature quantum bit storage exceeding 39 minutes using ionized donors in silicon-28}},\ }\href {https://doi.org/10.1126/science.1239584} {\bibfield  {journal} {\bibinfo  {journal} {Science}\ }\textbf {\bibinfo {volume} {342}},\ \bibinfo {pages} {830} (\bibinfo {year} {2013})}\BibitemShut
  {NoStop}%
\bibitem [{\citenamefont {Chartrand}\ \emph {et~al.}(2018)\citenamefont {Chartrand}, \citenamefont {Bergeron}, \citenamefont {Morse}, \citenamefont {Riemann}, \citenamefont {Abrosimov}, \citenamefont {Becker}, \citenamefont {Pohl}, \citenamefont {Simmons},\ and\ \citenamefont {Thewalt}}]{Chartrand_2018_CGW}%
  \BibitemOpen
  \bibfield  {author} {\bibinfo {author} {\bibfnamefont {C.}~\bibnamefont {Chartrand}}, \bibinfo {author} {\bibfnamefont {L.}~\bibnamefont {Bergeron}}, \bibinfo {author} {\bibfnamefont {K.~J.}\ \bibnamefont {Morse}}, \bibinfo {author} {\bibfnamefont {H.}~\bibnamefont {Riemann}}, \bibinfo {author} {\bibfnamefont {N.~V.}\ \bibnamefont {Abrosimov}}, \bibinfo {author} {\bibfnamefont {P.}~\bibnamefont {Becker}}, \bibinfo {author} {\bibfnamefont {H.-J.}\ \bibnamefont {Pohl}}, \bibinfo {author} {\bibfnamefont {S.}~\bibnamefont {Simmons}},\ and\ \bibinfo {author} {\bibfnamefont {M.~L.~W.}\ \bibnamefont {Thewalt}},\ }\bibfield  {title} {\bibinfo {title} {Highly enriched $^{28}\mathrm{Si}$ reveals remarkable optical linewidths and fine structure for well-known damage centers},\ }\href {https://doi.org/10.1103/PhysRevB.98.195201} {\bibfield  {journal} {\bibinfo  {journal} {Phys. Rev. B}\ }\textbf {\bibinfo {volume} {98}},\ \bibinfo {pages} {195201} (\bibinfo {year} {2018})}\BibitemShut {NoStop}%
\bibitem [{\citenamefont {Bergeron}\ \emph {et~al.}(2020)\citenamefont {Bergeron}, \citenamefont {Chartrand}, \citenamefont {Kurkjian}, \citenamefont {Morse}, \citenamefont {Riemann}, \citenamefont {Abrosimov}, \citenamefont {Becker}, \citenamefont {Pohl}, \citenamefont {Thewalt},\ and\ \citenamefont {Simmons}}]{Bergeron:2020_PRX}%
  \BibitemOpen
  \bibfield  {author} {\bibinfo {author} {\bibfnamefont {L.}~\bibnamefont {Bergeron}}, \bibinfo {author} {\bibfnamefont {C.}~\bibnamefont {Chartrand}}, \bibinfo {author} {\bibfnamefont {A.~T.~K.}\ \bibnamefont {Kurkjian}}, \bibinfo {author} {\bibfnamefont {K.~J.}\ \bibnamefont {Morse}}, \bibinfo {author} {\bibfnamefont {H.}~\bibnamefont {Riemann}}, \bibinfo {author} {\bibfnamefont {N.~V.}\ \bibnamefont {Abrosimov}}, \bibinfo {author} {\bibfnamefont {P.}~\bibnamefont {Becker}}, \bibinfo {author} {\bibfnamefont {H.-J.}\ \bibnamefont {Pohl}}, \bibinfo {author} {\bibfnamefont {M.~L.~W.}\ \bibnamefont {Thewalt}},\ and\ \bibinfo {author} {\bibfnamefont {S.}~\bibnamefont {Simmons}},\ }\bibfield  {title} {\bibinfo {title} {{Silicon-Integrated Telecommunications Photon-Spin Interface}},\ }\href {https://doi.org/10.1103/prxquantum.1.020301} {\bibfield  {journal} {\bibinfo  {journal} {PRX Quantum}\ }\textbf {\bibinfo {volume} {1}},\ \bibinfo {pages} {20301} (\bibinfo {year} {2020})}\BibitemShut {NoStop}%
\bibitem [{\citenamefont {DeAbreu}\ \emph {et~al.}(2023)\citenamefont {DeAbreu}, \citenamefont {Bowness}, \citenamefont {Alizadeh}, \citenamefont {Chartrand}, \citenamefont {Brunelle}, \citenamefont {MacQuarrie}, \citenamefont {Lee-Hone}, \citenamefont {Ruether}, \citenamefont {Kazemi}, \citenamefont {Kurkjian}, \citenamefont {Roorda}, \citenamefont {Abrosimov}, \citenamefont {Pohl}, \citenamefont {Thewalt}, \citenamefont {Higginbottom},\ and\ \citenamefont {Simmons}}]{DeAbreu_2023_WaveguideIntegratedCenters}%
  \BibitemOpen
  \bibfield  {author} {\bibinfo {author} {\bibfnamefont {A.}~\bibnamefont {DeAbreu}}, \bibinfo {author} {\bibfnamefont {C.}~\bibnamefont {Bowness}}, \bibinfo {author} {\bibfnamefont {A.}~\bibnamefont {Alizadeh}}, \bibinfo {author} {\bibfnamefont {C.}~\bibnamefont {Chartrand}}, \bibinfo {author} {\bibfnamefont {N.~A.}\ \bibnamefont {Brunelle}}, \bibinfo {author} {\bibfnamefont {E.~R.}\ \bibnamefont {MacQuarrie}}, \bibinfo {author} {\bibfnamefont {N.~R.}\ \bibnamefont {Lee-Hone}}, \bibinfo {author} {\bibfnamefont {M.}~\bibnamefont {Ruether}}, \bibinfo {author} {\bibfnamefont {M.}~\bibnamefont {Kazemi}}, \bibinfo {author} {\bibfnamefont {A.~T.~K.}\ \bibnamefont {Kurkjian}}, \bibinfo {author} {\bibfnamefont {S.}~\bibnamefont {Roorda}}, \bibinfo {author} {\bibfnamefont {N.~V.}\ \bibnamefont {Abrosimov}}, \bibinfo {author} {\bibfnamefont {H.-J.}\ \bibnamefont {Pohl}}, \bibinfo {author} {\bibfnamefont {M.~L.~W.}\ \bibnamefont {Thewalt}}, \bibinfo {author} {\bibfnamefont {D.~B.}\ \bibnamefont {Higginbottom}},\ and\
  \bibinfo {author} {\bibfnamefont {S.}~\bibnamefont {Simmons}},\ }\bibfield  {title} {\bibinfo {title} {Waveguide-integrated silicon t centres},\ }\href {https://doi.org/10.1364/OE.482008} {\bibfield  {journal} {\bibinfo  {journal} {Opt. Express}\ }\textbf {\bibinfo {volume} {31}},\ \bibinfo {pages} {15045} (\bibinfo {year} {2023})}\BibitemShut {NoStop}%
\bibitem [{\citenamefont {Barrett}\ and\ \citenamefont {Kok}(2005)}]{Barrett2005}%
  \BibitemOpen
  \bibfield  {author} {\bibinfo {author} {\bibfnamefont {S.~D.}\ \bibnamefont {Barrett}}\ and\ \bibinfo {author} {\bibfnamefont {P.}~\bibnamefont {Kok}},\ }\bibfield  {title} {\bibinfo {title} {{Efficient high-fidelity quantum computation using matter qubits and linear optics}},\ }\href {https://doi.org/10.1103/PhysRevA.71.060310} {\bibfield  {journal} {\bibinfo  {journal} {Physical Review A}\ }\textbf {\bibinfo {volume} {71}},\ \bibinfo {pages} {060310} (\bibinfo {year} {2005})}\BibitemShut {NoStop}%
\bibitem [{\citenamefont {Cabrillo}\ \emph {et~al.}(1999)\citenamefont {Cabrillo}, \citenamefont {Cirac}, \citenamefont {Garc{\'{i}}a-Fern{\'{a}}ndez},\ and\ \citenamefont {Zoller}}]{Cabrillo1999}%
  \BibitemOpen
  \bibfield  {author} {\bibinfo {author} {\bibfnamefont {C.}~\bibnamefont {Cabrillo}}, \bibinfo {author} {\bibfnamefont {J.}~\bibnamefont {Cirac}}, \bibinfo {author} {\bibfnamefont {P.}~\bibnamefont {Garc{\'{i}}a-Fern{\'{a}}ndez}},\ and\ \bibinfo {author} {\bibfnamefont {P.}~\bibnamefont {Zoller}},\ }\bibfield  {title} {\bibinfo {title} {{Creation of entangled states of distant atoms by interference}},\ }\href {https://doi.org/10.1103/PhysRevA.59.1025} {\bibfield  {journal} {\bibinfo  {journal} {Physical Review A}\ }\textbf {\bibinfo {volume} {59}},\ \bibinfo {pages} {1025} (\bibinfo {year} {1999})}\BibitemShut {NoStop}%
\bibitem [{\citenamefont {Brevoord}\ \emph {et~al.}(2024)\citenamefont {Brevoord}, \citenamefont {De~Santis}, \citenamefont {Yamamoto}, \citenamefont {Pasini}, \citenamefont {Codreanu}, \citenamefont {Turan}, \citenamefont {Beukers}, \citenamefont {Waas},\ and\ \citenamefont {Hanson}}]{Brevoord_2024_heraldinitializationDiamondSnV}%
  \BibitemOpen
  \bibfield  {author} {\bibinfo {author} {\bibfnamefont {J.~M.}\ \bibnamefont {Brevoord}}, \bibinfo {author} {\bibfnamefont {L.}~\bibnamefont {De~Santis}}, \bibinfo {author} {\bibfnamefont {T.}~\bibnamefont {Yamamoto}}, \bibinfo {author} {\bibfnamefont {M.}~\bibnamefont {Pasini}}, \bibinfo {author} {\bibfnamefont {N.}~\bibnamefont {Codreanu}}, \bibinfo {author} {\bibfnamefont {T.}~\bibnamefont {Turan}}, \bibinfo {author} {\bibfnamefont {H.~K.}\ \bibnamefont {Beukers}}, \bibinfo {author} {\bibfnamefont {C.}~\bibnamefont {Waas}},\ and\ \bibinfo {author} {\bibfnamefont {R.}~\bibnamefont {Hanson}},\ }\bibfield  {title} {\bibinfo {title} {Heralded initialization of charge state and optical-transition frequency of diamond tin-vacancy centers},\ }\href {https://doi.org/10.1103/PhysRevApplied.21.054047} {\bibfield  {journal} {\bibinfo  {journal} {Phys. Rev. Appl.}\ }\textbf {\bibinfo {volume} {21}},\ \bibinfo {pages} {054047} (\bibinfo {year} {2024})}\BibitemShut {NoStop}%
\bibitem [{\citenamefont {Hermans}\ \emph {et~al.}(2023)\citenamefont {Hermans}, \citenamefont {Pompili}, \citenamefont {Santos~Martins}, \citenamefont {R-P~Montblanch}, \citenamefont {Beukers}, \citenamefont {Baier}, \citenamefont {Borregaard},\ and\ \citenamefont {Hanson}}]{Hermans2023EntanglingRemote}%
  \BibitemOpen
  \bibfield  {author} {\bibinfo {author} {\bibfnamefont {S.~L.}\ \bibnamefont {Hermans}}, \bibinfo {author} {\bibfnamefont {M.}~\bibnamefont {Pompili}}, \bibinfo {author} {\bibfnamefont {L.~D.}\ \bibnamefont {Santos~Martins}}, \bibinfo {author} {\bibfnamefont {A.}~\bibnamefont {R-P~Montblanch}}, \bibinfo {author} {\bibfnamefont {H.~K.}\ \bibnamefont {Beukers}}, \bibinfo {author} {\bibfnamefont {S.}~\bibnamefont {Baier}}, \bibinfo {author} {\bibfnamefont {J.}~\bibnamefont {Borregaard}},\ and\ \bibinfo {author} {\bibfnamefont {R.}~\bibnamefont {Hanson}},\ }\bibfield  {title} {\bibinfo {title} {{Entangling remote qubits using the single-photon protocol: an in-depth theoretical and experimental study}},\ }\href {https://doi.org/10.1088/1367-2630/ACB004} {\bibfield  {journal} {\bibinfo  {journal} {New Journal of Physics}\ }\textbf {\bibinfo {volume} {25}},\ \bibinfo {pages} {013011} (\bibinfo {year} {2023})}\BibitemShut {NoStop}%
\bibitem [{\citenamefont {Schmidgall}\ \emph {et~al.}(2018)\citenamefont {Schmidgall}, \citenamefont {Chakravarthi}, \citenamefont {Gould}, \citenamefont {Christen}, \citenamefont {Hestroffer}, \citenamefont {Hatami},\ and\ \citenamefont {Fu}}]{Schmidgall_2018_frequencyControlNVSD}%
  \BibitemOpen
  \bibfield  {author} {\bibinfo {author} {\bibfnamefont {E.~R.}\ \bibnamefont {Schmidgall}}, \bibinfo {author} {\bibfnamefont {S.}~\bibnamefont {Chakravarthi}}, \bibinfo {author} {\bibfnamefont {M.}~\bibnamefont {Gould}}, \bibinfo {author} {\bibfnamefont {I.~R.}\ \bibnamefont {Christen}}, \bibinfo {author} {\bibfnamefont {K.}~\bibnamefont {Hestroffer}}, \bibinfo {author} {\bibfnamefont {F.}~\bibnamefont {Hatami}},\ and\ \bibinfo {author} {\bibfnamefont {K.-M.~C.}\ \bibnamefont {Fu}},\ }\bibfield  {title} {\bibinfo {title} {Frequency control of single quantum emitters in integrated photonic circuits},\ }\href {https://doi.org/10.1021/acs.nanolett.7b04717} {\bibfield  {journal} {\bibinfo  {journal} {Nano Letters}\ }\textbf {\bibinfo {volume} {18}},\ \bibinfo {pages} {1175} (\bibinfo {year} {2018})}\BibitemShut {NoStop}%
\bibitem [{\citenamefont {Acosta}\ \emph {et~al.}(2012)\citenamefont {Acosta}, \citenamefont {Santori}, \citenamefont {Faraon}, \citenamefont {Huang}, \citenamefont {Fu}, \citenamefont {Stacey}, \citenamefont {Simpson}, \citenamefont {Ganesan}, \citenamefont {Tomljenovic-Hanic}, \citenamefont {Greentree}, \citenamefont {Prawer},\ and\ \citenamefont {Beausoleil}}]{Acosta_2012_DynamicStabilizationNVcenterOpticalResonances}%
  \BibitemOpen
  \bibfield  {author} {\bibinfo {author} {\bibfnamefont {V.~M.}\ \bibnamefont {Acosta}}, \bibinfo {author} {\bibfnamefont {C.}~\bibnamefont {Santori}}, \bibinfo {author} {\bibfnamefont {A.}~\bibnamefont {Faraon}}, \bibinfo {author} {\bibfnamefont {Z.}~\bibnamefont {Huang}}, \bibinfo {author} {\bibfnamefont {K.-M.~C.}\ \bibnamefont {Fu}}, \bibinfo {author} {\bibfnamefont {A.}~\bibnamefont {Stacey}}, \bibinfo {author} {\bibfnamefont {D.~A.}\ \bibnamefont {Simpson}}, \bibinfo {author} {\bibfnamefont {K.}~\bibnamefont {Ganesan}}, \bibinfo {author} {\bibfnamefont {S.}~\bibnamefont {Tomljenovic-Hanic}}, \bibinfo {author} {\bibfnamefont {A.~D.}\ \bibnamefont {Greentree}}, \bibinfo {author} {\bibfnamefont {S.}~\bibnamefont {Prawer}},\ and\ \bibinfo {author} {\bibfnamefont {R.~G.}\ \bibnamefont {Beausoleil}},\ }\bibfield  {title} {\bibinfo {title} {Dynamic stabilization of the optical resonances of single nitrogen-vacancy centers in diamond},\ }\href {https://doi.org/10.1103/PhysRevLett.108.206401} {\bibfield
  {journal} {\bibinfo  {journal} {Phys. Rev. Lett.}\ }\textbf {\bibinfo {volume} {108}},\ \bibinfo {pages} {206401} (\bibinfo {year} {2012})}\BibitemShut {NoStop}%
\bibitem [{\citenamefont {Pompili}\ \emph {et~al.}(2021)\citenamefont {Pompili}, \citenamefont {Hermans}, \citenamefont {Baier}, \citenamefont {Beukers}, \citenamefont {Humphreys}, \citenamefont {Schouten}, \citenamefont {Vermeulen}, \citenamefont {Tiggelman}, \citenamefont {dos Santos~Martins}, \citenamefont {Dirkse}, \citenamefont {Wehner},\ and\ \citenamefont {Hanson}}]{Pompili2021a}%
  \BibitemOpen
  \bibfield  {author} {\bibinfo {author} {\bibfnamefont {M.}~\bibnamefont {Pompili}}, \bibinfo {author} {\bibfnamefont {S.~L.}\ \bibnamefont {Hermans}}, \bibinfo {author} {\bibfnamefont {S.}~\bibnamefont {Baier}}, \bibinfo {author} {\bibfnamefont {H.~K.}\ \bibnamefont {Beukers}}, \bibinfo {author} {\bibfnamefont {P.~C.}\ \bibnamefont {Humphreys}}, \bibinfo {author} {\bibfnamefont {R.~N.}\ \bibnamefont {Schouten}}, \bibinfo {author} {\bibfnamefont {R.~F.}\ \bibnamefont {Vermeulen}}, \bibinfo {author} {\bibfnamefont {M.~J.}\ \bibnamefont {Tiggelman}}, \bibinfo {author} {\bibfnamefont {L.}~\bibnamefont {dos Santos~Martins}}, \bibinfo {author} {\bibfnamefont {B.}~\bibnamefont {Dirkse}}, \bibinfo {author} {\bibfnamefont {S.}~\bibnamefont {Wehner}},\ and\ \bibinfo {author} {\bibfnamefont {R.}~\bibnamefont {Hanson}},\ }\bibfield  {title} {\bibinfo {title} {{Realization of a multinode quantum network of remote solid-state qubits}},\ }\href {https://doi.org/10.1126/science.abg1919} {\bibfield  {journal} {\bibinfo
  {journal} {Science}\ }\textbf {\bibinfo {volume} {372}},\ \bibinfo {pages} {259} (\bibinfo {year} {2021})}\BibitemShut {NoStop}%
\bibitem [{\citenamefont {Sallen}\ \emph {et~al.}(2010)\citenamefont {Sallen}, \citenamefont {Tribu}, \citenamefont {Aichele}, \citenamefont {Andr{\'e}}, \citenamefont {Besombes}, \citenamefont {Bougerol}, \citenamefont {Richard}, \citenamefont {Tatarenko}, \citenamefont {Kheng},\ and\ \citenamefont {Poizat}}]{Sallen_2010_subnanosecondSDquantumdots}%
  \BibitemOpen
  \bibfield  {author} {\bibinfo {author} {\bibfnamefont {G.}~\bibnamefont {Sallen}}, \bibinfo {author} {\bibfnamefont {A.}~\bibnamefont {Tribu}}, \bibinfo {author} {\bibfnamefont {T.}~\bibnamefont {Aichele}}, \bibinfo {author} {\bibfnamefont {R.}~\bibnamefont {Andr{\'e}}}, \bibinfo {author} {\bibfnamefont {L.}~\bibnamefont {Besombes}}, \bibinfo {author} {\bibfnamefont {C.}~\bibnamefont {Bougerol}}, \bibinfo {author} {\bibfnamefont {M.}~\bibnamefont {Richard}}, \bibinfo {author} {\bibfnamefont {S.}~\bibnamefont {Tatarenko}}, \bibinfo {author} {\bibfnamefont {K.}~\bibnamefont {Kheng}},\ and\ \bibinfo {author} {\bibfnamefont {J.-P.}\ \bibnamefont {Poizat}},\ }\bibfield  {title} {\bibinfo {title} {Subnanosecond spectral diffusion measurement using photon correlation},\ }\href {https://doi.org/10.1038/nphoton.2010.174} {\bibfield  {journal} {\bibinfo  {journal} {Nature Photonics}\ }\textbf {\bibinfo {volume} {4}},\ \bibinfo {pages} {696} (\bibinfo {year} {2010})}\BibitemShut {NoStop}%
\bibitem [{\citenamefont {Lee}\ \emph {et~al.}(2023)\citenamefont {Lee}, \citenamefont {Islam}, \citenamefont {Harper}, \citenamefont {Buyukkaya}, \citenamefont {Higginbottom}, \citenamefont {Simmons},\ and\ \citenamefont {Waks}}]{Lee2023HighEfficiency}%
  \BibitemOpen
  \bibfield  {author} {\bibinfo {author} {\bibfnamefont {C.~M.}\ \bibnamefont {Lee}}, \bibinfo {author} {\bibfnamefont {F.}~\bibnamefont {Islam}}, \bibinfo {author} {\bibfnamefont {S.}~\bibnamefont {Harper}}, \bibinfo {author} {\bibfnamefont {M.~A.}\ \bibnamefont {Buyukkaya}}, \bibinfo {author} {\bibfnamefont {D.}~\bibnamefont {Higginbottom}}, \bibinfo {author} {\bibfnamefont {S.}~\bibnamefont {Simmons}},\ and\ \bibinfo {author} {\bibfnamefont {E.}~\bibnamefont {Waks}},\ }\bibfield  {title} {\bibinfo {title} {{High-Efficiency Single Photon Emission from a Silicon T-Center in a Nanobeam}},\ }\href {https://doi.org/10.1021/ACSPHOTONICS.3C01142/SUPPL{\_}FILE/PH3C01142{\_}SI{\_}001.PDF} {\bibfield  {journal} {\bibinfo  {journal} {ACS Photonics}\ }\textbf {\bibinfo {volume} {10}},\ \bibinfo {pages} {3844} (\bibinfo {year} {2023})}\BibitemShut {NoStop}%
\bibitem [{\citenamefont {Islam}\ \emph {et~al.}(2023)\citenamefont {Islam}, \citenamefont {Lee}, \citenamefont {Harper}, \citenamefont {Rahaman}, \citenamefont {Zhao}, \citenamefont {Vij},\ and\ \citenamefont {Waks}}]{Islam2023cavityenhanced}%
  \BibitemOpen
  \bibfield  {author} {\bibinfo {author} {\bibfnamefont {F.}~\bibnamefont {Islam}}, \bibinfo {author} {\bibfnamefont {C.-M.}\ \bibnamefont {Lee}}, \bibinfo {author} {\bibfnamefont {S.}~\bibnamefont {Harper}}, \bibinfo {author} {\bibfnamefont {M.~H.}\ \bibnamefont {Rahaman}}, \bibinfo {author} {\bibfnamefont {Y.}~\bibnamefont {Zhao}}, \bibinfo {author} {\bibfnamefont {N.~K.}\ \bibnamefont {Vij}},\ and\ \bibinfo {author} {\bibfnamefont {E.}~\bibnamefont {Waks}},\ }\bibfield  {title} {\bibinfo {title} {{Cavity enhanced emission from a silicon T center}},\ }\bibfield  {journal} {\bibinfo  {journal} {Nano Letters}\ }\href {https://doi.org/10.1021/acs.nanolett.3c04056} {10.1021/acs.nanolett.3c04056} (\bibinfo {year} {2023})\BibitemShut {NoStop}%
\bibitem [{\citenamefont {Dobinson}\ \emph {et~al.}(2025)\citenamefont {Dobinson}, \citenamefont {Bowness}, \citenamefont {Meynell}, \citenamefont {Chartrand}, \citenamefont {Hoffmann}, \citenamefont {Gascoine}, \citenamefont {MacGilp}, \citenamefont {Afzal}, \citenamefont {Dangel}, \citenamefont {Jahed}, \citenamefont {Thewalt}, \citenamefont {Simmons},\ and\ \citenamefont {Higginbottom}}]{Dobinson_2025_electricallytriggeredspinphotondevicessilicon}%
  \BibitemOpen
  \bibfield  {author} {\bibinfo {author} {\bibfnamefont {M.}~\bibnamefont {Dobinson}}, \bibinfo {author} {\bibfnamefont {C.}~\bibnamefont {Bowness}}, \bibinfo {author} {\bibfnamefont {S.~A.}\ \bibnamefont {Meynell}}, \bibinfo {author} {\bibfnamefont {C.}~\bibnamefont {Chartrand}}, \bibinfo {author} {\bibfnamefont {E.}~\bibnamefont {Hoffmann}}, \bibinfo {author} {\bibfnamefont {M.}~\bibnamefont {Gascoine}}, \bibinfo {author} {\bibfnamefont {I.}~\bibnamefont {MacGilp}}, \bibinfo {author} {\bibfnamefont {F.}~\bibnamefont {Afzal}}, \bibinfo {author} {\bibfnamefont {C.}~\bibnamefont {Dangel}}, \bibinfo {author} {\bibfnamefont {N.}~\bibnamefont {Jahed}}, \bibinfo {author} {\bibfnamefont {M.~L.~W.}\ \bibnamefont {Thewalt}}, \bibinfo {author} {\bibfnamefont {S.}~\bibnamefont {Simmons}},\ and\ \bibinfo {author} {\bibfnamefont {D.~B.}\ \bibnamefont {Higginbottom}},\ }\href {https://arxiv.org/abs/2501.10597} {\bibinfo {title} {Electrically-triggered spin-photon devices in silicon}} (\bibinfo {year} {2025}),\ \Eprint
  {https://arxiv.org/abs/2501.10597} {arXiv:2501.10597 [quant-ph]} \BibitemShut {NoStop}%
\bibitem [{\citenamefont {Day}\ \emph {et~al.}(2025)\citenamefont {Day}, \citenamefont {Zhang}, \citenamefont {Jin}, \citenamefont {Song}, \citenamefont {Sutula}, \citenamefont {Sipahigil}, \citenamefont {Bhaskar},\ and\ \citenamefont {Hu}}]{Day_2025_probingnegativedifferentialresistance}%
  \BibitemOpen
  \bibfield  {author} {\bibinfo {author} {\bibfnamefont {A.~M.}\ \bibnamefont {Day}}, \bibinfo {author} {\bibfnamefont {C.}~\bibnamefont {Zhang}}, \bibinfo {author} {\bibfnamefont {C.}~\bibnamefont {Jin}}, \bibinfo {author} {\bibfnamefont {H.}~\bibnamefont {Song}}, \bibinfo {author} {\bibfnamefont {M.}~\bibnamefont {Sutula}}, \bibinfo {author} {\bibfnamefont {A.}~\bibnamefont {Sipahigil}}, \bibinfo {author} {\bibfnamefont {M.~K.}\ \bibnamefont {Bhaskar}},\ and\ \bibinfo {author} {\bibfnamefont {E.~L.}\ \bibnamefont {Hu}},\ }\href {https://arxiv.org/abs/2501.11888} {\bibinfo {title} {Probing negative differential resistance in silicon with a p-i-n diode-integrated t center ensemble}} (\bibinfo {year} {2025}),\ \Eprint {https://arxiv.org/abs/2501.11888} {arXiv:2501.11888} \BibitemShut {NoStop}%
\bibitem [{\citenamefont {MacQuarrie}\ \emph {et~al.}(2021)\citenamefont {MacQuarrie}, \citenamefont {Chartrand}, \citenamefont {Higginbottom}, \citenamefont {Morse}, \citenamefont {Karasyuk}, \citenamefont {Roorda},\ and\ \citenamefont {Simmons}}]{MacQuarrie_2021_GeneratingT}%
  \BibitemOpen
  \bibfield  {author} {\bibinfo {author} {\bibfnamefont {E.~R.}\ \bibnamefont {MacQuarrie}}, \bibinfo {author} {\bibfnamefont {C.}~\bibnamefont {Chartrand}}, \bibinfo {author} {\bibfnamefont {D.~B.}\ \bibnamefont {Higginbottom}}, \bibinfo {author} {\bibfnamefont {K.~J.}\ \bibnamefont {Morse}}, \bibinfo {author} {\bibfnamefont {V.~A.}\ \bibnamefont {Karasyuk}}, \bibinfo {author} {\bibfnamefont {S.}~\bibnamefont {Roorda}},\ and\ \bibinfo {author} {\bibfnamefont {S.}~\bibnamefont {Simmons}},\ }\bibfield  {title} {\bibinfo {title} {Generating t centres in photonic silicon-on-insulator material by ion implantation},\ }\href {https://doi.org/10.1088/1367-2630/ac291f} {\bibfield  {journal} {\bibinfo  {journal} {New Journal of Physics}\ }\textbf {\bibinfo {volume} {23}},\ \bibinfo {pages} {103008} (\bibinfo {year} {2021})}\BibitemShut {NoStop}%
\bibitem [{Sup()}]{Supplement}%
  \BibitemOpen
  \href@noop {} {}\bibinfo {note} {See Supplemental Material at [URL will be inserted by publisher] for: a summary of T centre device properties; off-resonant effects; evidence of spin-state symmetry in laser-induced mixing; laser-induced mixing on another device; dark time resonance check PLE; spectral diffusion at a second temperature point; a discussion of the Ornsteinâ Uhlenbeck spectral diffusion model; analysis of the charge resonance check efficiency; and which includes references \cite{Taminiau_2019_NVmatdevSDimprove, Uhlenbeck_1930_TheoryOfBrownianMotion, Gardiner_2004_HandbookOfStochasticMethods,vanKampen_2007_StochasticProcesses}}\BibitemShut {NoStop}%
\bibitem [{\citenamefont {Delteil}\ \emph {et~al.}(2024)\citenamefont {Delteil}, \citenamefont {Buil},\ and\ \citenamefont {Hermier}}]{Delteil_2024_photonstatisticsdiffusiveemittersOUMarkov}%
  \BibitemOpen
  \bibfield  {author} {\bibinfo {author} {\bibfnamefont {A.}~\bibnamefont {Delteil}}, \bibinfo {author} {\bibfnamefont {S.}~\bibnamefont {Buil}},\ and\ \bibinfo {author} {\bibfnamefont {J.-P.}\ \bibnamefont {Hermier}},\ }\bibfield  {title} {\bibinfo {title} {Photon statistics of resonantly driven spectrally diffusive quantum emitters},\ }\href {https://doi.org/10.1103/PhysRevB.109.155308} {\bibfield  {journal} {\bibinfo  {journal} {Phys. Rev. B}\ }\textbf {\bibinfo {volume} {109}},\ \bibinfo {pages} {155308} (\bibinfo {year} {2024})}\BibitemShut {NoStop}%
\bibitem [{\citenamefont {van~de Stolpe}\ \emph {et~al.}(2025)\citenamefont {van~de Stolpe}, \citenamefont {Feije}, \citenamefont {Loenen}, \citenamefont {Das}, \citenamefont {Timmer}, \citenamefont {de~Jong},\ and\ \citenamefont {Taminiau}}]{vandeStolpe_2025_checkprobeSiC}%
  \BibitemOpen
  \bibfield  {author} {\bibinfo {author} {\bibfnamefont {G.~L.}\ \bibnamefont {van~de Stolpe}}, \bibinfo {author} {\bibfnamefont {L.~J.}\ \bibnamefont {Feije}}, \bibinfo {author} {\bibfnamefont {S.~J.~H.}\ \bibnamefont {Loenen}}, \bibinfo {author} {\bibfnamefont {A.}~\bibnamefont {Das}}, \bibinfo {author} {\bibfnamefont {G.~M.}\ \bibnamefont {Timmer}}, \bibinfo {author} {\bibfnamefont {T.~W.}\ \bibnamefont {de~Jong}},\ and\ \bibinfo {author} {\bibfnamefont {T.~H.}\ \bibnamefont {Taminiau}},\ }\bibfield  {title} {\bibinfo {title} {Check-probe spectroscopy of lifetime-limited emitters in bulk-grown silicon carbide},\ }\href {https://doi.org/10.1038/s41534-025-00985-3} {\bibfield  {journal} {\bibinfo  {journal} {npj Quantum Information}\ }\textbf {\bibinfo {volume} {11}},\ \bibinfo {pages} {31} (\bibinfo {year} {2025})}\BibitemShut {NoStop}%
\bibitem [{\citenamefont {G{\"o}rlitz}\ \emph {et~al.}(2022)\citenamefont {G{\"o}rlitz}, \citenamefont {Herrmann}, \citenamefont {Fuchs}, \citenamefont {Iwasaki}, \citenamefont {Taniguchi}, \citenamefont {Rogalla}, \citenamefont {Hardeman}, \citenamefont {Colard}, \citenamefont {Markham}, \citenamefont {Hatano},\ and\ \citenamefont {Becher}}]{Gorlitz_2022_SnVSDunderlaserillumination}%
  \BibitemOpen
  \bibfield  {author} {\bibinfo {author} {\bibfnamefont {J.}~\bibnamefont {G{\"o}rlitz}}, \bibinfo {author} {\bibfnamefont {D.}~\bibnamefont {Herrmann}}, \bibinfo {author} {\bibfnamefont {P.}~\bibnamefont {Fuchs}}, \bibinfo {author} {\bibfnamefont {T.}~\bibnamefont {Iwasaki}}, \bibinfo {author} {\bibfnamefont {T.}~\bibnamefont {Taniguchi}}, \bibinfo {author} {\bibfnamefont {D.}~\bibnamefont {Rogalla}}, \bibinfo {author} {\bibfnamefont {D.}~\bibnamefont {Hardeman}}, \bibinfo {author} {\bibfnamefont {P.-O.}\ \bibnamefont {Colard}}, \bibinfo {author} {\bibfnamefont {M.}~\bibnamefont {Markham}}, \bibinfo {author} {\bibfnamefont {M.}~\bibnamefont {Hatano}},\ and\ \bibinfo {author} {\bibfnamefont {C.}~\bibnamefont {Becher}},\ }\bibfield  {title} {\bibinfo {title} {Coherence of a charge stabilised tin-vacancy spin in diamond},\ }\href {https://doi.org/10.1038/s41534-022-00552-0} {\bibfield  {journal} {\bibinfo  {journal} {npj Quantum Information}\ }\textbf {\bibinfo {volume} {8}},\ \bibinfo {pages} {45} (\bibinfo
  {year} {2022})}\BibitemShut {NoStop}%
\bibitem [{\citenamefont {Azuma}\ \emph {et~al.}(2015)\citenamefont {Azuma}, \citenamefont {Tamaki},\ and\ \citenamefont {Lo}}]{Azuma_2015_AllPhotonicQuantumRepeaters}%
  \BibitemOpen
  \bibfield  {author} {\bibinfo {author} {\bibfnamefont {K.}~\bibnamefont {Azuma}}, \bibinfo {author} {\bibfnamefont {K.}~\bibnamefont {Tamaki}},\ and\ \bibinfo {author} {\bibfnamefont {H.-K.}\ \bibnamefont {Lo}},\ }\bibfield  {title} {\bibinfo {title} {All-photonic quantum repeaters},\ }\href {https://doi.org/10.1038/ncomms7787} {\bibfield  {journal} {\bibinfo  {journal} {Nature Communications}\ }\textbf {\bibinfo {volume} {6}},\ \bibinfo {pages} {6787} (\bibinfo {year} {2015})}\BibitemShut {NoStop}%
\bibitem [{\citenamefont {Hilaire}\ \emph {et~al.}(2021)\citenamefont {Hilaire}, \citenamefont {Barnes}, \citenamefont {Economou},\ and\ \citenamefont {Grosshans}}]{Hilaire_2021_ErrorCorrectingEntanglement}%
  \BibitemOpen
  \bibfield  {author} {\bibinfo {author} {\bibfnamefont {P.}~\bibnamefont {Hilaire}}, \bibinfo {author} {\bibfnamefont {E.}~\bibnamefont {Barnes}}, \bibinfo {author} {\bibfnamefont {S.~E.}\ \bibnamefont {Economou}},\ and\ \bibinfo {author} {\bibfnamefont {F.}~\bibnamefont {Grosshans}},\ }\bibfield  {title} {\bibinfo {title} {Error-correcting entanglement swapping using a practical logical photon encoding},\ }\href {https://doi.org/10.1103/PhysRevA.104.052623} {\bibfield  {journal} {\bibinfo  {journal} {Phys. Rev. A}\ }\textbf {\bibinfo {volume} {104}},\ \bibinfo {pages} {052623} (\bibinfo {year} {2021})}\BibitemShut {NoStop}%
\bibitem [{\citenamefont {Zhang}\ \emph {et~al.}(2022)\citenamefont {Zhang}, \citenamefont {Liu}, \citenamefont {Li}, \citenamefont {Fei}, \citenamefont {Yin}, \citenamefont {Li}, \citenamefont {Liu}, \citenamefont {Mao}, \citenamefont {Chen},\ and\ \citenamefont {Pan}}]{Zhang_2022_AllPhotonicRepeater}%
  \BibitemOpen
  \bibfield  {author} {\bibinfo {author} {\bibfnamefont {R.}~\bibnamefont {Zhang}}, \bibinfo {author} {\bibfnamefont {L.-Z.}\ \bibnamefont {Liu}}, \bibinfo {author} {\bibfnamefont {Z.-D.}\ \bibnamefont {Li}}, \bibinfo {author} {\bibfnamefont {Y.-Y.}\ \bibnamefont {Fei}}, \bibinfo {author} {\bibfnamefont {X.-F.}\ \bibnamefont {Yin}}, \bibinfo {author} {\bibfnamefont {L.}~\bibnamefont {Li}}, \bibinfo {author} {\bibfnamefont {N.-L.}\ \bibnamefont {Liu}}, \bibinfo {author} {\bibfnamefont {Y.}~\bibnamefont {Mao}}, \bibinfo {author} {\bibfnamefont {Y.-A.}\ \bibnamefont {Chen}},\ and\ \bibinfo {author} {\bibfnamefont {J.-W.}\ \bibnamefont {Pan}},\ }\bibfield  {title} {\bibinfo {title} {Loss-tolerant all-photonic quantum repeater with generalized shor code},\ }\href {https://doi.org/10.1364/OPTICA.439170} {\bibfield  {journal} {\bibinfo  {journal} {Optica}\ }\textbf {\bibinfo {volume} {9}},\ \bibinfo {pages} {152} (\bibinfo {year} {2022})}\BibitemShut {NoStop}%
\bibitem [{\citenamefont {Wein}\ \emph {et~al.}(2024)\citenamefont {Wein}, \citenamefont {de~BrugiÃ¨re}, \citenamefont {Music}, \citenamefont {Senellart}, \citenamefont {Bourdoncle},\ and\ \citenamefont {Mansfield}}]{Wein_2024_minimizingresourceoverheadfusionbased}%
  \BibitemOpen
  \bibfield  {author} {\bibinfo {author} {\bibfnamefont {S.~C.}\ \bibnamefont {Wein}}, \bibinfo {author} {\bibfnamefont {T.~G.}\ \bibnamefont {de~BrugiÃ¨re}}, \bibinfo {author} {\bibfnamefont {L.}~\bibnamefont {Music}}, \bibinfo {author} {\bibfnamefont {P.}~\bibnamefont {Senellart}}, \bibinfo {author} {\bibfnamefont {B.}~\bibnamefont {Bourdoncle}},\ and\ \bibinfo {author} {\bibfnamefont {S.}~\bibnamefont {Mansfield}},\ }\href {https://arxiv.org/abs/2412.08611} {\bibinfo {title} {Minimizing resource overhead in fusion-based quantum computation using hybrid spin-photon devices}} (\bibinfo {year} {2024}),\ \Eprint {https://arxiv.org/abs/2412.08611} {arXiv:2412.08611} \BibitemShut {NoStop}%
\bibitem [{\citenamefont {Borregaard}\ \emph {et~al.}(2020)\citenamefont {Borregaard}, \citenamefont {Pichler}, \citenamefont {Schr\"oder}, \citenamefont {Lukin}, \citenamefont {Lodahl},\ and\ \citenamefont {S\o{}rensen}}]{Borregaard_2020_OneWayQuantumRepeater}%
  \BibitemOpen
  \bibfield  {author} {\bibinfo {author} {\bibfnamefont {J.}~\bibnamefont {Borregaard}}, \bibinfo {author} {\bibfnamefont {H.}~\bibnamefont {Pichler}}, \bibinfo {author} {\bibfnamefont {T.}~\bibnamefont {Schr\"oder}}, \bibinfo {author} {\bibfnamefont {M.~D.}\ \bibnamefont {Lukin}}, \bibinfo {author} {\bibfnamefont {P.}~\bibnamefont {Lodahl}},\ and\ \bibinfo {author} {\bibfnamefont {A.~S.}\ \bibnamefont {S\o{}rensen}},\ }\bibfield  {title} {\bibinfo {title} {One-way quantum repeater based on near-deterministic photon-emitter interfaces},\ }\href {https://doi.org/10.1103/PhysRevX.10.021071} {\bibfield  {journal} {\bibinfo  {journal} {Phys. Rev. X}\ }\textbf {\bibinfo {volume} {10}},\ \bibinfo {pages} {021071} (\bibinfo {year} {2020})}\BibitemShut {NoStop}%
\bibitem [{\citenamefont {Pingenot}\ \emph {et~al.}(2011)\citenamefont {Pingenot}, \citenamefont {Pryor},\ and\ \citenamefont {Flatt\'e}}]{Pingenot_2011_EfieldCouplingHolegFactor}%
  \BibitemOpen
  \bibfield  {author} {\bibinfo {author} {\bibfnamefont {J.}~\bibnamefont {Pingenot}}, \bibinfo {author} {\bibfnamefont {C.~E.}\ \bibnamefont {Pryor}},\ and\ \bibinfo {author} {\bibfnamefont {M.~E.}\ \bibnamefont {Flatt\'e}},\ }\bibfield  {title} {\bibinfo {title} {Electric-field manipulation of the land\'e $g$ tensor of a hole in an in${}_{0.5}$ga${}_{0.5}$as/gaas self-assembled quantum dot},\ }\href {https://doi.org/10.1103/PhysRevB.84.195403} {\bibfield  {journal} {\bibinfo  {journal} {Phys. Rev. B}\ }\textbf {\bibinfo {volume} {84}},\ \bibinfo {pages} {195403} (\bibinfo {year} {2011})}\BibitemShut {NoStop}%
\bibitem [{\citenamefont {Kuhlmann}\ \emph {et~al.}(2013)\citenamefont {Kuhlmann}, \citenamefont {Houel}, \citenamefont {Ludwig}, \citenamefont {Greuter}, \citenamefont {Reuter}, \citenamefont {Wieck}, \citenamefont {Poggio},\ and\ \citenamefont {Warburton}}]{Kuhlmann_2013_ChargeNoiseSpinNoise}%
  \BibitemOpen
  \bibfield  {author} {\bibinfo {author} {\bibfnamefont {A.~V.}\ \bibnamefont {Kuhlmann}}, \bibinfo {author} {\bibfnamefont {J.}~\bibnamefont {Houel}}, \bibinfo {author} {\bibfnamefont {A.}~\bibnamefont {Ludwig}}, \bibinfo {author} {\bibfnamefont {L.}~\bibnamefont {Greuter}}, \bibinfo {author} {\bibfnamefont {D.}~\bibnamefont {Reuter}}, \bibinfo {author} {\bibfnamefont {A.~D.}\ \bibnamefont {Wieck}}, \bibinfo {author} {\bibfnamefont {M.}~\bibnamefont {Poggio}},\ and\ \bibinfo {author} {\bibfnamefont {R.~J.}\ \bibnamefont {Warburton}},\ }\bibfield  {title} {\bibinfo {title} {Charge noise and spin noise in a semiconductor quantum device},\ }\href {https://doi.org/10.1038/nphys2688} {\bibfield  {journal} {\bibinfo  {journal} {Nature Physics}\ }\textbf {\bibinfo {volume} {9}},\ \bibinfo {pages} {570} (\bibinfo {year} {2013})}\BibitemShut {NoStop}%
\bibitem [{\citenamefont {Studenikin}\ \emph {et~al.}(2019)\citenamefont {Studenikin}, \citenamefont {Korkusinski}, \citenamefont {Takahashi}, \citenamefont {Ducatel}, \citenamefont {Padawer-Blatt}, \citenamefont {Bogan}, \citenamefont {Austing}, \citenamefont {Gaudreau}, \citenamefont {Zawadzki}, \citenamefont {Sachrajda}, \citenamefont {Hirayama}, \citenamefont {Tracy}, \citenamefont {Reno},\ and\ \citenamefont {Hargett}}]{Studenikin_2019_TunablegfactorHole}%
  \BibitemOpen
  \bibfield  {author} {\bibinfo {author} {\bibfnamefont {S.}~\bibnamefont {Studenikin}}, \bibinfo {author} {\bibfnamefont {M.}~\bibnamefont {Korkusinski}}, \bibinfo {author} {\bibfnamefont {M.}~\bibnamefont {Takahashi}}, \bibinfo {author} {\bibfnamefont {J.}~\bibnamefont {Ducatel}}, \bibinfo {author} {\bibfnamefont {A.}~\bibnamefont {Padawer-Blatt}}, \bibinfo {author} {\bibfnamefont {A.}~\bibnamefont {Bogan}}, \bibinfo {author} {\bibfnamefont {D.~G.}\ \bibnamefont {Austing}}, \bibinfo {author} {\bibfnamefont {L.}~\bibnamefont {Gaudreau}}, \bibinfo {author} {\bibfnamefont {P.}~\bibnamefont {Zawadzki}}, \bibinfo {author} {\bibfnamefont {A.}~\bibnamefont {Sachrajda}}, \bibinfo {author} {\bibfnamefont {Y.}~\bibnamefont {Hirayama}}, \bibinfo {author} {\bibfnamefont {L.}~\bibnamefont {Tracy}}, \bibinfo {author} {\bibfnamefont {J.}~\bibnamefont {Reno}},\ and\ \bibinfo {author} {\bibfnamefont {T.}~\bibnamefont {Hargett}},\ }\bibfield  {title} {\bibinfo {title} {Electrically tunable effective g-factor of a single hole
  in a lateral gaas/algaas quantum dot},\ }\href {https://doi.org/10.1038/s42005-019-0262-1} {\bibfield  {journal} {\bibinfo  {journal} {Communications Physics}\ }\textbf {\bibinfo {volume} {2}},\ \bibinfo {pages} {159} (\bibinfo {year} {2019})}\BibitemShut {NoStop}%
\bibitem [{\citenamefont {Clear}\ \emph {et~al.}(2024)\citenamefont {Clear}, \citenamefont {Hosseini}, \citenamefont {AlizadehKhaledi}, \citenamefont {Brunelle}, \citenamefont {Woolverton}, \citenamefont {Kanaganayagam}, \citenamefont {Kazemi}, \citenamefont {Chartrand}, \citenamefont {Keshavarz}, \citenamefont {Xiong}, \citenamefont {Alaerts}, \citenamefont {Soykal}, \citenamefont {Hautier}, \citenamefont {Karassiouk}, \citenamefont {Thewalt}, \citenamefont {Higginbottom},\ and\ \citenamefont {Simmons}}]{Clear_2024_OpticalTransitionParameters}%
  \BibitemOpen
  \bibfield  {author} {\bibinfo {author} {\bibfnamefont {C.}~\bibnamefont {Clear}}, \bibinfo {author} {\bibfnamefont {S.}~\bibnamefont {Hosseini}}, \bibinfo {author} {\bibfnamefont {A.}~\bibnamefont {AlizadehKhaledi}}, \bibinfo {author} {\bibfnamefont {N.}~\bibnamefont {Brunelle}}, \bibinfo {author} {\bibfnamefont {A.}~\bibnamefont {Woolverton}}, \bibinfo {author} {\bibfnamefont {J.}~\bibnamefont {Kanaganayagam}}, \bibinfo {author} {\bibfnamefont {M.}~\bibnamefont {Kazemi}}, \bibinfo {author} {\bibfnamefont {C.}~\bibnamefont {Chartrand}}, \bibinfo {author} {\bibfnamefont {M.}~\bibnamefont {Keshavarz}}, \bibinfo {author} {\bibfnamefont {Y.}~\bibnamefont {Xiong}}, \bibinfo {author} {\bibfnamefont {L.}~\bibnamefont {Alaerts}}, \bibinfo {author} {\bibfnamefont {O.~O.}\ \bibnamefont {Soykal}}, \bibinfo {author} {\bibfnamefont {G.}~\bibnamefont {Hautier}}, \bibinfo {author} {\bibfnamefont {V.}~\bibnamefont {Karassiouk}}, \bibinfo {author} {\bibfnamefont {M.}~\bibnamefont {Thewalt}}, \bibinfo {author} {\bibfnamefont
  {D.}~\bibnamefont {Higginbottom}},\ and\ \bibinfo {author} {\bibfnamefont {S.}~\bibnamefont {Simmons}},\ }\bibfield  {title} {\bibinfo {title} {Optical-transition parameters of the silicon $t$ center},\ }\href {https://doi.org/10.1103/PhysRevApplied.22.064014} {\bibfield  {journal} {\bibinfo  {journal} {Phys. Rev. Appl.}\ }\textbf {\bibinfo {volume} {22}},\ \bibinfo {pages} {064014} (\bibinfo {year} {2024})}\BibitemShut {NoStop}%
\bibitem [{\citenamefont {Zhang}\ \emph {et~al.}(2025)\citenamefont {Zhang}, \citenamefont {Fiaschi}, \citenamefont {Komza}, \citenamefont {Song}, \citenamefont {Schenkel},\ and\ \citenamefont {Sipahigil}}]{Zhang_2025_SD}%
  \BibitemOpen
  \bibfield  {author} {\bibinfo {author} {\bibfnamefont {X.}~\bibnamefont {Zhang}}, \bibinfo {author} {\bibfnamefont {N.}~\bibnamefont {Fiaschi}}, \bibinfo {author} {\bibfnamefont {L.}~\bibnamefont {Komza}}, \bibinfo {author} {\bibfnamefont {H.}~\bibnamefont {Song}}, \bibinfo {author} {\bibfnamefont {T.}~\bibnamefont {Schenkel}},\ and\ \bibinfo {author} {\bibfnamefont {A.}~\bibnamefont {Sipahigil}},\ }\bibfield  {title} {\bibinfo {title} {Laser-induced spectral diffusion of {T} centers in silicon nanophotonic devices}} (\bibinfo {year} {2025}),\ \bibinfo {note} {{M}anuscript in preparation}\BibitemShut {NoStop}%
\bibitem [{\citenamefont {Taherizadegan}\ \emph {et~al.}(2025)\citenamefont {Taherizadegan}, \citenamefont {Asadi}, \citenamefont {Ji}, \citenamefont {Higginbottom},\ and\ \citenamefont {Simon}}]{taherizadegan_2025_gates}%
  \BibitemOpen
  \bibfield  {author} {\bibinfo {author} {\bibfnamefont {S.}~\bibnamefont {Taherizadegan}}, \bibinfo {author} {\bibfnamefont {F.~K.}\ \bibnamefont {Asadi}}, \bibinfo {author} {\bibfnamefont {J.-W.}\ \bibnamefont {Ji}}, \bibinfo {author} {\bibfnamefont {D.}~\bibnamefont {Higginbottom}},\ and\ \bibinfo {author} {\bibfnamefont {C.}~\bibnamefont {Simon}},\ }\href {https://arxiv.org/abs/2508.06474} {\bibinfo {title} {Exploring the feasibility of probabilistic and deterministic quantum gates between t centers in silicon}} (\bibinfo {year} {2025}),\ \Eprint {https://arxiv.org/abs/2508.06474} {arXiv:2508.06474 [quant-ph]} \BibitemShut {NoStop}%
\bibitem [{\citenamefont {Robinson}\ and\ \citenamefont {Goldberg}(2000)}]{Robinson_2000_laser-inducedSD}%
  \BibitemOpen
  \bibfield  {author} {\bibinfo {author} {\bibfnamefont {H.~D.}\ \bibnamefont {Robinson}}\ and\ \bibinfo {author} {\bibfnamefont {B.~B.}\ \bibnamefont {Goldberg}},\ }\bibfield  {title} {\bibinfo {title} {Light-induced spectral diffusion in single self-assembled quantum dots},\ }\href {https://doi.org/10.1103/PhysRevB.61.R5086} {\bibfield  {journal} {\bibinfo  {journal} {Phys. Rev. B}\ }\textbf {\bibinfo {volume} {61}},\ \bibinfo {pages} {R5086} (\bibinfo {year} {2000})}\BibitemShut {NoStop}%
\bibitem [{\citenamefont {Holmes}\ \emph {et~al.}(2015)\citenamefont {Holmes}, \citenamefont {Kako}, \citenamefont {Choi}, \citenamefont {Arita},\ and\ \citenamefont {Arakawa}}]{Holmes_2015_SDemissionLWGaN}%
  \BibitemOpen
  \bibfield  {author} {\bibinfo {author} {\bibfnamefont {M.}~\bibnamefont {Holmes}}, \bibinfo {author} {\bibfnamefont {S.}~\bibnamefont {Kako}}, \bibinfo {author} {\bibfnamefont {K.}~\bibnamefont {Choi}}, \bibinfo {author} {\bibfnamefont {M.}~\bibnamefont {Arita}},\ and\ \bibinfo {author} {\bibfnamefont {Y.}~\bibnamefont {Arakawa}},\ }\bibfield  {title} {\bibinfo {title} {Spectral diffusion and its influence on the emission linewidths of site-controlled gan nanowire quantum dots},\ }\href {https://doi.org/10.1103/PhysRevB.92.115447} {\bibfield  {journal} {\bibinfo  {journal} {Phys. Rev. B}\ }\textbf {\bibinfo {volume} {92}},\ \bibinfo {pages} {115447} (\bibinfo {year} {2015})}\BibitemShut {NoStop}%
\bibitem [{\citenamefont {Gao}\ \emph {et~al.}(2017)\citenamefont {Gao}, \citenamefont {Solovev}, \citenamefont {Holmes}, \citenamefont {Arita},\ and\ \citenamefont {Arakawa}}]{Gao_2017_NanosecondGaNQD-Photo-inducedSD}%
  \BibitemOpen
  \bibfield  {author} {\bibinfo {author} {\bibfnamefont {K.~é.}\ \bibnamefont {Gao}}, \bibinfo {author} {\bibfnamefont {I.}~\bibnamefont {Solovev}}, \bibinfo {author} {\bibfnamefont {M.}~\bibnamefont {Holmes}}, \bibinfo {author} {\bibfnamefont {M.~æ.}\ \bibnamefont {Arita}},\ and\ \bibinfo {author} {\bibfnamefont {Y.~è.}\ \bibnamefont {Arakawa}},\ }\bibfield  {title} {\bibinfo {title} {Nanosecond-scale spectral diffusion in the single photon emission of a gan quantum dot},\ }\href {https://doi.org/10.1063/1.4997117} {\bibfield  {journal} {\bibinfo  {journal} {AIP Advances}\ }\textbf {\bibinfo {volume} {7}},\ \bibinfo {pages} {125216} (\bibinfo {year} {2017})}\BibitemShut {NoStop}%
\bibitem [{\citenamefont {Ha}\ \emph {et~al.}(2015)\citenamefont {Ha}, \citenamefont {Mano}, \citenamefont {Chou}, \citenamefont {Wu}, \citenamefont {Cheng}, \citenamefont {Bocquel}, \citenamefont {Koenraad}, \citenamefont {Ohtake}, \citenamefont {Sakuma}, \citenamefont {Sakoda},\ and\ \citenamefont {Kuroda}}]{Ha_2015_SizeDependenceLineBroadening}%
  \BibitemOpen
  \bibfield  {author} {\bibinfo {author} {\bibfnamefont {N.}~\bibnamefont {Ha}}, \bibinfo {author} {\bibfnamefont {T.}~\bibnamefont {Mano}}, \bibinfo {author} {\bibfnamefont {Y.-L.}\ \bibnamefont {Chou}}, \bibinfo {author} {\bibfnamefont {Y.-N.}\ \bibnamefont {Wu}}, \bibinfo {author} {\bibfnamefont {S.-J.}\ \bibnamefont {Cheng}}, \bibinfo {author} {\bibfnamefont {J.}~\bibnamefont {Bocquel}}, \bibinfo {author} {\bibfnamefont {P.~M.}\ \bibnamefont {Koenraad}}, \bibinfo {author} {\bibfnamefont {A.}~\bibnamefont {Ohtake}}, \bibinfo {author} {\bibfnamefont {Y.}~\bibnamefont {Sakuma}}, \bibinfo {author} {\bibfnamefont {K.}~\bibnamefont {Sakoda}},\ and\ \bibinfo {author} {\bibfnamefont {T.}~\bibnamefont {Kuroda}},\ }\bibfield  {title} {\bibinfo {title} {Size-dependent line broadening in the emission spectra of single gaas quantum dots: Impact of surface charge on spectral diffusion},\ }\href {https://doi.org/10.1103/PhysRevB.92.075306} {\bibfield  {journal} {\bibinfo  {journal} {Phys. Rev. B}\ }\textbf {\bibinfo
  {volume} {92}},\ \bibinfo {pages} {075306} (\bibinfo {year} {2015})}\BibitemShut {NoStop}%
\bibitem [{\citenamefont {Manna}\ \emph {et~al.}(2020)\citenamefont {Manna}, \citenamefont {Huang}, \citenamefont {{da Silva}}, \citenamefont {Schimpf}, \citenamefont {Rota}, \citenamefont {Lehner}, \citenamefont {Reindl}, \citenamefont {Trotta},\ and\ \citenamefont {Rastelli}}]{Manna_2020_SurfacePassivationQD}%
  \BibitemOpen
  \bibfield  {author} {\bibinfo {author} {\bibfnamefont {S.}~\bibnamefont {Manna}}, \bibinfo {author} {\bibfnamefont {H.}~\bibnamefont {Huang}}, \bibinfo {author} {\bibfnamefont {S.~F.~C.}\ \bibnamefont {{da Silva}}}, \bibinfo {author} {\bibfnamefont {C.}~\bibnamefont {Schimpf}}, \bibinfo {author} {\bibfnamefont {M.~B.}\ \bibnamefont {Rota}}, \bibinfo {author} {\bibfnamefont {B.}~\bibnamefont {Lehner}}, \bibinfo {author} {\bibfnamefont {M.}~\bibnamefont {Reindl}}, \bibinfo {author} {\bibfnamefont {R.}~\bibnamefont {Trotta}},\ and\ \bibinfo {author} {\bibfnamefont {A.}~\bibnamefont {Rastelli}},\ }\bibfield  {title} {\bibinfo {title} {Surface passivation and oxide encapsulation to improve optical properties of a single gaas quantum dot close to the surface},\ }\href {https://doi.org/https://doi.org/10.1016/j.apsusc.2020.147360} {\bibfield  {journal} {\bibinfo  {journal} {Applied Surface Science}\ }\textbf {\bibinfo {volume} {532}},\ \bibinfo {pages} {147360} (\bibinfo {year} {2020})}\BibitemShut {NoStop}%
\bibitem [{\citenamefont {Xiong}\ \emph {et~al.}(2024)\citenamefont {Xiong}, \citenamefont {Zheng}, \citenamefont {McBride}, \citenamefont {Zhang}, \citenamefont {Griffin},\ and\ \citenamefont {Hautier}}]{Xiong2024Discovery}%
  \BibitemOpen
  \bibfield  {author} {\bibinfo {author} {\bibfnamefont {Y.}~\bibnamefont {Xiong}}, \bibinfo {author} {\bibfnamefont {J.}~\bibnamefont {Zheng}}, \bibinfo {author} {\bibfnamefont {S.}~\bibnamefont {McBride}}, \bibinfo {author} {\bibfnamefont {X.}~\bibnamefont {Zhang}}, \bibinfo {author} {\bibfnamefont {S.~M.}\ \bibnamefont {Griffin}},\ and\ \bibinfo {author} {\bibfnamefont {G.}~\bibnamefont {Hautier}},\ }\bibfield  {title} {\bibinfo {title} {{Discovery of T center-like quantum defects in silicon}},\ }\href {https://doi.org/10.1021/JACS.4C06613/ASSET/IMAGES/LARGE/JA4C06613{\_}0007.JPEG} {\bibfield  {journal} {\bibinfo  {journal} {Journal of the American Chemical Society}\ }\textbf {\bibinfo {volume} {146}},\ \bibinfo {pages} {30046} (\bibinfo {year} {2024})}\BibitemShut {NoStop}%
\bibitem [{\citenamefont {Quan}\ and\ \citenamefont {Loncar}(2011)}]{Loncar_2011_quadraticL0nanobeam}%
  \BibitemOpen
  \bibfield  {author} {\bibinfo {author} {\bibfnamefont {Q.}~\bibnamefont {Quan}}\ and\ \bibinfo {author} {\bibfnamefont {M.}~\bibnamefont {Loncar}},\ }\bibfield  {title} {\bibinfo {title} {Deterministic design of wavelength scale, ultra-high q photonic crystal nanobeam cavities},\ }\href {https://doi.org/10.1364/OE.19.018529} {\bibfield  {journal} {\bibinfo  {journal} {Opt. Express}\ }\textbf {\bibinfo {volume} {19}},\ \bibinfo {pages} {18529} (\bibinfo {year} {2011})}\BibitemShut {NoStop}%
\bibitem [{\citenamefont {van Dam}\ \emph {et~al.}(2019)\citenamefont {van Dam}, \citenamefont {Walsh}, \citenamefont {Degen}, \citenamefont {Bersin}, \citenamefont {Mouradian}, \citenamefont {Galiullin}, \citenamefont {Ruf}, \citenamefont {IJspeert}, \citenamefont {Taminiau}, \citenamefont {Hanson},\ and\ \citenamefont {Englund}}]{Taminiau_2019_NVmatdevSDimprove}%
  \BibitemOpen
  \bibfield  {author} {\bibinfo {author} {\bibfnamefont {S.~B.}\ \bibnamefont {van Dam}}, \bibinfo {author} {\bibfnamefont {M.}~\bibnamefont {Walsh}}, \bibinfo {author} {\bibfnamefont {M.~J.}\ \bibnamefont {Degen}}, \bibinfo {author} {\bibfnamefont {E.}~\bibnamefont {Bersin}}, \bibinfo {author} {\bibfnamefont {S.~L.}\ \bibnamefont {Mouradian}}, \bibinfo {author} {\bibfnamefont {A.}~\bibnamefont {Galiullin}}, \bibinfo {author} {\bibfnamefont {M.}~\bibnamefont {Ruf}}, \bibinfo {author} {\bibfnamefont {M.}~\bibnamefont {IJspeert}}, \bibinfo {author} {\bibfnamefont {T.~H.}\ \bibnamefont {Taminiau}}, \bibinfo {author} {\bibfnamefont {R.}~\bibnamefont {Hanson}},\ and\ \bibinfo {author} {\bibfnamefont {D.~R.}\ \bibnamefont {Englund}},\ }\bibfield  {title} {\bibinfo {title} {Optical coherence of diamond nitrogen-vacancy centers formed by ion implantation and annealing},\ }\href {https://doi.org/10.1103/PhysRevB.99.161203} {\bibfield  {journal} {\bibinfo  {journal} {Phys. Rev. B}\ }\textbf {\bibinfo {volume} {99}},\
  \bibinfo {pages} {161203} (\bibinfo {year} {2019})}\BibitemShut {NoStop}%
\bibitem [{\citenamefont {Uhlenbeck}\ and\ \citenamefont {Ornstein}(1930)}]{Uhlenbeck_1930_TheoryOfBrownianMotion}%
  \BibitemOpen
  \bibfield  {author} {\bibinfo {author} {\bibfnamefont {G.~E.}\ \bibnamefont {Uhlenbeck}}\ and\ \bibinfo {author} {\bibfnamefont {L.~S.}\ \bibnamefont {Ornstein}},\ }\bibfield  {title} {\bibinfo {title} {On the theory of the brownian motion},\ }\href {https://doi.org/10.1103/PhysRev.36.823} {\bibfield  {journal} {\bibinfo  {journal} {Phys. Rev.}\ }\textbf {\bibinfo {volume} {36}},\ \bibinfo {pages} {823} (\bibinfo {year} {1930})}\BibitemShut {NoStop}%
\bibitem [{\citenamefont {Gardiner}(2004)}]{Gardiner_2004_HandbookOfStochasticMethods}%
  \BibitemOpen
  \bibfield  {author} {\bibinfo {author} {\bibfnamefont {C.~W.}\ \bibnamefont {Gardiner}},\ }\href@noop {} {\emph {\bibinfo {title} {Handbook of stochastic methods for physics, chemistry and the natural sciences}}},\ \bibinfo {edition} {3rd}\ ed.,\ \bibinfo {series} {Springer Series in Synergetics}, Vol.~\bibinfo {volume} {13}\ (\bibinfo  {publisher} {Springer-Verlag},\ \bibinfo {address} {Berlin},\ \bibinfo {year} {2004})\ pp.\ \bibinfo {pages} {xviii+415}\BibitemShut {NoStop}%
\bibitem [{\citenamefont {Kampen}(2007)}]{vanKampen_2007_StochasticProcesses}%
  \BibitemOpen
  \bibfield  {author} {\bibinfo {author} {\bibfnamefont {N.~V.}\ \bibnamefont {Kampen}},\ }\bibfield  {title} {\bibinfo {title} {Chapter iii - stochastic processes},\ }in\ \href {https://doi.org/https://doi.org/10.1016/B978-044452965-7/50006-4} {\emph {\bibinfo {booktitle} {Stochastic Processes in Physics and Chemistry (Third Edition)}}},\ \bibinfo {series and number} {North-Holland Personal Library},\ \bibinfo {editor} {edited by\ \bibinfo {editor} {\bibfnamefont {N.~V.}\ \bibnamefont {Kampen}}}\ (\bibinfo  {publisher} {Elsevier},\ \bibinfo {address} {Amsterdam},\ \bibinfo {year} {2007})\ \bibinfo {edition} {third edition}\ ed.,\ pp.\ \bibinfo {pages} {52--72}\BibitemShut {NoStop}%
\end{thebibliography}

\begin{thebibliography}{6}%
\makeatletter
\providecommand \@ifxundefined [1]{%
 \@ifx{#1\undefined}
}%
\providecommand \@ifnum [1]{%
 \ifnum #1\expandafter \@firstoftwo
 \else \expandafter \@secondoftwo
 \fi
}%
\providecommand \@ifx [1]{%
 \ifx #1\expandafter \@firstoftwo
 \else \expandafter \@secondoftwo
 \fi
}%
\providecommand \natexlab [1]{#1}%
\providecommand \enquote  [1]{``#1''}%
\providecommand \bibnamefont  [1]{#1}%
\providecommand \bibfnamefont [1]{#1}%
\providecommand \citenamefont [1]{#1}%
\providecommand \href@noop [0]{\@secondoftwo}%
\providecommand \href [0]{\begingroup \@sanitize@url \@href}%
\providecommand \@href[1]{\@@startlink{#1}\@@href}%
\providecommand \@@href[1]{\endgroup#1\@@endlink}%
\providecommand \@sanitize@url [0]{\catcode `\\12\catcode `\$12\catcode `\&12\catcode `\#12\catcode `\^12\catcode `\_12\catcode `\%12\relax}%
\providecommand \@@startlink[1]{}%
\providecommand \@@endlink[0]{}%
\providecommand \url  [0]{\begingroup\@sanitize@url \@url }%
\providecommand \@url [1]{\endgroup\@href {#1}{\urlprefix }}%
\providecommand \urlprefix  [0]{URL }%
\providecommand \Eprint [0]{\href }%
\providecommand \doibase [0]{https://doi.org/}%
\providecommand \selectlanguage [0]{\@gobble}%
\providecommand \bibinfo  [0]{\@secondoftwo}%
\providecommand \bibfield  [0]{\@secondoftwo}%
\providecommand \translation [1]{[#1]}%
\providecommand \BibitemOpen [0]{}%
\providecommand \bibitemStop [0]{}%
\providecommand \bibitemNoStop [0]{.\EOS\space}%
\providecommand \EOS [0]{\spacefactor3000\relax}%
\providecommand \BibitemShut  [1]{\csname bibitem#1\endcsname}%
\let\auto@bib@innerbib\@empty
\bibitem [{\citenamefont {DeAbreu}\ \emph {et~al.}(2023)\citenamefont {DeAbreu}, \citenamefont {Bowness}, \citenamefont {Alizadeh}, \citenamefont {Chartrand}, \citenamefont {Brunelle}, \citenamefont {MacQuarrie}, \citenamefont {Lee-Hone}, \citenamefont {Ruether}, \citenamefont {Kazemi}, \citenamefont {Kurkjian}, \citenamefont {Roorda}, \citenamefont {Abrosimov}, \citenamefont {Pohl}, \citenamefont {Thewalt}, \citenamefont {Higginbottom},\ and\ \citenamefont {Simmons}}]{Deabreu2023waveguide}%
  \BibitemOpen
  \bibfield  {author} {\bibinfo {author} {\bibfnamefont {A.}~\bibnamefont {DeAbreu}}, \bibinfo {author} {\bibfnamefont {C.}~\bibnamefont {Bowness}}, \bibinfo {author} {\bibfnamefont {A.}~\bibnamefont {Alizadeh}}, \bibinfo {author} {\bibfnamefont {C.}~\bibnamefont {Chartrand}}, \bibinfo {author} {\bibfnamefont {N.~A.}\ \bibnamefont {Brunelle}}, \bibinfo {author} {\bibfnamefont {E.~R.}\ \bibnamefont {MacQuarrie}}, \bibinfo {author} {\bibfnamefont {N.~R.}\ \bibnamefont {Lee-Hone}}, \bibinfo {author} {\bibfnamefont {M.}~\bibnamefont {Ruether}}, \bibinfo {author} {\bibfnamefont {M.}~\bibnamefont {Kazemi}}, \bibinfo {author} {\bibfnamefont {A.~T.~K.}\ \bibnamefont {Kurkjian}}, \bibinfo {author} {\bibfnamefont {S.}~\bibnamefont {Roorda}}, \bibinfo {author} {\bibfnamefont {N.~V.}\ \bibnamefont {Abrosimov}}, \bibinfo {author} {\bibfnamefont {H.-J.}\ \bibnamefont {Pohl}}, \bibinfo {author} {\bibfnamefont {M.~L.~W.}\ \bibnamefont {Thewalt}}, \bibinfo {author} {\bibfnamefont {D.~B.}\ \bibnamefont {Higginbottom}},\ and\
  \bibinfo {author} {\bibfnamefont {S.}~\bibnamefont {Simmons}},\ }\bibfield  {title} {\bibinfo {title} {{Waveguide-integrated silicon T centres}},\ }\href {https://doi.org/10.1364/OE.482008} {\bibfield  {journal} {\bibinfo  {journal} {Optics Express}\ }\textbf {\bibinfo {volume} {31}},\ \bibinfo {pages} {15045} (\bibinfo {year} {2023})}\BibitemShut {NoStop}%
\bibitem [{\citenamefont {van Dam}\ \emph {et~al.}(2019)\citenamefont {van Dam}, \citenamefont {Walsh}, \citenamefont {Degen}, \citenamefont {Bersin}, \citenamefont {Mouradian}, \citenamefont {Galiullin}, \citenamefont {Ruf}, \citenamefont {IJspeert}, \citenamefont {Taminiau}, \citenamefont {Hanson},\ and\ \citenamefont {Englund}}]{Taminiau_2019_NVmatdevSDimprove}%
  \BibitemOpen
  \bibfield  {author} {\bibinfo {author} {\bibfnamefont {S.~B.}\ \bibnamefont {van Dam}}, \bibinfo {author} {\bibfnamefont {M.}~\bibnamefont {Walsh}}, \bibinfo {author} {\bibfnamefont {M.~J.}\ \bibnamefont {Degen}}, \bibinfo {author} {\bibfnamefont {E.}~\bibnamefont {Bersin}}, \bibinfo {author} {\bibfnamefont {S.~L.}\ \bibnamefont {Mouradian}}, \bibinfo {author} {\bibfnamefont {A.}~\bibnamefont {Galiullin}}, \bibinfo {author} {\bibfnamefont {M.}~\bibnamefont {Ruf}}, \bibinfo {author} {\bibfnamefont {M.}~\bibnamefont {IJspeert}}, \bibinfo {author} {\bibfnamefont {T.~H.}\ \bibnamefont {Taminiau}}, \bibinfo {author} {\bibfnamefont {R.}~\bibnamefont {Hanson}},\ and\ \bibinfo {author} {\bibfnamefont {D.~R.}\ \bibnamefont {Englund}},\ }\bibfield  {title} {\bibinfo {title} {Optical coherence of diamond nitrogen-vacancy centers formed by ion implantation and annealing},\ }\href {https://doi.org/10.1103/PhysRevB.99.161203} {\bibfield  {journal} {\bibinfo  {journal} {Phys. Rev. B}\ }\textbf {\bibinfo {volume} {99}},\
  \bibinfo {pages} {161203} (\bibinfo {year} {2019})}\BibitemShut {NoStop}%
\bibitem [{\citenamefont {Uhlenbeck}\ and\ \citenamefont {Ornstein}(1930)}]{Uhlenbeck_1930_TheoryOfBrownianMotion}%
  \BibitemOpen
  \bibfield  {author} {\bibinfo {author} {\bibfnamefont {G.~E.}\ \bibnamefont {Uhlenbeck}}\ and\ \bibinfo {author} {\bibfnamefont {L.~S.}\ \bibnamefont {Ornstein}},\ }\bibfield  {title} {\bibinfo {title} {On the theory of the brownian motion},\ }\href {https://doi.org/10.1103/PhysRev.36.823} {\bibfield  {journal} {\bibinfo  {journal} {Phys. Rev.}\ }\textbf {\bibinfo {volume} {36}},\ \bibinfo {pages} {823} (\bibinfo {year} {1930})}\BibitemShut {NoStop}%
\bibitem [{\citenamefont {Delteil}\ \emph {et~al.}(2024)\citenamefont {Delteil}, \citenamefont {Buil},\ and\ \citenamefont {Hermier}}]{Delteil_2024_photonstatisticsdiffusiveemittersOUMarkov}%
  \BibitemOpen
  \bibfield  {author} {\bibinfo {author} {\bibfnamefont {A.}~\bibnamefont {Delteil}}, \bibinfo {author} {\bibfnamefont {S.}~\bibnamefont {Buil}},\ and\ \bibinfo {author} {\bibfnamefont {J.-P.}\ \bibnamefont {Hermier}},\ }\bibfield  {title} {\bibinfo {title} {Photon statistics of resonantly driven spectrally diffusive quantum emitters},\ }\href {https://doi.org/10.1103/PhysRevB.109.155308} {\bibfield  {journal} {\bibinfo  {journal} {Phys. Rev. B}\ }\textbf {\bibinfo {volume} {109}},\ \bibinfo {pages} {155308} (\bibinfo {year} {2024})}\BibitemShut {NoStop}%
\bibitem [{\citenamefont {Gardiner}(2004)}]{Gardiner_2004_HandbookOfStochasticMethods}%
  \BibitemOpen
  \bibfield  {author} {\bibinfo {author} {\bibfnamefont {C.~W.}\ \bibnamefont {Gardiner}},\ }\href@noop {} {\emph {\bibinfo {title} {Handbook of stochastic methods for physics, chemistry and the natural sciences}}},\ \bibinfo {edition} {3rd}\ ed.,\ \bibinfo {series} {Springer Series in Synergetics}, Vol.~\bibinfo {volume} {13}\ (\bibinfo  {publisher} {Springer-Verlag},\ \bibinfo {address} {Berlin},\ \bibinfo {year} {2004})\ pp.\ \bibinfo {pages} {xviii+415}\BibitemShut {NoStop}%
\bibitem [{\citenamefont {Kampen}(2007)}]{vanKampen_2007_StochasticProcesses}%
  \BibitemOpen
  \bibfield  {author} {\bibinfo {author} {\bibfnamefont {N.~V.}\ \bibnamefont {Kampen}},\ }\bibfield  {title} {\bibinfo {title} {Chapter iii - stochastic processes},\ }in\ \href {https://doi.org/https://doi.org/10.1016/B978-044452965-7/50006-4} {\emph {\bibinfo {booktitle} {Stochastic Processes in Physics and Chemistry (Third Edition)}}},\ \bibinfo {series and number} {North-Holland Personal Library},\ \bibinfo {editor} {edited by\ \bibinfo {editor} {\bibfnamefont {N.~V.}\ \bibnamefont {Kampen}}}\ (\bibinfo  {publisher} {Elsevier},\ \bibinfo {address} {Amsterdam},\ \bibinfo {year} {2007})\ \bibinfo {edition} {third edition}\ ed.,\ pp.\ \bibinfo {pages} {52--72}\BibitemShut {NoStop}%
\end{thebibliography}
\end{document}